\documentclass[reprint,aps,prb]{revtex4-2}
\usepackage{graphicx}
\usepackage{dcolumn}
\usepackage{bm}
\usepackage{amsmath,amsfonts,amssymb,braket}
\usepackage{color}
\usepackage{xtab,afterpage}
\usepackage{lipsum}
\usepackage{here}
\usepackage{longtable}

\everymath{\displaystyle}

\makeatletter
\def\LT@LR@e{\LTleft\z@   \LTright\z@}%
\makeatother

\begin{document}

\title{Classification of multiorbital superconducting state based on augmented multipoles}

\author{Akimitsu Kirikoshi}
\author{Satoru Hayami}
\affiliation{Department of Physics, Hokkaido University, Sapporo 060-0810, Japan}

\date{\today}

\begin{abstract}
  The pairing interactions between electrons play an essential role in determining the properties in superconducting states. Recently, a plethora of unconventional superconducting states has been extensively explored, which often emerge owing to multipole fluctuations in the vicinity of multipole orders.
  We classify such superconducting states from the viewpoint of the multipole degrees of freedom by extending its representation to
  Nambu space. 
  We clarify that under the crystallographic point group, arbitrary Cooper pairs between electrons with any angular momenta are systematically classified by four types of multipoles: electric, magnetic, magnetic toroidal, and electric toroidal.
  As examples, we apply our formulation to an $sp$-orbital electron system, which potentially exhibits exotic Cooper pairs under polar and axial point groups. Our systematic classification will be
  useful in characterizing unconventional superconducting states in multiorbital systems.
\end{abstract}

\maketitle

\section{Introduction}
\label{sec:intro}

Exploring a new type of pairings in superconductivity (SC) is one of the challenging issues in condensed matter physics.
Thanks to its internal degrees of freedom including the node structure and spin dependence, various physical properties can happen in the supercurrent and magnetoelectric responses.
In classifying such physical properties, the symmetry argument is useful; the nature of the pair potential is characterized by the irreducible representations under the crystallographic point group according to its wavenumber and spin dependence. 
The pioneering studies on the symmetry classification of the pair potential have been done by Volovik-Gor'kov~\cite{JETPLett.39.674} and Sigrist-Ueda~\cite{RevModPhys.63.239}, where the orbital degree of freedom is neglected.

The type of the SC state is dependent on the mechanism of the Cooper pair formation. In the conventional SC described by BCS theory~\cite{PhysRev.108.1175}, the Cooper pair is constructed through the electron-phonon coupling. Meanwhile, the anisotropic SC state under the strong electron correlation is thought to be realized as a consequence of spin fluctuations by the short-range repulsive interaction. 
In U-based ferromagnetic SCs~\cite{UGe2, URhGe_1, URhGe_2, UCoGe, UIr, JPSJ.88.022001}, a Cooper pairing is generated by ferromagnetic spin fluctuations analogous to superfluid ${}^{3}\mathrm{He}$~\cite{3He-A}. Antiferromagnetic spin fluctuations develop in d-wave SC in high-$T_{\mathrm{c}}$ cuprates, and heavy-fermion systems containing Ce~\cite{CeCu2Si2, CeCu2Ge2, Ce2Rh2Si2, Ce2Pd2Si2, CeIn3, CeNi2Ge2, CeRhIn5, CeCoIn5, Ce2RhIn8, Ce2Ni3Ge5, CeNiGe3, CeCoGe3, CePd5Al2} or actinoids~\cite{PuCoGa5, PuRhGa5, NpPd5Al2}.
In addition, there are other mechanisms forming the SC states, such as charge fluctuations arising from valence fluctuations in heavy-fermion systems~\cite{PhysRevB.30.1182, J_Phys_Condens_Matt_19.125201, PhysRevB.69.024508, PhysicaB.259.1, science.1091648} and electric quadrupole (orbital) fluctuations inducing an $s_{++}$-wave SC state in iron-based pnictides~\cite{PhysRevLett.104.157001,PhysRevB.81.054518,PhysRevLett.109.137001}.
In this way, the correlation between SC and other electric or magnetic degrees of freedom can happen in strongly correlated electron systems, which has attracted much attention from researchers in the field for many years.

More recently, various unconventional SC phases accompanying complicated spin and orbital degrees of freedom, i.e., multipole degrees of freedom, have been discovered.
For example, $\mathrm{CeCu}_{2}\mathrm{Si}_{2}$ is one of the typical examples to exhibit unconventional $s$-wave superconductivity~\cite{CeCu2Si2,PhysRevLett.82.5353,RevModPhys.81.1551,PhysRevLett.112.067002,PhysRevB.94.054514,sciadv.1601667,pnas.1720291115} through multipole fluctuations~\cite{JPSJ.88.063701}. 
In addition, in $\mathrm{Pr}T_{2}\mathrm{Zn}_{20}$($T=\mathrm{Ru,Ir}$)~\cite{JPSJ.79.033704} and $\mathrm{Pr}T_{2}\mathrm{Al}_{20}$($T=\mathrm{V,Ti}$)~\cite{JPSJ.80.063701,PhysRevLett.113.267001}, unconventional SC states through quadrupole fluctuations have been investigated in both experiments and theory. 
Besides, various unconventional SC states as a consequence of the entanglement of spin and orbital degrees of freedom have been investigated; noncentrosymmetric SC $\mathrm{Sr}_{1-x}\mathrm{Ca}_{x}\mathrm{TiO}_{\delta}$ shows a ferroelectric SC state~\cite{nphys4085}.
For $\mathrm{Ba}(\mathrm{Fe}_{1-x}\mathrm{Co}_{x})\mathrm{As}_{2}$ in the overdoped region, the coexistence of electric hexadecapole ordering and SC owing to electric quadrupole fluctuations has been pointed out~\cite{JPSJ.86.064706}.
In the time-reversal broken system, the possibility of the Bogoliubov Fermi surface, which is topologically protected~\cite{PhysRevLett.118.127001, PhysRevB.98.224509}, has been discussed in $\mathrm{FeSe}_{1-x}\mathrm{S}_{x}$~\cite{pnas.2208276120, nature.11.523, PhysRevB.102.064504}, where the role of the magnetic toroidal dipole on the formation of the nematic Bogoliubov Fermi surface has been suggested~\cite{wu2023nematic}.
Furthermore, new pairings have been proposed for systems with strong spin--orbit interactions.
In half-Heusler semimetals $R\mathrm{PtBi}$ $(R=\mathrm{La}, \mathrm{Y}, \mathrm{Lu})$ and $R\mathrm{PdBi}$ $(R=\mathrm{Er}, \mathrm{Lu}, \mathrm{Ho}, \mathrm{Y}, \mathrm{Sm}, \mathrm{Tb}, \mathrm{Dy}, \mathrm{Tm})$, it has been considered that $J=2$ quintet states and $J=3$ septet states can be realized by pairing between $j=3/2$ fermion states (we denote $J$ as a total angular momentum of Cooper pairs, whereas $j$ as that of electrons)~\cite{PhysRevLett.116.177001}.
The FFLO state is another unconventional SC state, whose appearance has been discussed under magnetic ordering~\cite{PhysRevB.93.224507, Comm.Phys.5.39}. 
Meanwhile, the systematic classification of these unconventional SC states has not been fully elucidated.

In order to achieve a unified description of the unconventional SC states, we introduce the concept of electronic augmented multipoles, which has been developed in describing unconventional parity-breaking states and spin--orbital entangled states in the normal space~\cite{JPSJ.87.033709, PhysRevB.98.165110, PhysRevB.98.245129, PhysRevB.107.195118}.
According to the spatial inversion (SI) and time-reversal (TR) parities, four types of multipoles, electric, magnetic, magnetic toroidal, and electric toroidal, are defined, which constitute a complete basis set~\cite{JPSJ.89.104704}.
The advantage of using these multipole bases is to systematically classify complex electronic order parameters under crystallographic (magnetic) point groups, which provides possible cross-correlated responses and transports~\cite{PhysRevB.98.165110, PhysRevB.104.054412}.
In this sense, the systematic classification of SC order parameters based on augmented multipoles is useful to not only organize unconventional SC states but also uncover further intriguing physical responses.

In this paper, we apply the multipole basis to the order parameter space in the SC state, i.e., the Nambu space.
By deriving the expression in terms of the transformation of the pair potential, we systematically obtain the correspondence between multipoles and SC order parameters in Nambu space.
We clarify that arbitrary Cooper pairs between electrons with any angular momenta can be described by multipoles in a unified manner~\cite{SciPostPhys.12.2.057}. 
Our systematic classification of SC order parameters might be regarded as the extension of the previous study~\cite{PhysRevB.94.174513}, because our case includes the odd-parity Cooper pairing as well as the even-parity one. 
We also discuss the relation between multipole fluctuations and the pair potential in the multiorbital systems from the symmetry viewpoint~\cite{PhysRevResearch.2.033225}. 
Moreover, we present two examples by focusing on the Cooper pairs in the polar and axial crystal systems in order to demonstrate unconventional behaviors of the pair potential.

This paper is organized as follows.
In Sec.~\ref{sec:formalism}, we show how to transform the multipole description in the normal space to the Nambu space, which enables us to systematically describe the pair potential in multiorbital systems.
In Sec.~\ref{sec:application}, we apply the multipole basis to a specific $sp$-orbital system and classify the pair potential according to the irreducible representation.
We also present the model analysis for unconventional SC states and their stability on the two-dimensional triangular lattice with (i) a polar field and (ii) an axial field.
Section~\ref{sec:summary} is a summary of this paper.

\section{Description of Pair Potential based on Multipoles}
\label{sec:formalism}

In this section, we show a systematic way of describing the pair potential based on the symmetry-adapted multipole basis and discuss the relation to the interaction channel under multipole fluctuations.

\subsection{Complete multipole basis}

First, we introduce the complete multipole basis in the normal space.
As shown in Ref.~\cite{JPSJ.89.104704}, the multipole basis set is complete to describe the normal space; 
they can be expressed in terms of electric (E, $\mathcal{T}$-even polar), magnetic (M, $\mathcal{T}$-odd axial), magnetic toroidal (MT, $\mathcal{T}$-odd polar), and electric toroidal (ET, $\mathcal{T}$-even axial) multipole bases, $Q_{lm},M_{lm},T_{lm}$, and $G_{lm}$, respectively, in the rotation group, where $\mathcal{T}$ represents the TR operation. 
We here apply such multipole bases in the normal space to those in the Nambu space in order to characterize arbitrary SC order parameters.

The multipole basis including both spin and orbital degrees of freedom can be obtained by the direct product of the spinless multipole basis $X_{lm}^{(\mathrm{orb})}(X=Q,M,T,G)$ and Pauli matrices 
\begin{equation}
  \sigma_{sk}=
  \begin{cases}
    \sigma_{0} & (s,k)=(0,0)
    \\
    \sigma_{z} & (s,k)=(1,0)
    \\
    \mp \frac{\sigma_{x}\pm i\sigma_{y}}{\sqrt{2}} & (s,k)=(1,\pm 1)
    \\
  \end{cases}
  \label{eq:pauli}
\end{equation}
($\sigma_0$ is the identity matrix) and the addition rule of the angular momentum as 
\begin{equation}
  \hat{X}_{lm}(s,k)=i^{s+k}
  \sum_{n}\braket{l+k,m-n;sn|lm}
  X_{l+k,m-n}^{(\mathrm{orb})}\sigma_{sn},
  \label{eq:spinful_def}
\end{equation}
where $\braket{l_{1}m_{1};l_{2}m_{2}|lm}$ is the Clebsch--Gordon coefficient.
$l$ and $m$ represent the rank of the multipole and its component ($-l \leq m \leq l$), respectively.
Here, $(s,k)=(0,0)$ is the multipole basis for charge (or spinless) sector, whereas $(s,k)=(1,\pm 1)$ and $(1,0)$ are those for spin (or spinful) sector.
See Appendix~\ref{app:MP} for the expression of $X_{lm}^{(\mathrm{orb})}$ and the correspondence between $\hat{X}_{lm}(1,k)$ and $X_{lm}^{(\mathrm{orb})}$ in the case of $sp$-orbital space.
Each $\hat{X}_{lm}(s,k)$ has definite parity for SI and TR operations as follows: 
\begin{equation}
  \begin{aligned}
    &\, \mathcal{P}:\hat{X}_{lm}(s,k)\mapsto (-1)^{P_{\nu}}\hat{X}_{lm}(s,k),
    \\
    &\, \mathcal{T}:\hat{X}_{lm}(s,k)\mapsto (-1)^{T_{\nu}}\hat{X}_{lm}(s,k),
  \end{aligned}
  \label{eq:parity_MP}
\end{equation}
and orthogonal from each other.
Here, $(-1)^{P_{\nu}}=+1(-1)$ represents an even(odd)-parity multipole, whereas $(-1)^{T_{\nu}}=+1(-1)$ represents an electric-type (a magnetic-type) multipole;
the subscript $\nu$ denotes a set of $(l,m,s,k)$ and the types of multipoles.
 
By using the complete multipole basis set, any electronic degrees of freedom in the normal space are expressed by any of four multipoles.
In order to demonstrate that, we consider the $s$- and $p$-orbital systems with the total angular momenta $j=1/2$ and $j=3/2$ as an example.
The active multipoles in this Hilbert space are summarized in Table~\ref{tab:sp_multipole}, where higher-rank multipoles with $l\geq 2$ as well as toroidal-type multipoles are activated when the $p$ orbital is involved.
The active multipole means that the matrix elements of the corresponding multipole operator have nonzero values in the Hilbert space of the system of interest.
In the Hilbert space between different $j$ of the same orbitals (the fourth row), even-parity ET dipole and MT quadrupole are activated, whereas, in the Hilbert space between different orbitals of the same $j$ (the fifth row), odd-parity E and MT dipoles are activated.
We will see that these multipole degrees of freedom also constitute a complete set in the Nambu space, which is useful to describe arbitrary Cooper pairs in multiorbital systems.
In the following subsection, we will apply the multipole representations to the Nambu space, and then, expand the pair potential in terms of the complete multipole basis.

\begin{table*}[tbp]
  \caption{
  Active multipoles in $s$- and $p$-orbitals with $j=1/2$ and $j=3/2$.
  Here, E, M, MT, and ET represent electric, magnetic, magnetic toroidal, and electric toroidal multipoles, respectively.
  }
  \begin{ruledtabular}
    \begin{tabular}{ccccccc}
      $j-j$ & orbital & $l=0$ & $l=1$ & $l=2$ & $l=3$ 
      \\
      \hline
      $1/2-1/2$ & $s-s,p-p$ & E & M & -- & -- 
      \\
      $3/2-3/2$ & $p-p$ & E & M & E & M 
      \\
      $1/2-3/2$ & $p-p$ & -- & M/ET & E/MT & -- 
      \\
      $1/2-1/2$ & $s-p$ & M/ET & E/MT & -- & -- 
      \\
      $1/2-3/2$ & $s-p$ & -- & E/MT & M/ET & -- 
      \\
    \end{tabular}
  \end{ruledtabular}
  \label{tab:sp_multipole}
\end{table*}

Hereafter, we adopt the spin--orbital basis $\ket{\phi_{l}\sigma}$ (L-S coupling scheme), where $\phi_{l}$ represents the rank-$l$ orbital component.
It is noted that the following discussion can also be applied to the basis characterized by total angular momentum $\ket{jj_{z}}$ (j-j coupling scheme) by taking the appropriate unitary transformation.
For example, for the $p$-orbital system, the unitary transformation from the orbital--spin basis $\{\ket{p_{x}\uparrow}, \ket{p_{y}\uparrow}, \ket{p_{z}\uparrow}, \ket{p_{x}\downarrow}, \ket{p_{y}\downarrow}, \ket{p_{z}\downarrow}\}$ to total angular momentum basis $\{\ket{jj_{z}}\}$ is given by 
\begin{equation*}
  \hat{\mathcal{U}}=
  \begin{bmatrix}
    0 & 0 & -\frac{1}{\sqrt{3}} & -\frac{1}{\sqrt{3}} & \frac{i}{\sqrt{3}} & 0 
    \\
    -\frac{1}{\sqrt{3}} & -\frac{i}{\sqrt{3}} & 0 & 0 & 0 & \frac{1}{\sqrt{3}} 
    \\
    -\frac{1}{\sqrt{2}} & \frac{i}{\sqrt{2}} & 0 & 0 & 0 & 0 
    \\
    0 & 0 & \frac{2}{\sqrt{6}} & -\frac{1}{\sqrt{6}} & \frac{i}{\sqrt{6}} & 0 
    \\
    \frac{1}{\sqrt{6}} & \frac{i}{\sqrt{6}} & 0 & 0 & 0 & \frac{2}{\sqrt{6}} 
    \\
    0 & 0 & 0 & \frac{1}{\sqrt{2}} & \frac{i}{\sqrt{2}} & 0 
    \\
  \end{bmatrix}.
\end{equation*}

\subsection{Multipole-fluctuated interaction}

Next, we introduce a multipole order parameter $\hat{O}$ in the normal space~\cite{PhysRevResearch.2.033225}:
\begin{equation}
  \hat{O}=\sum_{\bm{k}}\sum_{\alpha\beta}\Lambda_{\alpha\beta}(\bm{k})\hat{c}_{\bm{k}\alpha}^{\dag}\hat{c}_{\bm{k}\beta},
  \label{eq:order_parameter}
\end{equation}
where $\hat{\Lambda}^{\dag}(\bm{k})=\hat{\Lambda}(\bm{k})$ with $[\hat{\Lambda}(\bm{k})]_{\alpha\beta}\equiv\Lambda_{\alpha\beta}(\bm{k})$ due to the hermiticity of $\hat{O}$.
$\hat{c}_{\bm{k}\alpha}^{\dag},\hat{c}_{\bm{k}\beta}$ is a creation/annihilation operator for electrons with the wavenumber $\bm{k}$ and $\alpha,\beta$ that represents a set of the internal degrees of the freedom of electrons, i.e., orbital and spin.
$\hat{\Lambda}(\bm{k})$ can be expanded by using the complete multipole basis as 
\begin{equation}
  \hat{\Lambda}(\bm{k})=\sum_{\nu=1}^{n^{2}}f_{\nu}(\bm{k})\hat{\chi}_{\nu}.
  \label{eq:decomposition_Lambda}
\end{equation}
Here, $n$ is the number of the electronic degrees of freedom (the dimension of the matrix), $\{\hat{\chi}_{\nu}\}_{\nu=1}^{n^{2}}$ are the multipole bases corresponding to $\hat{X}_{lm}(s,k)$ introduced in the previous subsection; $\nu$ is the index to distinguish the multipoles.
$f_{\nu}(\bm{k})$ is a real function due to the hermiticity of $\hat{\Lambda}(\bm{k})$ and $\hat{\chi}_{\nu}$.

It is noted that $\hat{O}$ can have several forms even for describing the same multipoles according to the representation of $\hat{\Lambda}(\bm{k})$, since the several choices of $f_{\nu}(\bm{k})$ and $\hat{\chi}_{\nu}$ can be possible when the multiorbital degree of freedom is considered.
For example, let us consider the case of the electric dipole corresponding to the time-reversal-even polar vector.
In this case, one of the representations of $\hat{\Lambda}(\bm{k})$ is given by $k_{x}\sigma_{y}-k_{y}\sigma_{x}$ by taking $f_{\nu}(\bm{k})=(-k_{y}, k_{x}, 0)$ and $\hat{\chi}_{\nu}=(\sigma_{x}, \sigma_{y}, \sigma_{z})$ for the single-orbital (orbital independent) system.
On the other hand, $\hat{\Lambda}(\bm{k})$ can be also expressed as $Q_{z}^{(\mathrm{a})}\sigma_{0}$ by taking $f_{\nu}(\bm{k})=1$ and $\hat{\chi}_{\nu}=Q_{z}^{(\mathrm{a})}$ when the $s$ and $p$ orbital degrees of freedom is considered; $Q_{z}^{(\mathrm{a})}$ is the spinless electric dipole activated in the $sp$-orbital space, as shown in Table~\ref{tab:sp_multipole}. 
Although both of them correspond to the same multipole order parameter $\hat{O}$ within Eq.~(\ref{eq:order_parameter}), they lead to different SC states:
The former leads to the $p$-wave SC state~\cite{PhysRevResearch.2.033225}, while the latter leads to the $s$-wave SC state; we will present the result for the latter case in Sec.~\ref{sec:application}. 
Such a situation occurs when the multipole can be active in both single-orbital space and multiorbital space;
even-rank and odd-rank E multipole and MT multipole (except for MT monopole), odd-rank M multipole, and even-rank ET multipole belong to this category~\cite{PhysRevB.98.165110}.

Meanwhile, the even-rank M multipoles and odd-rank ET multipoles cannot be described by the single-orbital space~\cite{PhysRevB.98.165110}. 
Thus, unconventional SC states with the even-rank M multipoles, odd-rank ET multipoles, and MT monopole only occur in the multiorbital system, which have never been captured within the single-orbital model~\cite{PhysRevResearch.2.033225,PhysRevLett.115.207002}.
The relation between $\hat{\Lambda}(\bm{k})$ and $(f_{\nu}(\bm{k}),\hat{\chi}_{\nu})$ is summarized as shown in Table~\ref{tab:multipole_symmetry}, where the expressions in the fifth column are brought about by the multiorbital degree of freedom.
From the table, one finds that $\hat{\chi}_{\nu}$ should be even-parity for the single-orbital system, while there is no constraint for the multiorbital system.

\begin{table}[htbp]
  \caption{
  The correspondence between $\hat{\Lambda}(\bm{k})$ and $(f_{\nu}(\bm{k}),\hat{\chi}_{\nu})$ in Eq.~(\ref{eq:decomposition_Lambda}).
  The second column shows the parity of spatial inversion for the real function $f_{\nu}(\bm{k})$: E(even) or O(odd).
  The third column shows the symmetry of an atomic multipole $\hat{\chi}_{\nu}$: EE(even-parity electric), OM(odd-parity magnetic), EM(even-parity magnetic), or OE(odd-parity electric), where electric (magnetic) means that the parity for time-reversal operation is even (odd).
  The last two columns show the example of the expressions for $\hat{\Lambda}(\bm{k})$ in the single-orbital and multiorbital systems, where $\bm{l}$ is an orbital angular momentum, $\bm{Q}(\bm{T})$ is an electric (a magnetic toroidal) dipole induced by the real (imaginary) hybridization between orbitals with different parity.
  }
  \begin{ruledtabular}
    \begin{tabular}{lccll}
      $\hat{\Lambda}(\bm{k})$ & $f_{\nu}(\bm{k})$ & $\hat{\chi}_{\nu}$ & single-orbital & multiorbital
      \\
      \hline
      EE & E & EE & $1, k^{2}$ & $\bm{l}\times\bm{\sigma}$
      \\
       & O & OM & -- & $\bm{k}\cdot\bm{T}$
      \\
      EM & E & EM & $\bm{\sigma}, (k_{x}^{2}-k_{y}^{2})\sigma_{z}$ & $\bm{l}$
      \\
       & O & OE & -- & $\bm{k}\cdot\bm{Q}$
      \\
      OE & E & OE & -- & $\bm{Q}$
      \\
       & O & EM & $\bm{k}\times\bm{\sigma}, \bm{k}\cdot\bm{\sigma}$ & $\bm{k}\times\bm{l}$ 
      \\
      OM & E & OM & -- & $\bm{Q}\cdot\bm{\sigma}$
      \\
       & O & EE & $\bm{k}$ & $\bm{k}\cdot(\bm{l}\times\bm{\sigma})$
      \\
    \end{tabular}
  \end{ruledtabular}
  \label{tab:multipole_symmetry}
\end{table}

Next, we introduce the interaction arising from the multipole fluctuations 
\begin{equation}
  \hat{\mathcal{H}}_{\mathrm{eff}}=\frac{1}{2N}\sum_{\bm{q}}V_{\bm{q}}\hat{O}(\bm{q})\hat{O}(-\bm{q}),
  \label{eq:fluctuation_interaction}
\end{equation}
where $\hat{O}(\bm{q})=\hat{O}^{\dag}(-\bm{q})$ is the Fourier transform in the momentum space of the order parameter:
\begin{equation}
  \hat{O}(\bm{q})=\frac{1}{2}\sum_{\bm{k}}\sum_{\alpha,\beta}[\Lambda_{\alpha\beta}(\bm{k}+\bm{q})+\Lambda_{\alpha\beta}(\bm{k})]\hat{c}_{\bm{k}+\bm{q},\alpha}^{\dag}\hat{c}_{\bm{k},\beta},
  \label{eq:order_parameter_q}
\end{equation}
and $N$ is the number of the site.
$\hat{O}(\bm{q}=\bm{0})$ corresponds to $\hat{O}$ introduced in Eq.~(\ref{eq:order_parameter}).
The multipole-fluctuated interaction in Eq.~(\ref{eq:fluctuation_interaction}) originates from the electron--electron Coulomb repulsive interaction or electron--phonon attractive interaction.
To simplify the discussion, we consider only the effective interaction in Eq.~(\ref{eq:fluctuation_interaction}) to pairing channels with zero center-of-mass momentum:
\begin{equation}
  \hat{\mathcal{H}}_{\mathrm{eff}}=\frac{1}{2N}\sum_{\bm{k},\bm{k}^{\prime}}\sum_{\alpha,\beta,\gamma,\delta}V_{\alpha\beta;\gamma\delta}(\bm{k},\bm{k}^{\prime})\hat{c}_{\bm{k}\alpha}^{\dag}\hat{c}_{-\bm{k}\beta}^{\dag}\hat{c}_{-\bm{k}^{\prime}\gamma}\hat{c}_{\bm{k}^{\prime}\delta}.
  \label{eq:interaction_pair}
\end{equation}
That is, we omit the possibility of the FFLO state and a helical SC state realized under the magnetic field~\cite{Agterberg2012}.
The pairing interaction vertex is given by  
\begin{equation}
  \begin{aligned}
    &\, V_{\alpha\beta;\gamma\delta}(\bm{k},\bm{k}^{\prime})
    \\
    =&\, \frac{1}{8}\{V_{\bm{k}-\bm{k}^{\prime}}[\Lambda_{\alpha\delta}(\bm{k})+\Lambda_{\alpha\delta}(\bm{k}^{\prime})][\Lambda_{\beta\gamma}(-\bm{k})+\Lambda_{\beta\gamma}(-\bm{k}^{\prime})]
    \\
    &\, -V_{\bm{k}+\bm{k}^{\prime}}[\Lambda_{\alpha\gamma}(\bm{k})+\Lambda_{\alpha\gamma}(-\bm{k}^{\prime})][\Lambda_{\beta\delta}(-\bm{k})+\Lambda_{\beta\delta}(\bm{k}^{\prime})]\}.
    \label{eq:vertex_pair}
  \end{aligned}
\end{equation}
The vertex satisfies the symmetry from the Pauli exclusion principle: $V_{\alpha\beta;\gamma\delta}(\bm{k},\bm{k}^{\prime})=-V_{\beta\alpha;\gamma\delta}(-\bm{k},\bm{k}^{\prime})=-V_{\alpha\beta;\delta\gamma}(\bm{k},-\bm{k}^{\prime})=V_{\gamma\delta;\alpha\beta}^{*}(\bm{k}^{\prime},\bm{k})$.

\subsection{Cooper channel in terms of multipoles}

Next, we describe the order parameter of the SC state based on the multipole basis.
Let us investigate the transformation property of the electron operator $\hat{c}_{\bm{k}\alpha}^{\dag}$ with $\alpha=(\phi_{l},\sigma)$.
Under the TR operation $\mathcal{T}$ and crystal symmetry operation $g\in G$ ($G$ represents the crystallographic point group): 
\begin{subequations}
  \label{eq:ransformation}
  \begin{align}
    &\, 
    \mathcal{T}\hat{c}_{\bm{k}\alpha}^{\dag}\hat{\mathcal{T}}^{-1}=(i\check{\mathcal{U}}_{T})_{\alpha\beta}\hat{c}_{-\bm{k}\beta}^{\dag},
    \\
    &\, 
    g\hat{c}_{\bm{k}\alpha}^{\dag}g^{-1}=[\hat{U}(g)]_{\alpha\beta}\hat{c}_{\bm{k}^{*}\beta}^{\dag},
  \end{align}
\end{subequations}
where summations over repeated arguments are implied.
The matrix $i\check{\mathcal{U}}_{T}$ is defined by 
\begin{equation}
  i\check{\mathcal{U}}_{T}=I_{0}\otimes(i\sigma_{y}),
  \label{eq:unitary_TR}
\end{equation}
with an identity operator in orbital space $I_{0}$, and the $y$ component of the Pauli matrix in spin space $\sigma_{y}$, whereas $\hat{U}(g)$ is the matrix that represents the acting of $g$ on the spin--orbital space and $\bm{k}^{*}=g\bm{k}$.
According to the transformation in Eq.~(\ref{eq:ransformation}), one finds that bilinear operators $\hat{\rho}_{\bm{k},\alpha\beta}\equiv \hat{c}_{\bm{k}\alpha}^{\dag}\hat{c}_{\bm{k}\beta}$ and $\hat{F}_{\bm{k},\alpha\beta}\equiv(i\check{\mathcal{U}}_{T})_{\beta\gamma}\hat{c}_{\bm{k}\alpha}^{\dag}\hat{c}_{-\bm{k}\gamma}^{\dag}$ transform such as 
\begin{equation}
  \begin{aligned}
    g\hat{\rho}_{\bm{k},\alpha\beta}g^{-1}=&\, [\hat{U}(g)\hat{\rho}_{\bm{k}^{*}}\hat{U}^{\dag}(g)]_{\alpha\beta},
    \\
    g\hat{F}_{\bm{k},\alpha\beta}g^{-1}=&\, [\hat{U}(g)\hat{F}_{\bm{k}^{*}}\hat{U}^{\dag}(g)]_{\alpha\beta},
    \\
  \end{aligned}
  \label{eq:transformation_op}
\end{equation}
respectively.
In other words, $\hat{\rho}_{\bm{k},\alpha\beta}$ and $\hat{F}_{\bm{k},\alpha\beta}$ transform in the same way~\cite{PhysRevLett.115.207002}.
Since the two-body interaction Hamiltonian can be described by the multipole--multipole interaction such as Eq.~(\ref{eq:fluctuation_interaction})~\cite{RevModPhys.81.807,PhysRevB.93.075101}, this result indicates that Eq.~(\ref{eq:interaction_pair}) can be rewritten in terms of the multipole operator in anomalous space defined by 
\begin{equation}
  \check{O}^{\dag}=\sum_{\bm{k}}\sum_{\alpha\beta\gamma}\Lambda_{\alpha\beta}(\bm{k})(i\check{\mathcal{U}}_{T})_{\beta\gamma}\hat{c}_{\bm{k}\alpha}^{\dag}\hat{c}_{-\bm{k}\gamma}^{\dag},
  \label{eq:pair_operator}
\end{equation}
and its hermitian conjugate $\check{O}$ such as
\begin{equation}
  \mathcal{H}_{\mathrm{eff}}=\frac{1}{2N}\sum_{\mu\nu}V_{\mu\nu}(\check{O}_{\mu}^{\dag}\check{O}_{\nu}+\check{O}_{\nu}^{\dag}\check{O}_{\mu}).
\end{equation}
Here and hereafter, we use the notation $\check{}$ for the operators, which implicitly indicates the quantities in anomalous space.

\subsection{Multipole description of pair potential}

Now, we apply the multipole basis to the pair potential in anomalous space.
In the following sections, we will see that our formalism gives the complete representation of the SC parameters in multiorbital systems.
To this end, we introduce the pair potential as 
\begin{equation}
  \begin{aligned}
    [\check{\Delta}(\bm{k})]_{\alpha\beta}
    \equiv&\, 
    \frac{1}{N}\sum_{\bm{k}^{\prime}}\sum_{\gamma\delta}V_{\alpha\beta;\gamma\delta}(\bm{k},\bm{k}^{\prime})\braket{\hat{c}_{-\bm{k}^{\prime}\gamma}\hat{c}_{\bm{k}^{\prime}\delta}}
    \\
    =&\, \frac{1}{N}\sum_{\bm{k}^{\prime}}\sum_{\gamma\delta}V_{\alpha\beta;\gamma\delta}(\bm{k},\bm{k}^{\prime})F_{\gamma\delta}(-\bm{k}^{\prime}),
  \end{aligned}
  \label{eq:pair_definition}
\end{equation}
where $\braket{\cdots}$ denotes the thermal average and $F_{\alpha\beta}(\bm{k})\equiv\braket{\hat{c}_{\bm{k}\alpha}\hat{c}_{-\bm{k}\beta}}=-F_{\beta\alpha}(-\bm{k})$.
From the definition, there is a restriction in terms of the pair potential to satisfy the fermionic antisymmetric property $[\check{\Delta}(\bm{k})]^{\mathrm{T}}=-\check{\Delta}(-\bm{k})$
.

The transformation in Eq.~(\ref{eq:transformation_op}) holds for the pair potential, which is represented as 
\begin{equation}
  \check{\Delta}(\bm{k})=\hat{\Delta}(\bm{k})(i\check{\mathcal{U}}_{T}).
  \label{eq:pair_potential_expression}
\end{equation}
Then, $\hat{\Delta}(\bm{k})$ is followed by the unitary transformation as in the normal space:
\begin{equation}
  g:\hat{\Delta}(\bm{k})\mapsto\hat{U}(g)\hat{\Delta}(\bm{k}^{*})\hat{U}^{\dag}(g).
  \label{eq:transformation_pair}
\end{equation}
In this notation, we directly apply the multipole basis in the normal space to the pair potential $\hat{\Delta}(\bm{k})$.

Next, we denote the pair potential $\hat{\Delta}(\bm{k})$ satisfying the transformation in Eq.~(\ref{eq:transformation_pair}).
Supposing the system with $n$ electronic degrees of freedom, the number of the independent matrix elements in anomalous space is given by $2n^{2}$, which consists of $n^{2}$ multipole bases $\{\hat{\chi}_{\nu}\}_{\nu=1}^{n^{2}}$ and a complex function (the number of degrees of freedom is 2).
In such a situation, $\hat{\Delta}(\bm{k})$ can be expanded in terms of the multipole basis $\{\hat{\chi}_{\nu}\}_{\nu=1}^{n^{2}}$ as 
\begin{equation}
  \hat{\Delta}(\bm{k})=\sum_{\nu=1}^{n^{2}}\Delta_{\nu}(\bm{k})\hat{\chi}_{\nu},
  \label{eq:decomposition_pair}
\end{equation}
where $\Delta_{\nu}(\bm{k})$ represents the expansion coefficient.

Let us investigate the transformation property of $\hat{\Delta}(\bm{k})$ in terms of the TR ($\mathcal{T}$) and SI ($\mathcal{P}$) operations.
In terms of the TR operation, momentum $\bm{k}$ and spin basis $\ket{\sigma}$ with $\sigma=\uparrow,\downarrow$ are transformed as follows: 
\begin{equation*}
  \begin{aligned}
    &\, \mathcal{T}:\bm{k}\mapsto -\bm{k},
    \\
    &\, \mathcal{T}:\ket{\uparrow}\mapsto \ket{\downarrow},
    \\
    &\, \mathcal{T}:\ket{\downarrow}\mapsto -\ket{\uparrow},
  \end{aligned}
\end{equation*}
whereas $\mathcal{T}$ does not affect the orbital $\phi_{l}$.
By further considering a complex conjugation operator $\mathcal{K}$, which should be included in $\mathcal{T}$, the operator is expressed as
\begin{equation}
  \mathcal{T}=-i\check{\mathcal{U}}_{T}\mathcal{K},
\end{equation}
where $\check{\mathcal{U}}_{T}$ is defined by Eq.~(\ref{eq:unitary_TR}).
Substituting the decomposition in Eq.~(\ref{eq:decomposition_pair}) into Eq.~(\ref{eq:pair_potential_expression}) and taking a transposition of both sides with considering Eq.~(\ref{eq:parity_MP}), we thereby obtain 
\begin{equation*}
  \begin{aligned}
    [\check{\Delta}(\bm{k})]^{\mathrm{T}}
    =&\, -\sum_{\nu}\Delta_{\nu}(\bm{k})(i\check{\mathcal{U}}_{T})\hat{\chi}_{\nu}^{*}
    \\
    =&\, -\sum_{\nu}(-1)^{T_{\nu}}\Delta_{\nu}(\bm{k})\hat{\chi}_{\nu}(i\check{\mathcal{U}}_{T}).
  \end{aligned}
\end{equation*}
We have used that $i\check{\mathcal{U}}_{T}^{\mathrm{T}}=-i\check{\mathcal{U}}_{T}$ from the definition in Eq.~(\ref{eq:unitary_TR}) and hermiticity and orthogonality of $\hat{\chi}_{\nu}$.
Since this is equivalent to $-\check{\Delta}(-\bm{k})=-\sum_{\nu}\Delta_{\nu}(-\bm{k})\hat{\chi}_{\nu}(i\check{\mathcal{U}}_{T})$ by the fermionic antisymmetry, we obtain 
\begin{equation}
  \Delta_{\nu}(-\bm{k})=(-1)^{T_{\nu}}\Delta_{\nu}(\bm{k}),
  \label{eq:parity_k}
\end{equation}
that is, the $\Delta_{\nu}(\bm{k})$ is an even (odd) function in terms of $\bm{k}$ if $\hat{\chi}_{\nu}$ is the E/ET (M/MT) multipole.
This result is the natural extention of the pair potential realized in the single-orbital system where the $s,d,\cdots$-wave SC is realized with the spin-singlet state, whereas the $p,f,\cdots$-wave SC with the spin-triplet state.
In Ref.~\cite{PhysRevB.94.174513}, the even (odd) function of $\Delta_{\nu}(\bm{k})$ is called as $m$-singlet ($m$-triplet), where $m$ means the multipole.
Acting $\mathcal{T}$ operator on $\hat{\Delta}(\bm{k})$, we obtain the transformation: 
\begin{equation}
  \mathcal{T}:\hat{\Delta}(\bm{k})
  =\sum_{\nu}(-1)^{T_{\nu}}\Delta_{\nu}^{*}(-\bm{k})\hat{\chi}_{\nu}
  =\sum_{\nu}\Delta_{\nu}^{*}(\bm{k})\hat{\chi}_{\nu}.
\end{equation}
Taking into account for arbitrariness of the global phase for the pair potential with respect to $U(1)$, the $\mathcal{T}$-symmetry broken SC state, e.g., the chiral SC, which is considered to be realized in some materials, e.g., $\mathrm{UPt_{3}}$~\cite{RevModPhys.74.235}, $\mathrm{Sr_{2}RuO_{4}}$~\cite{Nature.372532a0}, and $\mathrm{SrPtAs}$~\cite{PhysRevB.87.180503,PhysRevB.89.020509}, can be characterized by $\Delta_{\nu}(\bm{k})\in\mathbb{C}/\mathbb{R}$ in general.
For example, $\hat{\Delta}(\bm{k})=\hat{Q}_{x}+i\hat{Q}_{y}$, where $\hat{Q}_{x},\hat{Q}_{y}$ are E dipoles in orbital space, is one of the representations of the $\mathcal{T}$-symmetry broken SC.

Next, we investigate the transformation in terms of the SI operation $\mathcal{P}$.
$\mathcal{P}$ operator acts on momentum $\bm{k}$ and the orbital $l$ as 
\begin{equation*}
  \begin{aligned}
    &\, \mathcal{P}:\bm{k}\mapsto -\bm{k},
    \\
    &\, \mathcal{P}:\ket{\phi_{l}}\mapsto (-1)^{l}\ket{\phi_{l}},
  \end{aligned}
\end{equation*}
whereas it does not affect the spin $\ket{\sigma}$.
Therefore, $\hat{\Delta}(\bm{k})$ is transformed as 
\begin{equation}
  \begin{aligned}
    \mathcal{P}:\hat{\Delta}(\bm{k})
    =&\, \sum_{\nu}(-1)^{P_{\nu}}\Delta_{\nu}(-\bm{k})\hat{\chi}_{\nu}
    \\
    =&\, \sum_{\nu}(-1)^{P_{\nu}}(-1)^{T_{\nu}}\Delta_{\nu}(\bm{k})\hat{\chi}_{\nu},
  \end{aligned}
\end{equation}
where we have used Eq.~(\ref{eq:parity_k}).
When the system has SI symmetry, $\hat{\Delta}(\bm{k})$ has definite parity, i.e., $\mathcal{P}:\hat{\Delta}(\bm{k})\mapsto\pm \hat{\Delta}(\bm{k})$.
In other words, even (odd) parity of $\hat{\Delta}(\bm{k})$ leads to the pairing in terms of the even(odd)-parity E and ET multipoles, and odd(even)-parity M and MT multipoles are realized.

As an example, let us consider a spinful $p$-orbital electron system.
Active multipoles are E monopole $\hat{Q}_{0}$, ET dipole $\hat{G}_{1m}$, E quadrupole $\hat{Q}_{2m}$, M dipole $\hat{M}_{1m}$, MT quadrupole $T_{2m}$, and M octupole $M_{3m}$.
When the system has the SI symmetry, the pair potential has the following form 
\begin{subequations}
\begin{equation}
  \begin{aligned}
    \hat{\Delta}_{g}(\bm{k})=&\, \Delta_{Q_{0}}(\bm{k})\hat{Q}_{0}
    +\sum_{m=-2}^{2}\Delta_{Q_{2m}}(\bm{k})\hat{Q}_{2m}
    \\
    &\, +\sum_{m=-1}^{1}\Delta_{G_{1m}}(\bm{k})\hat{G}_{1m}
  \end{aligned}
  \label{eq:Delta_g}
\end{equation}
for even parity of $\hat{\Delta}(\bm{k})$ [$\hat{\Delta}_{g}(\bm{k})=\hat{\Delta}_{g}(-\bm{k})$], whereas 
\begin{equation}
  \begin{aligned}
    \hat{\Delta}_{u}(\bm{k})=&\, \sum_{m=-1}^{1}\Delta_{M_{1m}}(\bm{k})\hat{M}_{1m}
    +\sum_{m=-3}^{3}\Delta_{M_{3m}}(\bm{k})\hat{M}_{3m}
    \\
    &\, +\sum_{m=-2}^{2}\Delta_{T_{2m}}(\bm{k})\hat{T}_{2m}
  \end{aligned}
  \label{eq:Delta_u}
\end{equation}
for odd parity of $\hat{\Delta}(\bm{k})$ [$\hat{\Delta}_{u}(\bm{k})=-\hat{\Delta}_{u}(-\bm{k})$].
\end{subequations}
Here, $\hat{Q}_{0}(\hat{M}_{1m})$ expresses the spin-singlet $J=0$ (spin-triplet $J=1$) SC state.
It is noted that the term with $\hat{M}_{1m}$ corresponds to an extension of the $d$-vector representation in the form of $\bm{d}(\bm{k}) \cdot \bm{\sigma}$ [$\bm{d}(\bm{k})=-\bm{d}(-\bm{k})$] in the context of the spin-triple SC state in the single-orbital system.
$\hat{Q}_{2m}(\hat{M}_{3m})$ also represents spin-quintet $J=2$ (spin-septet $J=3$) state, as discussed in Ref.~\cite{PhysRevLett.116.177001}.
On the other hand, the pairing represented by the ET/MT multipoles in Eqs.~(\ref{eq:Delta_g}) and (\ref{eq:Delta_u}) leads to another unconventional SC state with the Cooper pairs constructed by $j=1/2$ and $j=3/2$ electrons that have been overlooked by previous classifications.
Such a pairing between electrons with different $j$ might be realized by a Kondo interaction mediated by Ruderman--Kittel--Kasuya--Yosida interaction~\cite{PhysRevX.10.041021, SciPostPhys.12.2.057}.

\begin{table*}[tbp]
  \caption{
    Basis functions for the 3D cubic lattice under the $O_{\mathrm{h}}$ symmetry.
    The superscript $+(-)$ in the first column represents even (odd) parity for the TR operation.
    The second column shows the lowest-order wavenumber basis function $\phi^{\Gamma}(\bm{k})$ of the irreducible representation (IR) $\Gamma$ for $V_{\bm{k}-\bm{k}^{\prime}}$ with $\bm{k}\to\bm{0}$ limit, where $V_{n}$ represents the $n$th-order nearest-neighbor pairing.
    For $A_{1g}^{+}$, we note that the zeroth-order (on-site) pairing exists $\phi^{A_{1g}}(\bm{k})=1$.
    We introduce the abbreviations $c_{i}\equiv \cos{k_{i}}, s_{i}\equiv\sin{k_{i}}$ for $i=x,y,z$, where the lattice constant is set to unity.
    The third column shows the active multipoles in atomic orbital space in the $sp$-orbital system.
    E and ET (M and MT) multipoles, which belong to IR $\Gamma^{+}$ $(\Gamma^{-})$, can couple to the basis functions $\phi^{\Gamma_{g}}(\bm{k})$ $[\phi^{\Gamma_{u}}(\bm{k})]$.
  }
  \begin{ruledtabular}
    \begin{tabular}{cll}
      IR$(\Gamma)$ & $\phi^{\Gamma}_{n,\gamma}(\bm{k})$ & Multipoles 
      \\
      \hline
      $A_{1g}^{+}$ & $V_{1}\sqrt{\frac{2}{3}}(c_{x}+c_{y}+c_{z})\sim k^{2}$ & $Q_{0}$ 
      \\
      $A_{2g}^{+}$ & $V_{5}2\sqrt{\frac{2}{3}}(c_{x}-c_{y})(c_{y}-c_{z})(c_{z}-c_{x})\sim (k_{x}^{2}-k_{y}^{2})(k_{y}^{2}-k_{z}^{2})(k_{z}^{2}-k_{x}^{2})$ & -- 
      \\
      $E_{g}^{+}$ & $V_{1}\left\{\frac{1}{\sqrt{3}}(2c_{z}-c_{x}-c_{y}),c_{x}-c_{y}\right\}\sim \{3k_{z}^{2}-k^{2},k_{x}^{2}-k_{y}^{2}\}$ & $\{Q_{u},Q_{v}\}$ 
      \\
      $T_{1g}^{+}$ & $V_{5}\{2\sqrt{2}s_{y}s_{z}(c_{y}-c_{z}),(\mathrm{cyclic})\}\sim \{k_{y}k_{z}(k_{y}^{2}-k_{z}^{2}),(\mathrm{cyclic})\}$ & $\{G_{x},G_{y},G_{z}\}$
      \\
      $T_{2g}^{+}$ & $V_{2}\{2s_{y}s_{z},(\mathrm{cyclic})\}\sim \{k_{y}k_{z},(\mathrm{cyclic})\}$ & $\{Q_{yz},Q_{zx},Q_{xy}\}$ 
      \\
      $A_{1u}^{+}$ & -- & $G_{0}$ 
      \\
      $A_{2u}^{+}$ & -- & -- 
      \\
      $E_{u}^{+}$ & -- & $\{G_{u},G_{v}\}$ 
      \\
      $T_{1u}^{+}$ & -- & $\{Q_{x},Q_{y},Q_{z}\}$ 
      \\
      $T_{2u}^{+}$ & -- & $\{G_{yz},G_{zx},G_{xy}\}$  
      \\
      $A_{1g}^{-}$ & -- & -- 
      \\
      $A_{2g}^{-}$ & -- & -- 
      \\
      $E_{g}^{-}$ & -- & $\{T_{u},T_{v}\}$ 
      \\
      $T_{1g}^{-}$ & -- & $\{M_{x},M_{y},M_{z}\},\{M_{x}^{\alpha},M_{y}^{\alpha},M_{z}^{\alpha}\}$ 
      \\
      $T_{2g}^{-}$ & -- & $\{M_{x}^{\beta},M_{y}^{\beta},M_{z}^{\beta}\}$, $\{T_{yz},T_{zx},T_{xy}\}$ 
      \\
      $A_{1u}^{-}$ & $V_{13}2\sqrt{\frac{2}{3}}s_{x}s_{y}s_{z}(c_{x}-c_{y})(c_{y}-c_{z})(c_{z}-c_{x})$ & $M_{0}$ 
      \\
      & $\sim k_{x}k_{y}k_{z}(k_{x}^{2}-k_{y}^{2})(k_{y}^{2}-k_{z}^{2})(k_{z}^{2}-k_{x}^{2})$ & 
      \\
      $A_{2u}^{-}$ & $V_{3}2\sqrt{2}s_{x}s_{y}s_{z}\sim k_{x}k_{y}k_{z}$ & -- 
      \\
      $E_{u}^{-}$ & $V_{6}\left\{4s_{x}s_{y}s_{z}(c_{x}-c_{y}),\frac{4}{\sqrt{3}}s_{x}s_{y}s_{z}(2c_{z}-c_{x}-c_{y})\right\}$ & $\{M_{u},M_{v}\}$ 
      \\
      & $\sim \left\{k_{x}k_{y}k_{z}(k_{x}^{2}-k_{y}^{2}),k_{x}k_{y}k_{z}(3k_{z}^{2}-k^{2})\right\}$ & 
      \\
      $T_{1u}^{-}$ & $V_{1}\{\sqrt{2}s_{x},\sqrt{2}s_{y},\sqrt{2}s_{z}\}\sim \{k_{x},k_{y},k_{z}\}$ & $\{T_{x},T_{y},T_{z}\}$ 
      \\
      $T_{2u}^{-}$ & $V_{2}\{\sqrt{2}s_{x}(c_{y}-c_{z}),(\mathrm{cyclic})\}\sim \{k_{x}(k_{y}^{2}-k_{z}^{2}),(\mathrm{cyclic})\}$ & $\{M_{yz},M_{zx},M_{xy}\}$ 
      \\
    \end{tabular}
  \end{ruledtabular}
  \label{tab:multipole_Oh}
\end{table*}

\begin{table*}[tbp]
  \caption{
    Basis functions for the 2D square lattice under the $D_{\mathrm{4h}}$ symmetry.
    The notations are the same as those in Table~\ref{tab:multipole_Oh}.
  }
  \begin{ruledtabular}
    \begin{tabular}{cll}
      IR$(\Gamma)$ & $\phi^{\Gamma}_{n,\gamma}(\bm{k})$ & Multipoles 
      \\
      \hline
      $A_{1g}^{+}$ & $V_{1}(c_{x}+c_{y})\sim k_{x}^{2}+k_{y}^{2}$ & $Q_{0},Q_{u}$ 
      \\
      $A_{2g}^{+}$ & $V_{4}2\sqrt{2}s_{x}s_{y}(c_{x}-c_{y})\sim k_{x}k_{y}(k_{x}^{2}-k_{y}^{2})$ & $G_{z}$ 
      \\
      $B_{1g}^{+}$ & $V_{1}(c_{x}-c_{y})\sim k_{x}^{2}-k_{y}^{2}$ & $Q_{v}$ 
      \\
      $B_{2g}^{+}$ & $V_{2}2s_{x}s_{y}\sim k_{x}k_{y}$ & $Q_{xy}$ 
      \\
      $E_{g}^{+}$ & -- & $\{Q_{yz},Q_{zx}\}$, $\{G_{x},G_{y}\}$ 
      \\
      $A_{1u}^{+}$ & -- & $G_{0},G_{u}$ 
      \\
      $A_{2u}^{+}$ & -- & $Q_{z}$ 
      \\
      $B_{1u}^{+}$ & -- & $G_{v}$ 
      \\
      $B_{2u}^{+}$ & -- & $G_{xy}$ 
      \\
      $E_{u}^{+}$ & -- & $\{Q_{x},Q_{y}\}$, $\{G_{yz},G_{zx}\}$ 
      \\
      $A_{1g}^{-}$ & -- & $T_{u}$ 
      \\
      $A_{2g}^{-}$ & -- & $M_{z},M_{z}^{\alpha}$ 
      \\
      $B_{1g}^{-}$ & -- & $M_{xyz}$, $T_{v}$ 
      \\
      $B_{2g}^{-}$ & -- & $M_{z}^{\beta}$, $T_{xy}$ 
      \\
      $E_{g}^{-}$ & -- & $\{M_{x},M_{y}\},\{M_{x}^{\alpha},M_{y}^{\alpha}\},\{M_{x}^{\beta},M_{y}^{\beta}\}$, $\{T_{yz},T_{zx}\}$ 
      \\
      $A_{1u}^{-}$ & -- & $M_{0},M_{u}$ 
      \\
      $A_{2u}^{-}$ & -- & $T_{z}$ 
      \\
      $B_{1u}^{-}$ & -- & $M_{v}$ 
      \\
      $B_{2u}^{-}$ & -- & $M_{xy}$ 
      \\
      $E_{u}^{-}$ & $V_{1}\{\sqrt{2}s_{x},\sqrt{2}s_{y}\}\sim \{k_{x},k_{y}\}$ & $\{M_{yz},M_{zx}\}$, $\{T_{x},T_{y}\}$ 
      \\
    \end{tabular}
  \end{ruledtabular}
  \label{tab:multipole_D4h}
\end{table*}

\begin{table*}[tbp]
  \caption{
    Basis functions for the 2D triangular lattice under the $D_{\mathrm{6h}}$ symmetry.
    We introduce the notations $c_{\tilde{i}}=\cos{\tilde{k}_{i}},c_{\tilde{i}}=\sin{\tilde{k}_{i}}$ for $i=x,y$ with $\tilde{k}_{x}=k_{x}/2,\tilde{k}_{y}=\sqrt{3}k_{y}/2$.
    The other notations are the same as those in Table~\ref{tab:multipole_Oh}.
  }
  \begin{ruledtabular}
    \begin{tabular}{cll}
      IR$(\Gamma)$ & $\phi^{\Gamma}_{n,\gamma}(\bm{k})$ & Multipoles 
      \\
      \hline
      $A_{1g}^{+}$ & $V_{1}\sqrt{\frac{2}{3}}(c_{x}+2c_{\tilde{x}}c_{\tilde{y}})\sim k_{x}^{2}+k_{y}^{2}$ & $Q_{0},Q_{u}$ 
      \\
      $A_{2g}^{+}$ & $V_{4}\frac{2}{\sqrt{3}}(s_{5\tilde{x}}s_{\tilde{y}}-s_{2x}s_{2\tilde{y}}+s_{\tilde{x}}s_{3\tilde{y}})\sim k_{x}k_{y}(3k_{x}^{2}-k_{y}^{2})(k_{x}^{2}-3k_{y}^{2})$ & $G_{z}$ 
      \\
      $B_{1g}^{+}$ & -- & -- 
      \\
      $B_{2g}^{+}$ & -- & -- 
      \\
      $E_{1g}^{+}$ & -- & $\{Q_{yz},Q_{zx}\}$, $\{G_{x},G_{y}\}$ 
      \\
      $E_{2g}^{+}$ & $V_{1}\left\{\frac{2}{\sqrt{3}}(c_{x}-c_{\tilde{x}}c_{\tilde{y}}, 2s_{\tilde{x}}s_{\tilde{y}})\right\}\sim \{k_{x}^{2}-k_{y}^{2}, k_{x}k_{y}\}$ & $\{Q_{v},Q_{xy}\}$ 
      \\
      $A_{1u}^{+}$ & -- & $G_{0},G_{u}$ 
      \\
      $A_{2u}^{+}$ & -- & $Q_{z}$ 
      \\
      $B_{1u}^{+}$ & -- & -- 
      \\
      $B_{2u}^{+}$ & -- & -- 
      \\
      $E_{1u}^{+}$ & -- & $\{Q_{x},Q_{y}\}$, $\{G_{yz},G_{zx}\}$ 
      \\
      $E_{2u}^{+}$ & -- & $\{G_{v},G_{xy}\}$ 
      \\
      $A_{1g}^{-}$ & -- & $T_{u}$ 
      \\
      $A_{2g}^{-}$ & -- & $M_{z},M_{z}^{\alpha}$ 
      \\
      $B_{1g}^{-}$ & -- & $M_{3a}$ 
      \\
      $B_{2g}^{-}$ & -- & $M_{3b}$ 
      \\
      $E_{1g}^{-}$ & -- & $\{M_{x},M_{y}\},\{M_{3u},M_{3v}\}$, $\{T_{yz},T_{zx}\}$ 
      \\
      $E_{2g}^{-}$ & -- & $\{M_{xyz},M_{z}^{\beta}\}$, $\{T_{v},T_{xy}\}$ 
      \\
      $A_{1u}^{-}$ & -- & $M_{0},M_{u}$ 
      \\
      $A_{2u}^{-}$ & -- & $T_{z}$ 
      \\
      $B_{1u}^{-}$ & $V_{2}\sqrt{\frac{2}{3}}(s_{2\tilde{y}}-2s_{\tilde{y}}c_{3\tilde{x}})\sim k_{y}(3k_{x}^{2}-k_{y}^{2})$ & -- 
      \\
      $B_{2u}^{-}$ & $V_{1}\sqrt{\frac{2}{3}}(s_{x}-2s_{\tilde{x}}c_{\tilde{y}})\sim k_{x}(k_{x}^{2}-3k_{y}^{2})$ & -- 
      \\
      $E_{1u}^{-}$ & $V_{1}\left\{\frac{2}{\sqrt{3}}(s_{x}+s_{\tilde{x}}c_{\tilde{y}}), 2s_{\tilde{y}}c_{\tilde{x}}\right\}\sim \{k_{x},k_{y}\}$ & $\{M_{yz},M_{zx}\}$, $\{T_{x},T_{y}\}$ 
      \\
      $E_{2u}^{-}$ & -- & $\{M_{v},M_{xy}\}$ 
      \\
    \end{tabular}
  \end{ruledtabular}
  \label{tab:multipole_D6h}
\end{table*}

Substituting Eq.~(\ref{eq:vertex_pair}) into Eq.~(\ref{eq:pair_definition}), we thereby obtain
\begin{equation}
  \begin{aligned}
    &\, [\check{\Delta}(\bm{k})]_{\alpha\beta}
    \\
    =&\, -\frac{1}{4N}\sum_{\bm{k}^{\prime}}\sum_{\gamma\delta}V_{\bm{k}-\bm{k}^{\prime}}
    [\Lambda_{\alpha\gamma}(\bm{k})F_{\gamma\delta}(\bm{k}^{\prime})\Lambda_{\beta\delta}(-\bm{k})
    \\
    &\, +\Lambda_{\alpha\gamma}(\bm{k})F_{\gamma\delta}(\bm{k}^{\prime})\Lambda_{\beta\delta}(-\bm{k}^{\prime})
    +\Lambda_{\alpha\gamma}(\bm{k}^{\prime})F_{\gamma\delta}(\bm{k}^{\prime})\Lambda_{\beta\delta}(-\bm{k})
    \\
    &\, +\Lambda_{\alpha\gamma}(\bm{k}^{\prime})F_{\gamma\delta}(\bm{k}^{\prime})\Lambda_{\beta\delta}(-\bm{k}^{\prime})],
  \end{aligned}
\end{equation}
under the mean-field approximation.
The wavenumber dependence of $\Delta_{\alpha\beta}(\bm{k})$ comes from two factors:
The one is the multipole $\Lambda_{\alpha\beta}(\bm{k})$, and the other is the interaction $V_{\bm{k}-\bm{k}^{\prime}}$.
In the lattice system, the latter can be expanded as 
\begin{equation}
  V_{\bm{k}-\bm{k}^{\prime}}=V_{0}+\sum_{n}V_{n}\sum_{\Gamma,\gamma}\phi^{\Gamma}_{n,\gamma}(\bm{k})\phi^{\Gamma*}_{n,\gamma}(\bm{k}^{\prime}),
\end{equation}
where $V_{n}$ is the coupling constant for the interaction between $n$-th-order nearest-neighboring sites, and $\phi^{\Gamma}_{n,\gamma}(\bm{k})$ is its $\gamma$-th basis belonging to the irreducible representation (IR) $\Gamma$ of the group $G$.
For example, in the 2D tetragonal lattice structure with the lattice constant $a$ under the $D_{\mathrm{4h}}$ symmetry, the nearest-neighbor ($n=1$) pairing interaction is expressed as $V_{\bm{q}}=2V_{1}[\cos{(q_{x}a)}+\cos{(q_{y}a)}]$, which can be decomposed into 
\begin{equation*}
  \begin{aligned}
    V_{\bm{k}-\bm{k}^{\prime}}=&\, V_{1}\phi^{A_{1g}}(\bm{k})\phi^{A_{1g}}(\bm{k}^{\prime})
    \\
    &\, +V_{1}\phi^{B_{1g}}(\bm{k})\phi^{B_{1g}}(\bm{k}^{\prime})
    \\
    &\, +V_{1}[\phi^{E_{u}}_{1}(\bm{k})\phi^{E_{u}}_{1}(\bm{k}^{\prime})+\phi^{E_{u}}_{2}(\bm{k})\phi^{E_{u}}_{2}(\bm{k}^{\prime})],
    \\
  \end{aligned}
\end{equation*}
where $\phi^{A_{1g}}(\bm{k})=\cos{(k_{x}a)}+\cos{(k_{y}a)}$, $\phi^{B_{1g}}(\bm{k})=\cos{(k_{x}a)}-\cos{(k_{y}a)}$, and $\{\phi^{E_{u}}_{1}(\bm{k}),\phi^{E_{u}}_{2}(\bm{k})\}=\{\sqrt{2}\sin{(k_{x}a)},\sqrt{2}\sin{(k_{y}a)}\}$. 
Thus, the nearest-neighbor pairing interaction can give rise to the pairing with the wavenumber dependence belonging to the IRs, $A_{1g}$, $B_{1g}$, and $E_{u}$, which are related to the E monopole, E quadrupole, and MT dipole, respectively~\cite{PhysRevB.107.195118}.
We show the lowest-order wavenumber basis function $\phi^{\Gamma}_{n,\gamma}(\bm{k})$ for each IR under $O_{\mathrm{h}}$, $D_{\mathrm{4h}}$, and $D_{\mathrm{6h}}$ in Tables~\ref{tab:multipole_Oh}, \ref{tab:multipole_D4h}, and \ref{tab:multipole_D6h}, respectively.

Finally, $\Delta_{\nu}(\bm{k})$ in Eq.~(\ref{eq:decomposition_pair}) is given by 
\begin{equation}
  \Delta_{\nu}(\bm{k})=
  \frac{1}{2}\mathrm{Tr}[\hat{\chi}_{\nu}\check{\Delta}(\bm{k})(i\check{\mathcal{U}}_{T})^{\dag}],
  \label{eq:gap_amplitude}
\end{equation}
where we normalize spinful multipoles as $\mathrm{Tr}[\hat{\chi}_{\mu}\hat{\chi}_{\nu}]=2\delta_{\mu\nu}$.
We note that the wavenumber dependence of $\Delta_{\nu}(\bm{k})$ is characterized by $\phi^{\Gamma}_{n,\gamma}(\bm{k})$.
We show several expressions of $\hat{\chi}_{\nu}$ in Eqs.~(\ref{eq:spinless_multipoles}), (\ref{eq:spinful_MPs}), and (\ref{eq:jjz_MPs}) in Appendices~\ref{app:MP} and \ref{app:jjz}.
We summarize the notations used in this paper in Table~\ref{tab:notations}.

\begin{table*}[tbp]
  \caption{
    Notations used in this paper.
  }
  \begin{ruledtabular}
    \begin{tabular}{lll}
       & normal space & anomalous space
      \\
      \hline
      Multipole operator & $\hat{O}=\sum_{\bm{k}}\sum_{\alpha\beta}[\hat{\Lambda}(\bm{k})]_{\alpha\beta}\hat{c}_{\bm{k}\alpha}^{\dag}\hat{c}_{\bm{k}\beta}$ & $\check{O}^{\dag}=\frac{1}{2}\sum_{\bm{k}}\sum_{\alpha\beta\gamma}[\hat{\Lambda}(\bm{k})]_{\alpha\beta}(i\check{\mathcal{U}}_{T})_{\beta\gamma}\hat{c}_{\bm{k}\alpha}^{\dag}\hat{c}_{-\bm{k}\gamma}^{\dag}$
      \\
      Multipole matrix & $\hat{X}_{lm}(s,k)$ or $\hat{\chi}_{\nu}$ & $\hat{\chi}_{\nu}(i\check{\mathcal{U}}_{T})$
      \\
      Order parameter & $\braket{\hat{O}}$ & $\check{\Delta}(\bm{k})=\hat{\Delta}(\bm{k})(i\check{\mathcal{U}}_{T})$
      \\
    \end{tabular}
  \end{ruledtabular}
  \label{tab:notations}
\end{table*}

\subsection{Application to on-site pairing}

In the following, we investigate the possibility of the unconventional SC state with the multipole degrees of freedom by considering specific $p$-orbital and $sp$-orbital systems under the assumption that the momentum dependence of the form factor $\hat{\Lambda}(\bm{k})$ is neglected, i.e., $\hat{\Lambda}(\bm{k})=\hat{\chi}_{\nu}$.
We also consider the uniform on-site interaction, i.e., $V_{\bm{q}}\approx V_{0}$.
In other words, we focus on the Cooper pairing resulting from the fluctuations in terms of the local electronic degrees of freedom.
Then, the corresponding vertex for the pairing interaction in Eq.~(\ref{eq:vertex_pair}) is simplified as 
\begin{equation}
  V_{\alpha\beta;\gamma\delta}(\bm{k},\bm{k}^{\prime})
  =\frac{V_{0}}{2}[(\hat{\chi}_{\nu})_{\alpha\delta}(\hat{\chi}_{\nu})_{\beta\gamma}
  -(\hat{\chi}_{\nu})_{\alpha\gamma}(\hat{\chi}_{\nu})_{\beta\delta}].
  \label{eq:local_vertex}
\end{equation}
In this case, the pairing interaction in Eq.~(\ref{eq:interaction_pair}) can be expressed as the product of multipoles~\cite{PhysRevLett.115.207002, PhysRevResearch.2.033225} as 
\begin{equation}
  \hat{\mathcal{H}}_{\mathrm{eff}}=\frac{V_{0}}{2N}\sum_{\mu,\mu^{\prime}}c_{\mu\mu^{\prime}}(\check{O}_{\mu}^{\dag}\check{O}_{\mu^{\prime}}+\check{O}_{\mu^{\prime}}^{\dag}\check{O}_{\mu}),
  \label{eq:cooper_pairing}
\end{equation}
where 
\begin{equation}
  \check{O}_{\mu}^{\dag}=\frac{1}{2}\sum_{\bm{k}}\sum_{\alpha\beta}[\hat{\chi}_{\mu}(i\check{\mathcal{U}}_{T})]_{\alpha\beta}\hat{c}_{\bm{k}\alpha}^{\dag}\hat{c}_{-\bm{k}\beta}^{\dag}
  \label{eq:local_pair_multipole}
\end{equation}
and $c_{\mu\mu^{\prime}}=c_{\mu^{\prime}\mu}=c_{\mu\mu^{\prime}}^{*}$.
Using them, $c_{\mu\mu^{\prime}}$ is given by 
\begin{equation}
  c_{\mu\mu^{\prime}}=(-1)^{T_{\Lambda}}\frac{1}{4}[1+(-1)^{T_{\mu}}]\mathrm{Tr}{[\hat{\chi}_{\mu}\hat{\chi}_{\nu}\hat{\chi}_{\mu^{\prime}}\hat{\chi}_{\nu}]},
  \label{eq:channel_coeeficient}
\end{equation}
where $(-1)^{T_{\Lambda}}$ and $(-1)^{T_{\mu}}=(-1)^{T_{
\mu^{\prime}}}$ are the parity for $\mathcal{T}$-operator of $\hat{\Lambda}=\hat{\chi}_{\nu}$ and $\hat{\chi}_{\mu},\hat{\chi}_{\mu^{\prime}}$, respectively.
From this expression, one finds that the M/MT-type multipole pairing with $(-1)^{T_{\mu}}=-1$ is prohibited.

In the crystal system, the rotational symmetry and inversion symmetry can be lost.
Under the point group symmetry, the components of the same rank split into subgroups according to the point group IRs.
The classification of multipoles has been already in many literature~\cite{PhysRevB.98.165110,PhysRevB.98.245129, PhysRevB.104.054412}, which can be applied to the SC order parameter.
We show the classification of multipoles in terms of IRs for $O_{\mathrm{h}}, D_{\mathrm{4h}}$, and $D_{\mathrm{6h}}$ in Tables~\ref{tab:multipole_Oh}, \ref{tab:multipole_D4h} and \ref{tab:multipole_D6h}, respectively.
The results for other point groups can be straightforwardly obtained by using the compatibility relation between a group and its subgroup.

As a demonstration, let us consider an $sp$-orbital system in the $D_{\mathrm{6h}}$ point group, which will be numerically analyzed in the next section.
First, we discuss SC accompanied by the ferroelectric fluctuation, which originates from the local spinless E dipole ($\hat{Q}_{z}$) fluctuation.
The corresponding vertex for the pairing interaction is given by 
\begin{equation}
  V_{\alpha\beta;\gamma\delta}(\bm{k},\bm{k}^{\prime})
  =\frac{V_{0}}{2}[(\hat{Q}_{z})_{\alpha\delta}(\hat{Q}_{z})_{\beta\gamma}
  -(\hat{Q}_{z})_{\alpha\gamma}(\hat{Q}_{z})_{\beta\delta}].
  \label{eq:FE_vertex}
\end{equation}
Substituting this expression into Eq.~(\ref{eq:interaction_pair}) and decomposing into the pairing interaction between multipoles as Eq.~(\ref{eq:cooper_pairing}), we obtain
\begin{widetext}
  \begin{equation}
    \begin{aligned}
      \hat{\mathcal{H}}_{\mathrm{eff}}=&\, \frac{V_{0}}{2N}
      \left[\check{Q}_{z}^{\dag}\check{Q}_{z}
      +\frac{1}{2\sqrt{3}}\left(\check{Q}_{0}^{s\dag}\check{Q}_{0}+\check{Q}_{0}^{\dag}\check{Q}_{0}^{s}\right)
      +\frac{1}{\sqrt{6}}\left(\check{Q}_{0}^{s\dag}\check{Q}_{u}+\check{Q}_{u}^{\dag}\check{Q}_{0}^{s}\right)\right.
      \\
      &\, -\frac{1}{3}\left(\check{G}_{0}^{\prime\dag}\check{G}_{0}^{\prime}+2\check{G}_{u}^{\prime\dag}\check{G}_{u}^{\prime}\right)
      -\frac{1}{3\sqrt{2}}\left(\check{G}_{0}^{\prime\dag}\check{G}_{u}^{\prime}+\check{G}_{u}^{\prime\dag}\check{G}_{0}^{\prime}\right)
      \\
      &\, \left.-\frac{1}{2}\left(\check{Q}_{x}^{\prime\dag}\check{Q}_{x}^{\prime}+\check{Q}_{y}^{\prime\dag}\check{Q}_{y}^{\prime}+\check{G}_{yz}^{\prime\dag}\check{G}_{yz}^{\prime}+\check{G}_{zx}^{\prime\dag}\check{G}_{zx}^{\prime}\right)
      +\frac{1}{4}\left(\check{Q}_{x}^{\prime\dag}\check{G}_{yz}^{\prime}+\check{G}_{yz}^{\prime\dag}\check{Q}_{x}^{\prime}\right)
      -\frac{1}{4}\left(\check{Q}_{y}^{\prime\dag}\check{G}_{zx}^{\prime}+\check{G}_{zx}^{\prime\dag}\check{Q}_{y}^{\prime}\right)\right].
    \end{aligned}
    \label{eq:FE_pairing}
  \end{equation}
\end{widetext}
We find that the some mixed channels such as $\check{Q}_{0}^{s\dag}\check{Q}_{u}+\check{Q}_{u}^{\dag}\check{Q}_{0}^{s}$ appear due to the symmetry lowering under crystallographic systems; they disappear in rorationally symmetric systems.
Although it is possible to make the quadratic form Hamiltonian fully diagonal, we avoid it because each multipole expresses pairing state with a different physical origin.
According to Table~\ref{tab:multipole_D6h} and Eq.~(\ref{eq:FE_pairing}), the expression of the pair potential $\check{\Delta}=\hat{\Delta}(i\check{\mathcal{U}}_{T})$ is given by 
\begin{equation}
  \begin{aligned}
    \hat{\Delta}=&\, \Delta_{0}^{s}\hat{Q}_{0}^{s}+\Delta_{0}\hat{Q}_{0}+\Delta_{u}\hat{Q}_{u} &\, A_{1g}^{+}
    \\
    &\, +\Delta_{z}\hat{Q}_{z} &\, A_{2u}^{+}
    \\
    &\, +\Delta_{0}^{\prime}\hat{G}_{0}^{\prime}+\Delta_{u}^{\prime}\hat{G}_{u}^{\prime} &\, A_{1u}^{+}
    \\
    &\, +\Delta_{x}^{\prime}\hat{Q}_{x}^{\prime}+\Delta_{y}^{\prime}\hat{Q}_{y}^{\prime}
    +\Delta_{yz}^{\prime}\hat{G}_{yz}^{\prime}+\Delta_{zx}^{\prime}\hat{G}_{zx}^{\prime}, &\, E_{1u}^{+}
  \end{aligned}
  \label{eq:FE_pair_potential}
\end{equation}
where $\hat{Q}_{0}^{s}$ is the E monopole activated in $s$-orbital space, $\hat{Q}_{0}(\hat{Q}_{u})$ is the E monopole (quadrupole) activated in $p$-orbital space, and the other E and ET multipoles, which are denoted as $\hat{Q}_{z},\hat{Q}_{x}^{\prime},\hat{Q}_{y}^{\prime},\hat{G}_{0}^{\prime},\hat{G}_{u}^{\prime},\hat{G}_{yz}^{\prime}$, and $\hat{G}_{zx}^{\prime}$, are activated in $sp$-orbital space.
See Appendix~\ref{app:MP} for the matrix form of multipoles.
Each coefficient $\Delta_{\nu}$ is determined self-consistently by solving the gap equation.
In the weak-coupling region, the $A_{1g}^{+}$-type pairing is favored in general, as demonstrated in the next section.
Thus, the electric monopoles $\hat{Q}_{0}^{s}$ and $\hat{Q}_{0}$ and electric quadrupole $\hat{Q}_{u}$ are favored under the $D_{\mathrm{6h}}$ point group.

Meanwhile, the other multipoles additionally contribute to the pair potential once the symmetry of the system is lowered from $D_{\mathrm{6h}}$.
When considering the situation where the ferroelectric order breaking the SI symmetry occurs, the symmetry of the system is reduced as $D_{\mathrm{6h}}\searrow C_{\mathrm{6v}}$.
In this case, the E dipole $\hat{Q}_{z}$ egrees of freedom is expected to contribute to the pair potential, since $\hat{Q}_{z}$ belongs to the totally symmetric IR under $C_{\mathrm{6v}}$.
Such a coexisting phase of ferroelectric ordering and SC was observed in noncentrosymmetric superconductors such as $\mathrm{SrTiO_{3}}$~\cite{PhysRevMaterials.3.091401, nphys4085}.

As another example, we suppose the ferro-axial fluctuation (We neglect the $s$-orbital degrees of freedom).
The ferro-axial (ferrorotational) order is characterized by the ET dipole, whose microscopic expression is given by~\cite{JPSJ.89.104704, JPSJ.91.113702}
\begin{equation}
  \hat{G}_{z}^{\prime}=\frac{1}{2}(l_{x}\sigma_{y}-l_{y}\sigma_{x}),
\end{equation}
where $l_{x}$ and $l_{y}$ are the orbital angular momentum operators.
By supposing the local fluctuations in terms of $\hat{G}_{z}^{\prime}$, the corresponding vertex for the pairing interaction is given by 
\begin{equation}
  V_{\alpha\beta;\gamma\delta}(\bm{k},\bm{k}^{\prime})
  =\frac{V_{0}}{2}[(\hat{G}_{z}^{\prime})_{\alpha\delta}(\hat{G}_{z}^{\prime})_{\beta\gamma}
  -(\hat{G}_{z}^{\prime})_{\alpha\gamma}(\hat{G}_{z}^{\prime})_{\beta\delta}].
  \label{eq:FA_vertex}
\end{equation}
Corresponding pairing Hamiltonian is given by 
\begin{widetext}
  \begin{equation}
    \begin{aligned}
      \mathcal{\hat{H}}_{\mathrm{eff}}=&\, \frac{V_{0}}{2N}
      \left[\frac{2}{3}\check{Q}_{0}^{\dag}\check{Q}_{0}-\frac{2}{3}\check{Q}_{u}^{\dag}\check{Q}_{u}
      +\frac{1}{6\sqrt{2}}\left(\check{Q}_{0}^{\dag}\check{Q}_{u}+\check{Q}_{u}^{\dag}\check{Q}_{0}\right)
      -\frac{1}{2}\left(\check{Q}_{yz}^{\dag}\check{Q}_{yz}+\check{Q}_{zx}^{\dag}\check{Q}_{zx}\right)\right.
      \\
      &\, -\frac{1}{3}\left(2\check{Q}_{0}^{\prime\dag}\check{Q}_{0}^{\prime}+\check{Q}_{u}^{\prime\dag}\check{Q}_{u}^{\prime}\right)
      +\frac{1}{3\sqrt{2}}\left(\check{Q}_{0}^{\prime\dag}\check{Q}_{u}^{\prime}+\check{Q}_{u}^{\prime\dag}\check{Q}_{0}^{\prime}\right)
      \\
      &\, +\check{G}_{z}^{\prime\dag}\check{G}_{z}^{\prime}-\frac{1}{4}\left(\check{Q}_{yz}^{\prime\dag}\check{Q}_{yz}^{\prime}+\check{Q}_{zx}^{\prime\dag}\check{Q}_{zx}^{\prime}+\check{G}_{x}^{\prime\dag}\check{G}_{x}^{\prime}+\check{G}_{y}^{\prime\dag}\check{G}_{y}^{\prime}\right)
      -\frac{1}{8}\left(\check{Q}_{yz}^{\prime\dag}\check{G}_{x}^{\prime}+\check{G}_{x}^{\prime\dag}\check{Q}_{yz}^{\prime}\right)
      +\frac{1}{8}\left(\check{Q}_{zx}^{\prime\dag}\check{G}_{y}^{\prime}+\check{G}_{y}^{\prime\dag}\check{Q}_{zx}^{\prime}\right)
      \\
      &\, -\frac{1}{6\sqrt{2}}\left(\check{Q}_{0}^{\dag}\check{Q}_{0}^{\prime}+\check{Q}_{0}^{\prime\dag}\check{Q}_{0}\right)
      -\frac{1}{6}\left(\check{Q}_{0}^{\dag}\check{Q}_{u}^{\prime}+\check{Q}_{u}^{\prime\dag}\check{Q}_{0}\right)
      -\frac{1}{6}\left(\check{Q}_{u}^{\dag}\check{Q}_{0}^{\prime}+\check{Q}_{0}^{\prime\dag}\check{Q}_{u}\right)
      -\frac{1}{3\sqrt{2}}\left(\check{Q}_{u}^{\dag}\check{Q}_{u}^{\prime}+\check{Q}_{u}^{\prime\dag}\check{Q}_{u}\right)
      \\
      &\, \left.-\frac{1}{4\sqrt{2}}\left(\check{Q}_{yz}^{\dag}\check{Q}_{yz}^{\prime}+\check{Q}_{yz}^{\prime\dag}\check{Q}_{yz}\right)
      -\frac{1}{4\sqrt{2}}\left(\check{Q}_{yz}^{\dag}\check{G}_{x}^{\prime}+\check{G}_{x}^{\prime\dag}\check{Q}_{yz}\right)
      -\frac{1}{4\sqrt{2}}\left(\check{Q}_{zx}^{\dag}\check{Q}_{zx}^{\prime}+\check{Q}_{zx}^{\prime\dag}\check{Q}_{zx}\right)
      +\frac{1}{4\sqrt{2}}\left(\check{Q}_{zx}^{\dag}\check{G}_{y}^{\prime}+\check{G}_{y}^{\prime\dag}\check{Q}_{zx}\right)
      \right].
    \end{aligned}
    \label{eq:FA_pairing}
  \end{equation}
\end{widetext}
From Eq.~(\ref{eq:FA_pairing}) with Table~\ref{tab:multipole_D6h} and by using the mean-field approximation, we obtain the pair potential as, 
\begin{equation}
  \begin{aligned}
    \hat{\Delta}=&\, \Delta_{0}\hat{Q}_{0}
    +\Delta_{u}\hat{Q}_{u}
    +\Delta_{0}^{\prime}\hat{Q}_{0}^{\prime}+\Delta_{u}^{\prime}\hat{Q}_{u}^{\prime} & A_{1g}^{+}
    \\
    &\, +\Delta_{z}^{\prime}\hat{G}_{z}^{\prime} & A_{2g}^{+}
    \\
    &\, +\Delta_{yz}\hat{Q}_{yz}+\Delta_{zx}\hat{Q}_{zx}
    +\Delta_{yz}^{\prime}\hat{Q}_{yz}^{\prime}+\Delta_{zx}^{\prime}\hat{Q}_{zx}^{\prime}
    \\
    &\, +\Delta_{x}^{\prime}\hat{G}_{x}^{\prime}+\Delta_{y}^{\prime}\hat{G}_{y}^{\prime}. & E_{1g}^{+}
  \end{aligned}
  \label{eq:FA_pair_potential}
\end{equation}
Here, $\hat{Q}_{0}^{\prime}$ $(\hat{Q}_{\alpha}^{\prime};\alpha=u,yz,zx)$ is the spinful E monopole (quadrupole), which represents the pairing state with orbital antisymmetric and spin triplet.
It is noted that these pairing states are distinguished from the spinless E pairing state, since they are activated in the different Hilbert space.
Similarly to the above ferroelectric case,  the ET dipole $\hat{G}_{z}$ degrees of freedom can contribute to the pair potential in the weak-coupling region when the IR $A_{2g}^{+}$ belongs to the totally symmetric IR through the symmetry lowering as $D_{\mathrm{6h}}\searrow C_{\mathrm{6h}}$.
We summarize the Cooper channel induced by other multipole-fluctuated interactions in the $sp$-orbital system in Appendix~\ref{app:cooper_channel}.

\section{Model Calculation}
\label{sec:application}

In this section, we investigate the stability of the SC state characterized by unconventional pair potentials based on the model analysis.
In the following, we choose the units of $k_{\mathrm{B}}=\hbar=1$, where $k_{\mathrm{B}}$ is the Boltzmann constant and $\hbar$ is the Dirac constant, respectively.
We consider the tight-binding model for $sp$-orbital electrons on a two-dimensional triangular lattice under the $D_{\mathrm{6h}}(6/mmm)$ symmetry.
The Hamiltonian is given by 
\begin{equation}
  \hat{H}=\hat{H}_{0}+\hat{H}_{\mathrm{eff}}.
\end{equation}
The first term represents the one-particle Hamiltonian and the second term is the effective interaction term given by Eq.~(\ref{eq:cooper_pairing}).
$\hat{H}_{0}$ is represented by 
\begin{equation}
  \hat{H}_{0}=\sum_{\bm{k}}\sum_{\alpha\alpha^{\prime}}h_{\alpha\alpha^{\prime}}(\bm{k})\hat{c}_{\bm{k}\alpha}^{\dag}\hat{c}_{\bm{k}\alpha^{\prime}}
\end{equation}
with $\alpha=(\phi,\sigma);\phi=s,p_{x},p_{y},p_{z};\sigma=\uparrow,\downarrow$, where the Hamiltonian matrix $\braket{\alpha|\hat{h}(\bm{k})|\alpha^{\prime}}=h_{\alpha\alpha^{\prime}}(\bm{k})$ is divided into four parts as follows:
\begin{equation}
  \hat{h}(\bm{k})=\hat{h}_{\mathrm{hop}}(\bm{k})+\hat{h}_{\mathrm{SOC}}+\hat{h}_{\mathrm{CEF}}+\hat{h}_{\mathrm{MF}}.
\end{equation}
The first term $\hat{h}_{\mathrm{hop}}(\bm{k})\propto \delta_{\sigma\sigma^{\prime}}$ stands for the hopping term.
We consider only the nearest-neighbor hopping: $t_{s}$ for the amplitude between $s$ orbitals, $t_{p\sigma},t_{p\pi}$ for that between $p$ orbitals of $\sigma$ and $\pi$ couplings, and $t_{sp}$ for that between $sp$ orbitals.
The second term $\hat{h}_{\mathrm{SOC}}$ represents the atomic spin--orbit coupling:
\begin{equation}
  \hat{h}_{\mathrm{SOC}}=\lambda\bm{l}\cdot\bm{\sigma},
\end{equation}
which divides the six degenerate $p$-orbital levels into a doublet $j=1/2$ and a quartet $j=3/2$.
The third term $\hat{h}_{\mathrm{CEF}}\propto\delta_{\sigma\sigma^{\prime}}$ denotes the crystalline electric field under the hexagonal symmetry.
Since $s\in A_{1g}$, $(p_{x},p_{y})\in E_{1u}$, and $p_{z}\in A_{2u}$, this term can be parametrized as 
\begin{equation}
  \braket{\phi|\hat{h}_{\mathrm{CEF}}|\phi^{\prime}}=\delta_{\phi,s}\delta_{\phi^{\prime},s}\Delta_{\mathrm{CEF}}^{s}+\delta_{\phi,p_{z}}\delta_{\phi^{\prime},p_{z}}\Delta_{\mathrm{CEF}}^{z},
\end{equation}
where we set the energy level of $(p_x, p_y)$ as the origin.
The fourth term gives the local molecular field to induce additional multipoles belonging to the IR different from the totally symmetric one in the normal state, which results in the symmetry lowering from $D_{\mathrm{6h}}$.
The expression of $\hat{h}_{\mathrm{MF}}$ is given by 
\begin{equation}
  \hat{h}_{\mathrm{MF}}=-g\hat{\chi}
  \label{eq:molecular_field}
\end{equation}
with the amplitude $g$ and the multipole matrix $\hat{\chi}$.

Taking the mean-field approximation in Eq.~(\ref{eq:cooper_pairing}) and using the expression for the pair potential in Eq.~(\ref{eq:pair_definition}), we thereby obtain as 
\begin{equation}
  \hat{\mathcal{H}}_{\mathrm{eff}}=\frac{1}{2}
  \sum_{\bm{k}}\sum_{\alpha\beta}[\check{\Delta}(\bm{k})]_{\alpha\beta}
  \hat{c}_{\bm{k}\alpha}^{\dag}\hat{c}_{-\bm{k}\beta}^{\dag}+\mathrm{H.C.},
  \label{{eq:pair_Hamiltonian}}
\end{equation}
where $\mathrm{H.C.}$ is the hermitian conjugated.
The stability of the SC state is investigated by taking the mean-field approximation for the pair potential.
We also assume the on-site attractive pairing interaction $V_{\bm{q}}=V_{0}<0$.
We set the unit of the energy is the absolute value of hopping between $s$-orbital electrons $t_{s}$, even though we neglect this contribution in Sec.~\ref{subsec:ferroaxial}.
In the following subsections, we use the parameters as 
\begin{equation}
  \begin{aligned}
    &\, t_{s}=-1,
    t_{p\sigma}=0.5,
    t_{p\pi}=0.2,
    t_{sp}=0.3,
    \\
    &\, \Delta_{\mathrm{CEF}}^{s}=-0.2,
    \Delta_{\mathrm{CEF}}^{z}=-0.1,
    \lambda=0.3,
    g=0.4,
    \\
    &\, 
    V_{0}=-2,
    N=128^{2},
  \end{aligned}
\end{equation}
where $N$ is the total number of sites.
We fixed the total density per an orbital $n$ ($0<n<2$) as $n=0.8$.

The pair potential at the temperature $T$ is obtained by solving the gap equation
\begin{equation}
  [\check{\Delta}]_{\alpha\beta}(T)
  =|V_{0}|\sum_{\gamma\delta}\Lambda_{\alpha\gamma}F_{\gamma\delta}(T)\Lambda_{\beta\delta}
  \label{eq:gap_eq}
\end{equation}
with $F_{\gamma\delta}(T)\equiv (1/N)\sum_{\bm{k}^{\prime}}F_{\gamma\delta}(\bm{k}^{\prime})$.
The amplitude $\Delta_{\nu}(T)$ for multipole $\hat{\chi}_{\nu}$ 
is obtained by substituting the solution of Eq.~(\ref{eq:gap_eq}) into $\check{\Delta}(\bm{k})$ of Eq.~(\ref{eq:gap_amplitude}).

\subsection{Ferroelectric fluctuation under $C_{\mathrm{6v}}$ symmetry}

\begin{figure}[t]
  \centering
  \includegraphics[width=\linewidth]{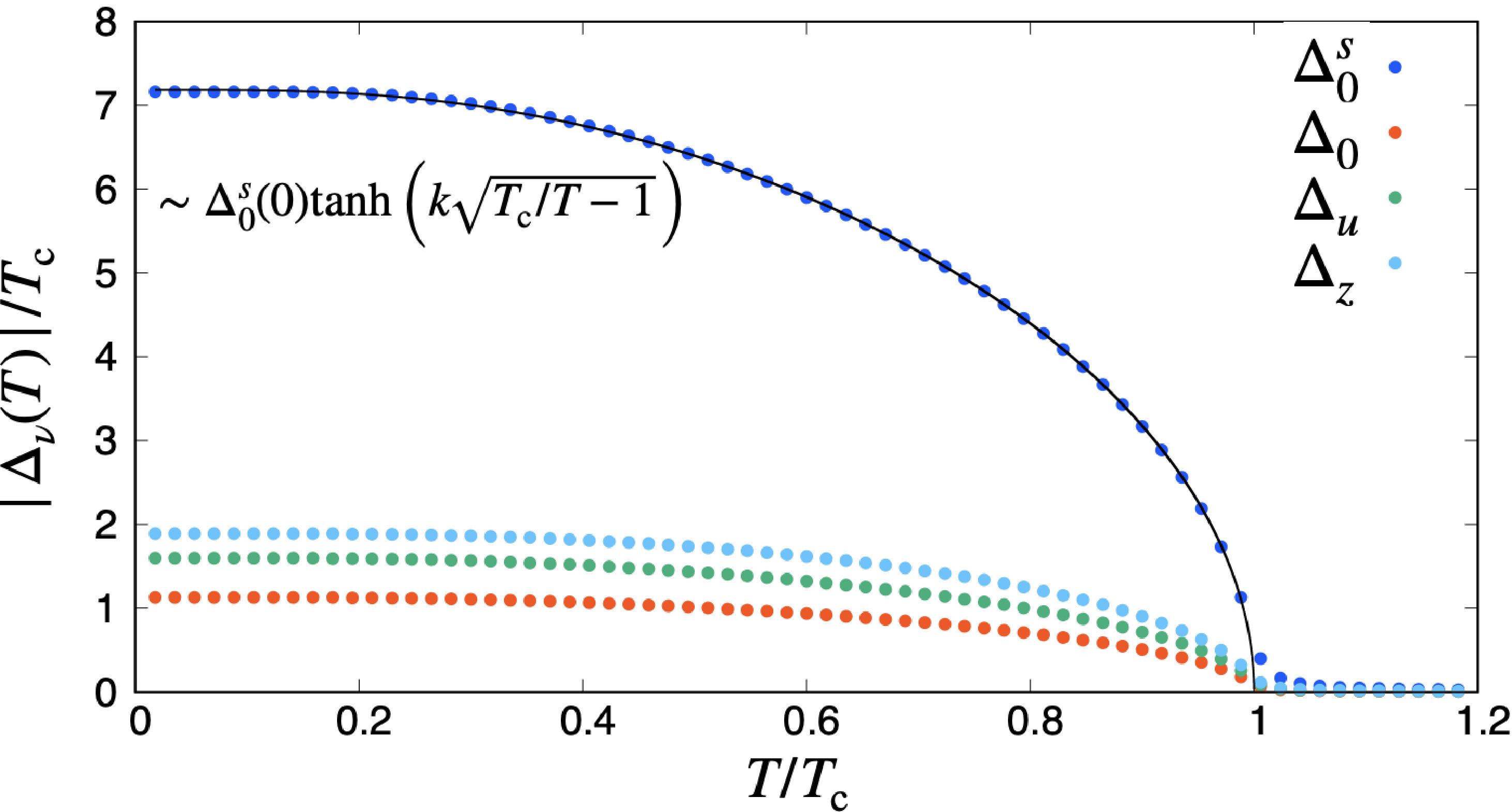}
  \caption{
    Temperature dependence of the pair potential in the polar system.
    The black solid line shows the fitting line of $\Delta_{0}^{s}(T)$ by $\Delta_{0}^{s}(0)\tanh(k\sqrt{T_{\mathrm{c}}/T-1})$ with the fitting parameter $k$.
    The critical temperature is $T_{\mathrm{c}}\simeq 2.85\times 10^{-3}$.
  }
  \label{fig:pair_fesc}
\end{figure}

First, we discuss the pairing state by ferroelectric fluctuation given by Eq.~(\ref{eq:FE_vertex}) under ferroelectric ordering with $\hat{\chi}=\hat{Q}_{z}$.
The general form of the pair potential is given by Eq.~(\ref{eq:FE_pair_potential}).
Figure~\ref{fig:pair_fesc} shows the temperature dependence of the amplitude $\Delta_{\nu}(T)$ in Eq.~(\ref{eq:FE_pair_potential}).
$\Delta_{0}\neq 0$ represents the isotropic pairing in the $p$-orbitals, whereas $\Delta_{u}\neq 0$ represents anisotropy of the pairings between $(p_x, p_y)$ or $p_z$ orbitals according to the hexagonal point-group symmetry.
One finds that $\Delta_{z}$ that arises from the E dipole degree of freedom $\hat{Q}_{z}$ becomes nonzero under the ferroelectric ordering.
We note that $\Delta_{z}$ vanishes if we turn off the molecular field $g=0$.

We also study the Pauli depairing effect due to the Zeeman coupling: 
\begin{equation}
  \hat{H}_{\mathrm{Zeeman}}=-\mu_{\mathrm{B}}\bm{H}\cdot\bm{\sigma},
\end{equation}
where $\mu_{\mathrm{B}}$ is the Bohr magneton.
Here, we consider the two directions of the magnetic field, i.e., the in-plane magnetic field $\bm{H}=(H_{x},0,0)$ and the perpendicular magnetic field $\bm{H}=(0,0,H_{z})$.
We neglect the orbital component of the magnetic field for simplicity.
We also set $\mu_{\mathrm{B}}=1$ in the following.

\begin{figure}[t]
  \centering
  \includegraphics[width=\linewidth]{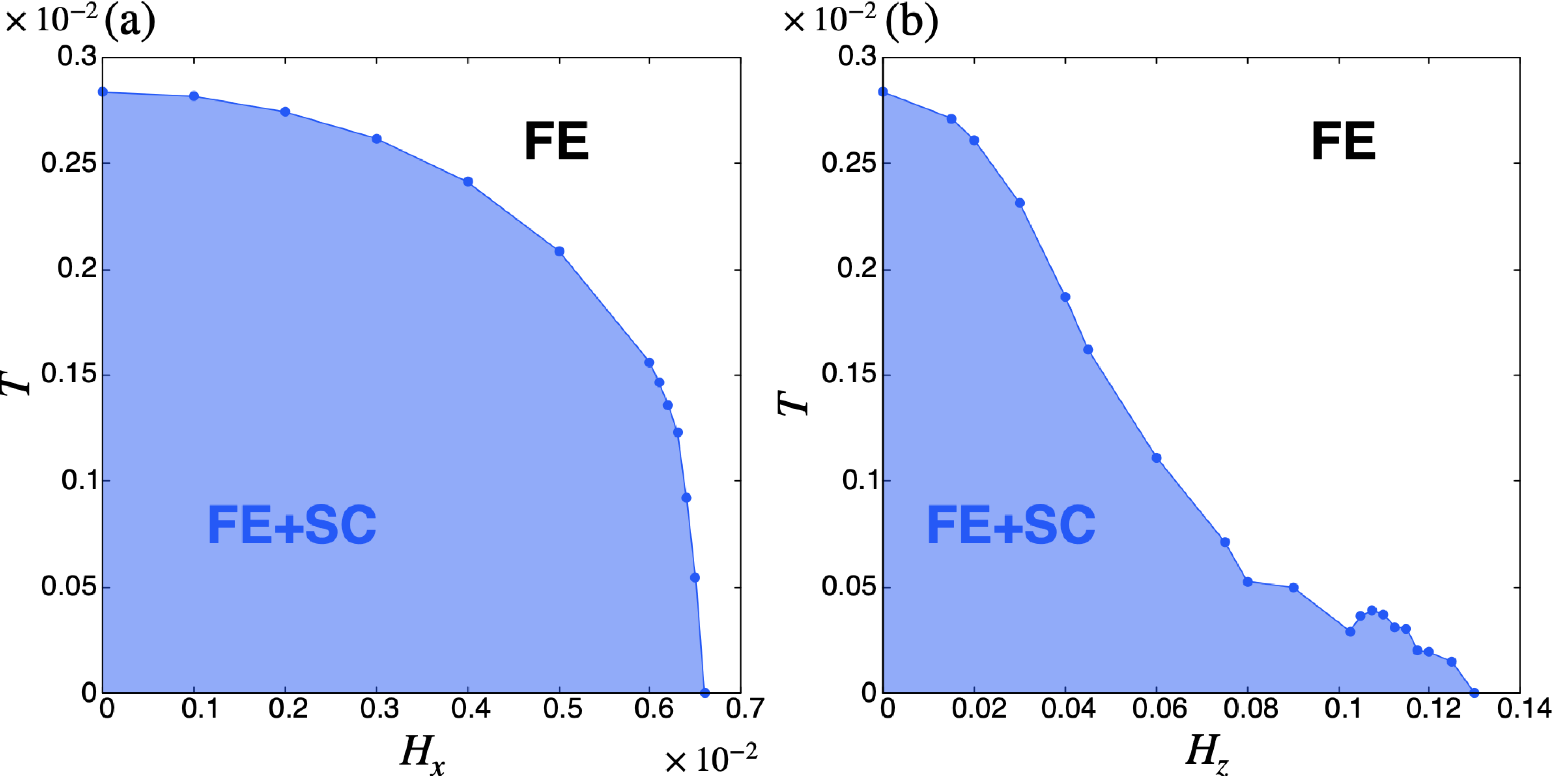}
  \caption{
    SC phase diagrams under the ferroelectric (FE) molecular field.
    The panels (a) and (b) show the result for applied in-plane and perpendicular magnetic fields, respectively.
  }
  \label{fig:phase_fesc}
\end{figure}

Figure~\ref{fig:phase_fesc} shows the $H$--$T$ phase diagram in the presence of the polar field $g=0.4$.
Comparing two phase diagrams, the SC state with the ferroelectric moment (nonzero $\braket{\hat{Q}_{z}}$) (FE+SC state) is stabilized by the applied perpendicular magnetic field.
This behavior has also been found in other noncentrosymmetric superconductors~\cite{JPSJ.76.051008, PhysRevB.98.024521}.
The different stability tendency against the field directions is attributed to the effective spin--orbit coupling in momentum space;
the Rashba-type antisymmetric spin--orbit coupling $(\bm{k}\times\bm{\sigma})_{z}$ is induced by the local molecular field $Q_{z}$, which locks the spin direction of the Cooper pairs in the $k_{x}$--$k_{y}$ plane.
In this case, the in-plane magnetic field deforms the Fermi surfaces asymmetrically, whereas the perpendicular magnetic field shifts the magnitude of the splitting of the Fermi surfaces.
Since the Pauli depairing effect does not affect the latter case, the Cooper pairs are rarely destroyed~\cite{NewJPhys.6.115}.
Thus, the Cooper pairs are robust against the perpendicular magnetic field.
We show the additional data for the molecular field dependence of the stability of the SC state against the magnetic field in Appendix~\ref{app:mf}.

It is noted that the state above the critical temperature $T_{\mathrm{c}}$ corresponds to the FE phase $\braket{\hat{Q}_{z}}\neq 0$ owing to the molecular-field term $g$.
Accordingly, the transition from the FE+SC state to the FE state is the second-order phase transition even in increasing the magnetic field in our calculation, which
is different from the previous study~\cite{PhysRevB.98.024521}.

\subsection{Ferroaxial fluctuation under $C_{\mathrm{6h}}$ symmetry}
\label{subsec:ferroaxial}

The second example is the SC arising from an ET dipole fluctuation by Eq.~(\ref{eq:FA_vertex}) under ferroaxial ordering.
When the symmetry of the system reduces from $D_{\mathrm{6h}}$ to $C_{\mathrm{6h}}$ as a consequence of the breaking of vertical mirror symmetry, the ET dipole $G_{z}$ is ordered.
We call the SC state with nonzero $\braket{\hat{G}_{z}}$ a ferroaxial SC (FA+SC) state.
From the previous model, we neglect the $s$-orbital degree of freedom, i.e., $t_{s}=0$ and $\Delta_{\mathrm{CEF}}^{s}=0$ in the following numerical calculation.
In addition, we take $\hat{\chi}=\hat{G}_{z}$ instead of $\hat{Q}_{z}$ in Eq.~(\ref{eq:molecular_field}).

\begin{figure}[t]
  \centering
  \includegraphics[width=\linewidth]{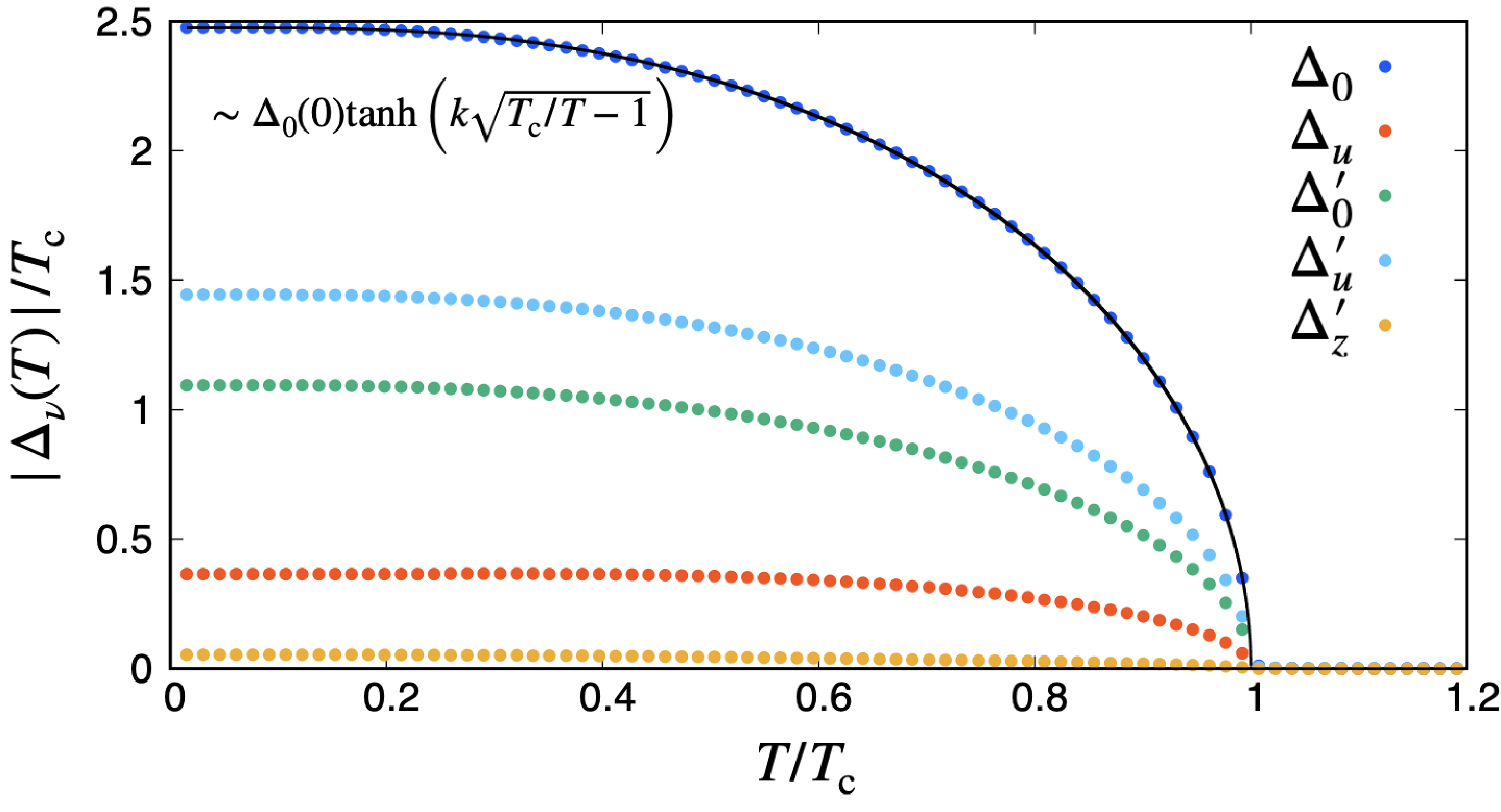}
  \caption{
    Temperature dependence of pair potential in ferroaxial ordering.
    The black solid line shows the fitting line of $\Delta_{0}(T)$ by $\Delta_{0}(0)\tanh(k\sqrt{T_{\mathrm{c}}/T-1})$.
    The critical temperature is $T_{\mathrm{c}}\simeq 6.55\times 10^{-2}$.
  }
  \label{fig:pair_fasc}
\end{figure}

\begin{figure}[t]
  \centering
  \includegraphics[width=\linewidth]{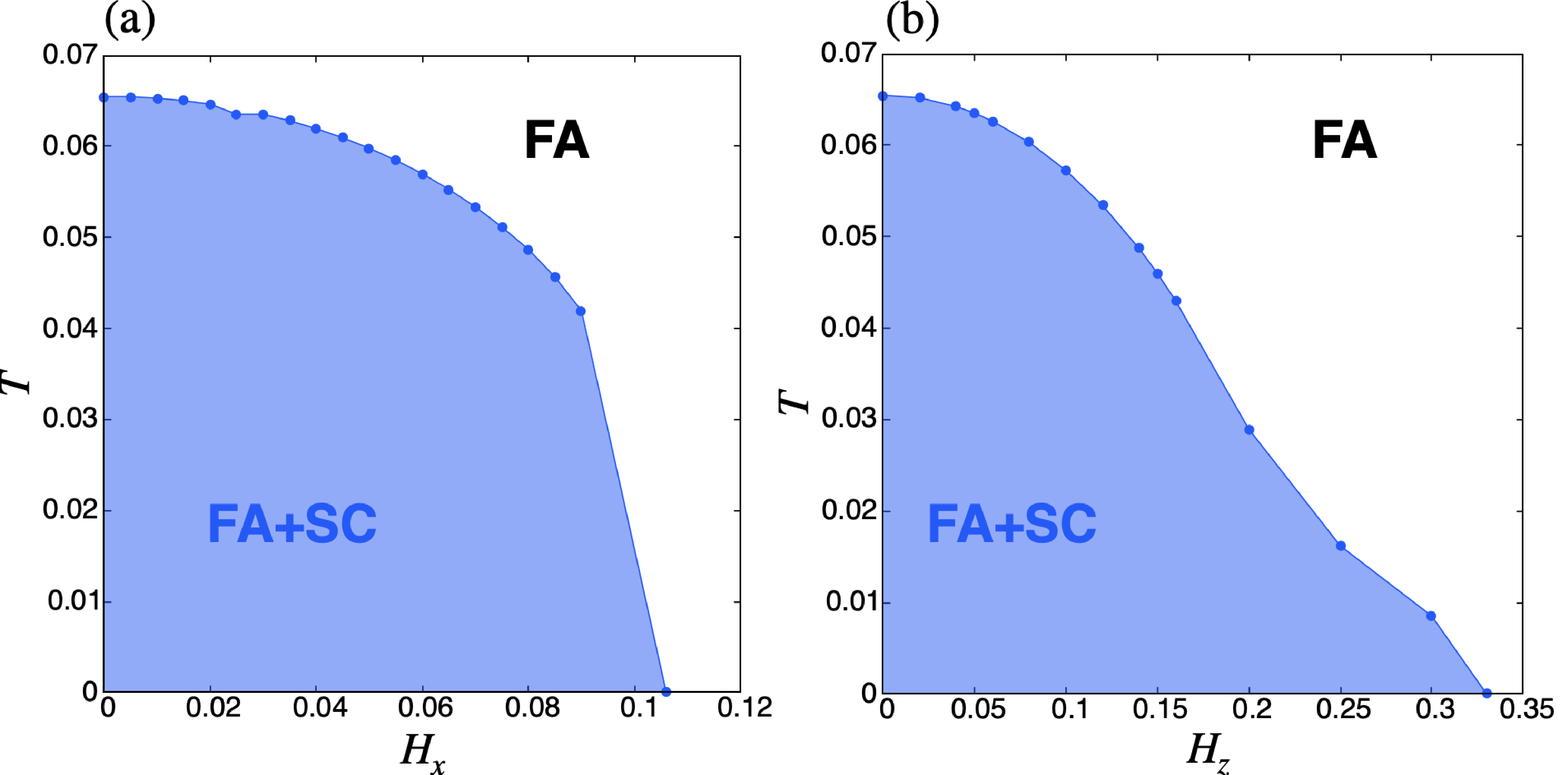}
  \caption{
    SC phase diagrams under the ferroaxial (FA) molecular field.
    The panels (a) and (b) show the result for applied in-plane and perpendicular magnetic fields, respectively.
  }
  \label{fig:phase_fasc}
\end{figure}

Figure~\ref{fig:pair_fasc} shows the temperature dependence of the pair potential when the ET dipole $\hat{G}_{z}$ emerges under the ferroaxial ordering.
$\Delta_{z}^{\prime}$ vanishes if we turn off the molecular field $g=0$.
Under the basis of $\ket{jj_{z}}$, the pair potential in this case can be expressed as 
\begin{equation}
  \begin{aligned}
    \hat{\Delta}^{(J)}=&\, \Delta_{0}^{(1)}\hat{Q}_{0}^{(1)}
    +\Delta_{0}^{(3)}\hat{Q}_{0}^{(3)}
    +\Delta_{u}^{(3)}\hat{Q}_{u}^{(3)}
    \\
    &\, +\Delta_{u}^{(2)}\hat{Q}_{u}^{(2)}
    +\Delta_{z}^{(2)}\hat{G}_{z}^{(2)},
  \end{aligned}
\end{equation}
\begin{subequations}
  \begin{equation}
    \Delta_{0}^{(1)}=\frac{\Delta_{0}-\sqrt{2}\Delta_{0}^{\prime}}{\sqrt{3}},
    \Delta_{0}^{(3)}=\frac{\sqrt{2}\Delta_{0}+\Delta_{0}^{\prime}}{\sqrt{3}}
  \end{equation}
  \begin{equation}
    \Delta_{u}^{(2)}=\frac{\sqrt{2}\Delta_{u}+\Delta_{u}^{\prime}}{\sqrt{3}},
    \Delta_{u}^{(3)}=\frac{\Delta_{u}-\sqrt{2}\Delta_{u}^{\prime}}{\sqrt{3}}
  \end{equation}
  \begin{equation}
    \Delta_{z}^{(2)}=\Delta_{z}^{\prime}
  \end{equation}
\end{subequations}
where the superscripts $(J)$ for $J=1,2,3$ represent the pairing between electrons with total angular momentum $(j_{1},j_{2})=(1/2,1/2), (1/2,3/2)$, and $(3/2,3/2)$, respectively(See Appendix~\ref{app:jjz}).
We also find that the emergence of $\hat{Q}_{u}^{\prime}$, which contributes to $\Delta_{u}^{(2)}$, results in the pairing between $j=1/2$ electrons and $j=3/2$ electrons.
Figures~\ref{fig:phase_fasc}(a) and \ref{fig:phase_fasc}(b) show the $H-T$ phase diagrams under the ferroaxial molecular field for in-plane and perpendicular magnetic fields, respectively.
In contrast to the result in Fig.~\ref{fig:phase_fesc}, the anisotropy against the in-plane and out-of-plane fields is small, which might be attributed to the fact that there is no spin splitting in the band structure under the ferroaxial ordering.
We show the molecular-field strength dependence of the pair potentials in Appendix~\ref{app:mf}.

\section{Summary and Disussion}
\label{sec:summary}

To summarize, we have extended the multipole basis in the normal space into the Nambu space to systematically characterize the SC order parameter in the framework of four types of multipoles.
In a multiorbital space, pairings between electrons with any angular momentum can be expressed by using ET and MT multipoles as well as E and M multipoles, which enables us to describe Cooper pairs formed by unconventional electronic degrees of freedom.
We have demonstrated that unconventional multipole SC states can be realized by multipole fluctuations in Sec.~\ref{sec:application}.
Our formalism includes all the SC order parameters in the multiorbital system that have been overlooked in the previous studies; the SC state in the ferroaxial system, characterized by the ET dipole, is a typical example.
In this way, our complete classification would open a path to examine the nature of the multiorbital SC state that has never been observed.

Another advantage based on multipole representation theory is that it can predict the cross-correlated phenomena, and such a study has been already done in a normal state~\cite{PhysRevB.98.165110,PhysRevB.98.245129}.
The appearance of odd-parity and toroidal-type multipole degrees of freedom in the pair potential indicates further cross-correlated responses in the SC state.
Indeed, in noncentrosymmetric SCs, the electromagnetic effects due to supercurrents have been studied as nonreciprocal phenomena~\cite{JPSJ.76.051008}, and recently the SC diode effect has been found~\cite{nature.584.373, PhysRevLett.131.027001}. 
Our results would serve as a reference to explore such intriguing cross-correlated responses driven by unconventional SC with multipole degrees of freedom.

\begin{acknowledgments}
  This research was supported by JSPS KAKENHI Grants Numbers JP21H01037, JP22H04468, JP22H00101, JP22H01183, JP23K03288, JP23H04869, JP23K20827, and by JST PRESTO (JPMJPR20L8) and JST CREST (JPMJCR23O4).
  Parts of the numerical calculations were performed in the supercomputing systems in ISSP, the University of Tokyo.
\end{acknowledgments}

\appendix

\section{Matrix elements of activated multipoles in $sp$-orbital space}
\label{app:MP}

In this appendix, we summarize the matrix elements of activated multipoles in $sp$-orbital space.
First, we give the matrix elements for the spinless multipoles.
For the spinless basis $\{\ket{s},\ket{p_{x}},\ket{p_{y}},\ket{p_{z}}\}$, the relevant matrix elements of the multipoles are given by~\cite{JPSJ.87.033709}
\begin{subequations}
  \label{eq:spinless_multipoles}
  \begin{equation}
    Q_{0}^{s}=
    \begin{pmatrix}
      1 & 0 & 0 & 0
      \\
      0 & 0 & 0 & 0
      \\
      0 & 0 & 0 & 0
      \\
      0 & 0 & 0 & 0
    \end{pmatrix},
    Q_{0}=\frac{1}{\sqrt{3}}
    \begin{pmatrix}
      0 & 0 & 0 & 0
      \\
      0 & 1 & 0 & 0
      \\
      0 & 0 & 1 & 0
      \\
      0 & 0 & 0 & 1
    \end{pmatrix},
  \end{equation}
  \begin{equation}
    \begin{aligned}
      &\, M_{x}=\frac{i}{\sqrt{2}}
      \begin{pmatrix}
        0 & 0 & 0 & 0 
        \\
        0 & 0 & 0 & 0 
        \\
        0 & 0 & 0 & -1 
        \\
        0 & 0 & 1 & 0 
      \end{pmatrix},
      M_{y}=\frac{i}{\sqrt{2}}
      \begin{pmatrix}
        0 & 0 & 0 & 0 
        \\
        0 & 0 & 0 & 1 
        \\
        0 & 0 & 0 & 0 
        \\
        0 & -1 & 0 & 0 
      \end{pmatrix},
      \\
      &\, M_{z}=\frac{i}{\sqrt{2}}
      \begin{pmatrix}
        0 & 0 & 0 & 0 
        \\
        0 & 0 & -1 & 0 
        \\
        0 & 1 & 0 & 0 
        \\
        0 & 0 & 0 & 0 
      \end{pmatrix},
      \\
    \end{aligned}
  \end{equation}
  \begin{equation}
    \begin{aligned}
      &\, Q_{u}=\frac{1}{\sqrt{6}}
      \begin{pmatrix}
        0 & 0 & 0 & 0
        \\
        0 & -1 & 0 & 0
        \\
        0 & 0 & -1 & 0
        \\
        0 & 0 & 0 & 2
      \end{pmatrix},
      Q_{v}=\frac{1}{\sqrt{2}}
      \begin{pmatrix}
        0 & 0 & 0 & 0
        \\
        0 & 1 & 0 & 0
        \\
        0 & 0 & -1 & 0
        \\
        0 & 0 & 0 & 0
      \end{pmatrix},
      \\
      &\, Q_{yz}=\frac{1}{\sqrt{2}}
      \begin{pmatrix}
        0 & 0 & 0 & 0
        \\
        0 & 0 & 0 & 0
        \\
        0 & 0 & 0 & 1
        \\
        0 & 0 & 1 & 0
      \end{pmatrix},
      Q_{zx}=\frac{1}{\sqrt{2}}
      \begin{pmatrix}
        0 & 0 & 0 & 0
        \\
        0 & 0 & 0 & 1
        \\
        0 & 0 & 0 & 0
        \\
        0 & 1 & 0 & 0
      \end{pmatrix},
      \\
      &\, Q_{xy}=\frac{1}{\sqrt{2}}
      \begin{pmatrix}
        0 & 0 & 0 & 0
        \\
        0 & 0 & 1 & 0
        \\
        0 & 1 & 0 & 0
        \\
        0 & 0 & 0 & 0
      \end{pmatrix},
    \end{aligned}
  \end{equation}
  \begin{equation}
    \begin{aligned}
      &\, Q_{x}=\frac{1}{\sqrt{2}}
      \begin{pmatrix}
        0 & 1 & 0 & 0
        \\
        1 & 0 & 0 & 0
        \\
        0 & 0 & 0 & 0
        \\
        0 & 0 & 0 & 0
      \end{pmatrix},
      Q_{y}=\frac{1}{\sqrt{2}}
      \begin{pmatrix}
        0 & 0 & 1 & 0 
        \\
        0 & 0 & 0 & 0 
        \\
        1 & 0 & 0 & 0 
        \\
        0 & 0 & 0 & 0 
      \end{pmatrix},
      \\
      &\, Q_{z}=\frac{1}{\sqrt{2}}
      \begin{pmatrix}
        0 & 0 & 0 & 1 
        \\
        0 & 0 & 0 & 0 
        \\
        0 & 0 & 0 & 0 
        \\
        1 & 0 & 0 & 0 
      \end{pmatrix},
      \\
    \end{aligned}
  \end{equation}
  and 
  \begin{equation}
    \begin{aligned}
      &\, T_{x}=\frac{i}{\sqrt{2}}
      \begin{pmatrix}
        0 & 1 & 0 & 0
        \\
        -1 & 0 & 0 & 0
        \\
        0 & 0 & 0 & 0
        \\
        0 & 0 & 0 & 0
      \end{pmatrix},
      T_{y}=\frac{i}{\sqrt{2}}
      \begin{pmatrix}
        0 & 0 & 1 & 0 
        \\
        0 & 0 & 0 & 0 
        \\
        -1 & 0 & 0 & 0 
        \\
        0 & 0 & 0 & 0 
      \end{pmatrix},
      \\
      &\, T_{z}=\frac{i}{\sqrt{2}}
      \begin{pmatrix}
        0 & 0 & 0 & 1 
        \\
        0 & 0 & 0 & 0 
        \\
        0 & 0 & 0 & 0 
        \\
        -1 & 0 & 0 & 0 
      \end{pmatrix},
    \end{aligned}
  \end{equation}
\end{subequations}
where $(Q_{0}^{s},Q_{0})$ represent the electric monopole, $(M_{x},M_{y},M_{z})$ represent the magnetic dipole, $(Q_{u},Q_{v},Q_{yz},Q_{zx},Q_{xy})$ represent the electric quadrupole, $(Q_{x},Q_{y},Q_{z})$ represent the electric dipole, and $(T_{x},T_{y},T_{z})$ represent the magnetic toroidal dipole.
Here, these spinless multipoles $\chi_{\nu}$ are orthogonal with each other and normalized such as $\mathrm{Tr}[\chi_{\nu}\chi_{\mu}]=\delta_{\nu\mu}$.
The angular momentum in $p$-orbital space can be expressed as $l_{j}=\sqrt{2}M_{j}(j=x,y,z)$.

According to Eq.~(\ref{eq:spinful_def}) with Eqs.~(\ref{eq:pauli}) and (\ref{eq:spinless_multipoles}), we obtain the spinful multipoles $\hat{X}_{lm}(s,k)$.
The relation between $\hat{X}_{lm}(s,k)$ and $X_{lm}^{(\mathrm{orb})}$ is summarized in Table~\ref{tab:spinful_spinless}~\cite{JPSJ.89.104704}.
We added the minus sign for $\hat{T}_{l}(1,0)$ and $\hat{G}_{l}(1,0)$ for convenience.
Note that $\hat{X}_{lm}(0,0)=X_{lm}^{(\mathrm{orb})}\sigma_{0}$.
\begin{table}[htbp]
  \caption{
  The relation between spinful multipoles $X_{lm}(s,k)$ and corresponding spinless multipoles $X_{lm}^{(\mathrm{orb})}$.
  }
  \begin{ruledtabular}
    \begin{tabular}{cccc}
      $\hat{X}_{l}(s,k)$ & $s=0,k=0$ & $s=1,k=0$ & $s=1,k=\pm 1$ 
      \\
      \hline
      $\hat{Q}_{l}(s,k)$ & $Q_{l}$ & $T_{l}$ & $M_{l\pm 1}$ 
      \\
      $\hat{M}_{l}(s,k)$ & $M_{l}$ & $G_{l}$ & $Q_{l\pm 1}$ 
      \\
      $\hat{T}_{l}(s,k)$ & $T_{l}$ & $-Q_{l}$ & $G_{l\pm 1}$ 
      \\
      $\hat{G}_{l}(s,k)$ & $G_{l}$ & $-M_{l}$ & $T_{l\pm 1}$ 
      \\
    \end{tabular}
  \end{ruledtabular}
  \label{tab:spinful_spinless}
\end{table}

Finally, the expressions of spinful multipoles $\hat{X}_{lm}(1,k)$ are given as follows: 
\begin{subequations}
  \label{eq:spinful_MPs}
  \begin{itemize}
    \item M dipole
      \begin{equation}
        \hat{\bm{M}}^{(s)}_{s}=Q_{0}^{s}\bm{\sigma}
      \end{equation}
  \end{itemize}
  in $s$-$s$ orbital space.
  \begin{itemize}
    \item E monopole
    \begin{equation}
      \hat{Q}_{0}^{\prime}=\frac{1}{\sqrt{3}}\bm{M}\cdot\bm{\sigma}
    \end{equation}
    \item E quadrupole
    \begin{equation}
      \begin{aligned}
        &\, \hat{Q}_{u}^{\prime}=\frac{1}{\sqrt{6}}(2M_{z}\sigma_{z}-M_{x}\sigma_{x}-M_{y}\sigma_{y})
        \\
        &\, \hat{Q}_{v}^{\prime}=\frac{1}{\sqrt{2}}(M_{x}\sigma_{x}-M_{y}\sigma_{y})
        \\
        &\, \hat{Q}_{yz}^{\prime},\hat{Q}_{zx}^{\prime}\hat{Q}_{xy}^{\prime}=\frac{1}{\sqrt{2}}(M_{y}\sigma_{z}+M_{z}\sigma_{y}), (\mathrm{cyclic})
      \end{aligned}
    \end{equation}
    \item M dipole
      \begin{equation}
        \hat{\bm{M}}^{(s)}_{p}=Q_{0}\bm{\sigma}
      \end{equation}
      and
      \begin{equation}
        \begin{aligned}
          &\, \hat{M}_{x}^{\prime}=-\frac{1}{\sqrt{10}}[(Q_{u}-\sqrt{3}Q_{v})\sigma_{x}-\sqrt{3}(Q_{zx}\sigma_{x}+Q_{xy}\sigma_{y})]
          \\
          &\, \hat{M}_{y}^{\prime}=-\frac{1}{\sqrt{10}}[(Q_{u}+\sqrt{3}Q_{v})\sigma_{y}-\sqrt{3}(Q_{xy}\sigma_{x}+Q_{yz}\sigma_{z})]
          \\
          &\, \hat{M}_{z}^{\prime}=-\frac{1}{\sqrt{10}}[-2Q_{u}\sigma_{z}-\sqrt{3}(Q_{yz}\sigma_{y}+Q_{zx}\sigma_{x})]
        \end{aligned}
      \end{equation}
      \item M octupole
      \begin{equation}
        \begin{aligned}
          &\, \hat{M}_{xyz}^{\prime}=\frac{1}{\sqrt{3}}(Q_{yz}\sigma_{x}+Q_{zx}\sigma_{y}+Q_{xy}\sigma_{z})
          \\
          &\, \hat{M}_{x}^{\alpha\prime}=-\frac{1}{\sqrt{5}}\left[\frac{\sqrt{3}}{2}(Q_{u}-\sqrt{3}Q_{v})\sigma_{x}+Q_{zx}\sigma_{z}+Q_{xy}\sigma_{y}\right]
          \\
          &\, \hat{M}_{y}^{\alpha\prime}=-\frac{1}{\sqrt{5}}\left[\frac{\sqrt{3}}{2}(Q_{u}+\sqrt{3}Q_{v})\sigma_{y}+Q_{xy}\sigma_{x}+Q_{yz}\sigma_{z}\right]
          \\
          &\, \hat{M}_{z}^{\alpha\prime}=-\frac{1}{\sqrt{5}}(-\sqrt{3}Q_{u}\sigma_{z}+Q_{yz}\sigma_{y}+Q_{zx}\sigma_{x})
          \\
          &\, \hat{M}_{x}^{\beta\prime}=-\frac{1}{\sqrt{3}}\left[\frac{1}{2}(\sqrt{3}Q_{u}+Q_{v})\sigma_{x}+Q_{zx}\sigma_{z}-Q_{xy}\sigma_{y}\right]
          \\
          &\, \hat{M}_{y}^{\beta\prime}=-\frac{1}{\sqrt{3}}\left[\frac{1}{2}(-\sqrt{3}Q_{u}+Q_{v})\sigma_{y}+Q_{xy}\sigma_{x}-Q_{yz}\sigma_{z}\right]
          \\
          &\, \hat{M}_{z}^{\beta\prime}=-\frac{1}{\sqrt{3}}(-Q_{v}\sigma_{z}+Q_{yz}\sigma_{y}-Q_{zx}\sigma_{x})
        \end{aligned}
      \end{equation}
      \item MT quadrupole
      \begin{equation}
        \begin{aligned}
          &\, \hat{T}_{u}^{\prime}=-\frac{1}{\sqrt{2}}(Q_{yz}\sigma_{x}-Q_{zx}\sigma_{y})
          \\
          &\, \hat{T}_{v}^{\prime}=\frac{1}{\sqrt{6}}(2Q_{xy}\sigma_{z}-Q_{yz}\sigma_{x}-Q_{zx}\sigma_{y})
          \\
          &\, \hat{T}_{yz}^{\prime}=\frac{1}{\sqrt{6}}[(\sqrt{3}Q_{u}+Q_{v})\sigma_{x}-(Q_{zx}\sigma_{z}-Q_{xy}\sigma_{y})]
          \\
          &\, \hat{T}_{zx}^{\prime}=\frac{1}{\sqrt{6}}[(-\sqrt{3}Q_{u}+Q_{v})\sigma_{y}-(Q_{xy}\sigma_{x}-Q_{yz}\sigma_{z})]
          \\
          &\, \hat{T}_{xy}^{\prime}=\frac{1}{\sqrt{6}}[-2Q_{v}\sigma_{z}-(Q_{yz}\sigma_{y}-Q_{zx}\sigma_{x})]
        \end{aligned}
      \end{equation}
      \item ET dipole
      \begin{equation}
        \hat{\bm{G}}^{\prime}=\frac{1}{\sqrt{2}}\bm{M}\times\bm{\sigma}
      \end{equation}
  \end{itemize}
  in $p$-$p$ orbital space.
  \begin{itemize}
    \item M/ET monopole
    \begin{equation}
      \hat{M}_{0}^{\prime}=\frac{1}{\sqrt{3}}\bm{Q}\cdot\bm{\sigma},
      \hat{G}_{0}^{\prime}=-\frac{1}{\sqrt{3}}\bm{T}\cdot\bm{\sigma}
    \end{equation}
    \item E/MT dipole
    \begin{equation}
      \hat{\bm{Q}}^{\prime}=-\frac{1}{\sqrt{2}}\bm{T}\times\bm{\sigma},
      \hat{\bm{T}}^{\prime}=\frac{1}{\sqrt{2}}\bm{Q}\times\bm{\sigma}
    \end{equation}
    \item M/ET quadrupole
  \end{itemize}
  \begin{equation}
    \begin{aligned}
      &\, \hat{M}_{u}^{\prime}=\frac{1}{\sqrt{6}}(2Q_{z}\sigma_{z}-Q_{x}\sigma_{x}-Q_{y}\sigma_{y})
      \\
      &\, \hat{M}_{v}^{\prime}=\frac{1}{\sqrt{2}}(Q_{x}\sigma_{x}-Q_{y}\sigma_{y})
      \\
      &\, \hat{M}_{yz}^{\prime},\hat{M}_{zx}^{\prime},\hat{M}_{xy}^{\prime}=\frac{1}{\sqrt{2}}(Q_{y}\sigma_{z}+Q_{z}\sigma_{y}), (\mathrm{cyclic})
      \\
      &\, \hat{G}_{u}^{\prime}=\frac{1}{\sqrt{6}}(2T_{z}\sigma_{z}-T_{x}\sigma_{x}-T_{y}\sigma_{y})
      \\
      &\, \hat{G}_{v}^{\prime}=\frac{1}{\sqrt{2}}(T_{x}\sigma_{x}-T_{y}\sigma_{y})
      \\
      &\, \hat{G}_{yz}^{\prime},\hat{G}_{zx}^{\prime},\hat{G}_{xy}^{\prime}=\frac{1}{\sqrt{2}}(T_{y}\sigma_{z}+T_{z}\sigma_{y}), (\mathrm{cyclic})
      \label{sp-orbital_spinful}
    \end{aligned}
  \end{equation}
  in $s$-$p$ orbital space.
\end{subequations}
We have introduced the abbreviations $\bm{X}=(X_{x},X_{y},X_{z})$ for $X=Q,M,T,G$, and $\sigma$.
The multipoles matrices in the spinful space $\hat{O}_{\alpha}$ are orthogonal with each other and normalized such as $\mathrm{Tr}[\hat{O}_{\alpha}\hat{O}_{\beta}]=2\delta_{\alpha\beta}$.
Since matrix elements of the multipoles for any angular momentum can be evaluated as well, we can classify the pair potential for any angular momentum.

\section{Cooper pairing derived from local multipole fluctuations}
\label{app:cooper_channel}

We derive the Cooper pairing induced by local multipole-fluctuated interactions in $sp$-orbital system.
We decompose Eq.~(\ref{eq:cooper_pairing}) as following:
\begin{equation}
  \hat{\mathcal{H}}_{\mathrm{eff}}=\hat{\mathcal{H}}^{(0)}+\hat{\mathcal{H}}^{(1)}+\hat{\mathcal{H}}^{(2)},
  \label{eq:channel_decomposition}
\end{equation}
where $\hat{\mathcal{H}}^{(0)} (\hat{\mathcal{H}}^{(1)})$ is the pairing interaction between spinless (spinful) multipoles, whereas $\hat{\mathcal{H}}^{(2)}$ is that of the product of spinless and spinful multipoles.
We summarize the possible pairing and corresponding coefficients $c_{\mu\nu}$ in Eq.~(\ref{eq:channel_coeeficient}) by local fluctuations for the $p$-orbital system in Tables~\ref{tab:Pairing_p_ls_E}, \ref{tab:Pairing_p_ls_M}, and $sp$-orbital system in Tables~\ref{tab:Pairing_sp_ls_E}, \ref{tab:Pairing_sp_ls_M}.

It is possible to read trends in which pairing may occur by rewriting the multipole-fluctuated interactions into the pairing interactions.
First, $c_{\nu\nu}>0$ $(c_{\nu\nu}<0)$ indicates that the corresponding Cooper pair is favored when the interaction is attractive $V_{0}<0$ (repulsive $V_{0}>0$).
In particular, $\check{Q}_{0}$-type pairing is favored in E/ET (M/MT) multipole-fluctuated interactions with attractive (repulsive) interactions.
Second, we see what types of pairing are prohibited.
For instance, the E quadrupole fluctuation in terms of the $\hat{Q}_{v}$-type component never leads to the $\check{Q}_{yz},\check{Q}_{zx}$-type pairings in the $p$-orbital system.
On the other hand, the spinful multipole fluctuation can lead to pairing expressed by spinless multipoles through $\hat{\mathcal{H}}^{(2)}$.
For example, from Table~\ref{tab:Pairing_p_ls_E}, although spinful E quadrupole fluctuation $\hat{\Lambda}=\hat{Q}_{u}^{\prime}$ prohibits pairing interaction $\check{Q}_{u}^{\dag}\check{Q}_{u}$, $\check{Q}_{u}$-type paring arises thurogh the pairing interaction $\check{Q}_{u}^{\dag}\check{Q}_{0}^{\prime}+\check{Q}_{0}^{\prime\dag}\check{Q}_{u}$ in $\hat{\mathcal{H}}_{2}$.
The negative sign of $\tilde{X}\tilde{Y}$ columns indicates that the relative phase between $\tilde{X}$ and $\tilde{Y}$ are different by $\pi$.
The similar analysis can be applied to the multipoles with the momentum dependent structure factor.

\begin{center}
  \begin{longtable*}{lccccccccccccccc}
    \caption{
      The Cooper channel $\hat{\mathcal{H}}_{\mathrm{eff}}$ in Eq.~(\ref{eq:channel_decomposition}) induced  by local E/ET-type multipole fluctuations in $p$-orbital space.
      The first column is the multipole fluctuation $\hat{\Lambda}$ in Eq.~(\ref{eq:local_vertex}).
      The cloumns $\check{O}_{\nu}$ give the coefficients $c_{\nu\nu}$ for the pairing term expressed as $(V_{0}/2N)c_{\nu\nu}\check{O}_{\nu}^{\dag}\check{O}_{\nu}$, whereas $\check{O}_{\mu}\check{O}_{\nu}$ columns represent the coefficients $c_{\mu\nu}$ of the pairing term $(V_{0}/2N)c_{\mu\nu}(\check{O}_{\mu}^{\dag}\check{O}_{\nu}+\check{O}_{\nu}^{\dag}\check{O}_{\mu})$.
    }
    \label{tab:Pairing_p_ls_E}
    \\
    
    \hline\hline
    \multicolumn{16}{c}
    {spinless Cooper pairing $\mathcal{H}^{(0)}$}
    \\
    \hline\hline
    $\hat{\Lambda}$ & $\check{Q}_{0}$ & $\check{Q}_{u}$ &  $\check{Q}_{v}$ & $\check{Q}_{yz}$ & $\check{Q}_{zx}$ & $\check{Q}_{xy}$ & $\check{Q}_{0}\check{Q}_{u}$ & $\check{Q}_{0}\check{Q}_{v}$ & $\check{Q}_{u}\check{Q}_{v}$ 
    \\
    \endfirsthead
    
    \multicolumn{16}{c}
    {{\tablename\ \thetable{} -- continued from previous page}}
    \\
    \endhead
    
    \multicolumn{16}{c}{{Continued on next page}}
    \\ 
    \endfoot
    
    \hline\hline
    \endlastfoot

    \hline
    $\hat{Q}_{0}$ & $\frac{2}{3}$ & $\frac{2}{3}$ & $\frac{2}{3}$ & $\frac{2}{3}$ & $\frac{2}{3}$ & $\frac{2}{3}$ & $0$ & $0$ & $0$ 
    \\
    $\hat{Q}_{u}$ & $\frac{2}{3}$ & $1$ & $\frac{1}{3}$ & $-\frac{2}{3}$ & $-\frac{2}{3}$ & $\frac{1}{3}$ & $\frac{\sqrt{2}}{6}$ & $0$ & $0$ 
    \\
    $\hat{Q}_{v}$ & $\frac{2}{3}$ & $\frac{1}{3}$ & $1$ & $0$ & $0$ & $-1$ & $-\frac{\sqrt{2}}{6}$ & $0$ & $0$ 
    \\
    $\hat{Q}_{yz}$ & $\frac{2}{3}$ & $-\frac{2}{3}$ & $0$ & $1$ & $0$ & $0$ & $\frac{\sqrt{2}}{12}$ & $-\frac{\sqrt{6}}{12}$ & $-\frac{\sqrt{3}}{6}$ 
    \\
    $\hat{Q}_{zx}$ & $\frac{2}{3}$ & $-\frac{2}{3}$ & $0$ & $0$ & $1$ & $0$ & $\frac{\sqrt{2}}{12}$ & $\frac{\sqrt{6}}{12}$ & $\frac{\sqrt{3}}{6}$ 
    \\
    $\hat{Q}_{xy}$ & $\frac{2}{3}$ & $\frac{1}{3}$ & $-1$ & $0$ & $0$ & $1$ & $-\frac{\sqrt{2}}{6}$ & $0$ & $0$ 
    \\
    $\hat{Q}_{0}^{\prime}$ & $\frac{2}{3}$ & $-\frac{1}{3}$ & $-\frac{1}{3}$ & $-\frac{1}{3}$ & $-\frac{1}{3}$ & $-\frac{1}{3}$ & $0$ & $0$ & $0$ 
    \\
    $\hat{Q}_{u}^{\prime}$ & $\frac{2}{3}$ & $0$ & $-\frac{2}{3}$ & $-\frac{1}{6}$ & $-\frac{1}{6}$ & $-\frac{2}{3}$ & $-\frac{\sqrt{2}}{12}$ & $0$ & $0$ 
    \\
    $\hat{Q}_{v}^{\prime}$ & $\frac{2}{3}$ & $-\frac{2}{3}$ & $0$ & $-\frac{1}{2}$ & $-\frac{1}{2}$ & $0$ & $\frac{\sqrt{2}}{12}$ & $0$ & $0$ 
    \\
    $\hat{Q}_{yz}^{\prime}$ & $\frac{2}{3}$ & $-\frac{1}{6}$ & $-\frac{1}{2}$ & $0$ & $-\frac{1}{2}$ & $-\frac{1}{2}$ & $-\frac{\sqrt{2}}{24}$ & $\frac{\sqrt{6}}{24}$ & $\frac{\sqrt{3}}{12}$ 
    \\
    $\hat{Q}_{zx}^{\prime}$ & $\frac{2}{3}$ & $-\frac{1}{6}$ & $-\frac{1}{2}$ & $-\frac{1}{2}$ & $0$ & $-\frac{1}{2}$ & $-\frac{\sqrt{2}}{24}$ & $-\frac{\sqrt{6}}{24}$ & $-\frac{\sqrt{3}}{12}$ 
    \\
    $\hat{Q}_{xy}^{\prime}$ & $\frac{2}{3}$ & $-\frac{2}{3}$ & $0$ & $-\frac{1}{2}$ & $-\frac{1}{2}$ & $0$ & $\frac{\sqrt{2}}{12}$ & $0$ & $0$ 
    \\
    $\hat{G}_{x}^{\prime}$ & $\frac{2}{3}$ & $-\frac{1}{6}$ & $-\frac{1}{2}$ & $0$ & $-\frac{1}{2}$ & $-\frac{1}{2}$ & $-\frac{\sqrt{2}}{24}$ & $\frac{\sqrt{6}}{24}$ & $\frac{\sqrt{3}}{12}$ 
    \\
    $\hat{G}_{y}^{\prime}$ & $\frac{2}{3}$ & $-\frac{1}{6}$ & $-\frac{1}{2}$ & $-\frac{1}{2}$ & $0$ & $-\frac{1}{2}$ & $-\frac{\sqrt{2}}{24}$ & $-\frac{\sqrt{6}}{24}$ & $-\frac{\sqrt{3}}{12}$ 
    \\
    $\hat{G}_{z}^{\prime}$ & $\frac{2}{3}$ & $-\frac{2}{3}$ & $0$ & $-\frac{1}{2}$ & $-\frac{1}{2}$ & $0$ & $\frac{\sqrt{2}}{12}$ & $0$ & $0$ 
    \\
    \hline\hline
    \multicolumn{16}{c}{spinful Cooper pairing $\hat{\mathcal{H}}_{1}$}
    \\
    \hline\hline
    $\hat{\Lambda}$ & $\check{Q}_{0}^{\prime}$ & $\check{Q}_{u}^{\prime}$ & $\check{Q}_{v}^{\prime}$ & $\check{Q}_{yz}^{\prime}$ & $\check{Q}_{zx}^{\prime}$ & $\check{Q}_{xy}^{\prime}$ & $\check{G}_{x}^{\prime}$ & $\check{G}_{y}^{\prime}$ & $\check{G}_{z}^{\prime}$ & $\check{Q}_{0}^{\prime}\check{Q}_{u}^{\prime}$ & $\check{Q}_{0}^{\prime}\check{Q}_{v}^{\prime}$ & $\check{Q}_{u}^{\prime}\check{Q}_{v}^{\prime}$ & $\check{Q}_{yz}^{\prime}\check{G}_{x}^{\prime}$ & $\check{Q}_{zx}^{\prime}\check{G}_{y}^{\prime}$ & $\check{Q}_{xy}^{\prime}\check{G}_{z}^{\prime}$ 
    \\
    \hline
    $\hat{Q}_{0}$ & $\frac{2}{3}$ & $\frac{2}{3}$ & $\frac{2}{3}$ & $\frac{2}{3}$ & $\frac{2}{3}$ & $\frac{2}{3}$ & $\frac{2}{3}$ & $\frac{2}{3}$ & $\frac{2}{3}$ & $0$ & $0$ & $0$ & $0$ & $0$ & $0$
    \\
    $\hat{Q}_{u}$ & $-\frac{1}{3}$ & $0$ & $-\frac{2}{3}$ & $-\frac{1}{6}$ & $-\frac{1}{6}$ & $-\frac{2}{3}$ & $-\frac{1}{6}$ & $-\frac{1}{6}$ & $-\frac{2}{3}$ & $\frac{\sqrt{2}}{6}$ & $0$ & $0$ & $-\frac{1}{4}$ & $\frac{1}{4}$ & $0$
    \\
    $\hat{Q}_{v}$ & $-\frac{1}{3}$ & $-\frac{2}{3}$ & $0$ & $-\frac{1}{2}$ & $-\frac{1}{2}$ & $0$ & $-\frac{1}{2}$ & $-\frac{1}{2}$ & $0$ & $-\frac{\sqrt{2}}{6}$ & $0$ & $0$ & $\frac{1}{4}$ & $-\frac{1}{4}$ & $0$
    \\
    $\hat{Q}_{yz}$ & $-\frac{1}{3}$ & $-\frac{1}{6}$ & $-\frac{1}{2}$ & $0$ & $-\frac{1}{2}$ & $-\frac{1}{2}$ & $0$ & $-\frac{1}{2}$ & $-\frac{1}{2}$ & $\frac{\sqrt{2}}{12}$ & $-\frac{\sqrt{6}}{12}$ & $\frac{\sqrt{3}}{12}$ & $0$ & $\frac{1}{4}$ & $-\frac{1}{4}$
    \\
    $\hat{Q}_{zx}$ & $-\frac{1}{3}$ & $-\frac{1}{6}$ & $-\frac{1}{2}$ & $-\frac{1}{2}$ & $0$ & $-\frac{1}{2}$ & $-\frac{1}{2}$ & $0$ & $-\frac{1}{2}$ & $\frac{\sqrt{2}}{12}$ & $\frac{\sqrt{6}}{12}$ & $-\frac{\sqrt{3}}{12}$ & $-\frac{1}{4}$ & $0$ & $\frac{1}{4}$
    \\
    $\hat{Q}_{xy}$ & $-\frac{1}{3}$ & $-\frac{2}{3}$ & $0$ & $-\frac{1}{2}$ & $-\frac{1}{2}$ & $0$ & $-\frac{1}{2}$ & $-\frac{1}{2}$ & $0$ & $-\frac{\sqrt{2}}{6}$ & $0$ & $0$ & $\frac{1}{4}$ & $-\frac{1}{4}$ & $0$
    \\
    $\hat{Q}_{0}^{\prime}$ & $1$ & $0$ & $0$ & $0$ & $0$ & $0$ & $-\frac{2}{3}$ & $-\frac{2}{3}$ & $-\frac{2}{3}$ & $0$ & $0$ & $0$ & $0$ & $0$ & $0$
    \\
    $\hat{Q}_{u}^{\prime}$ & $0$ & $1$ & $0$ & $-\frac{3}{4}$ & $-\frac{3}{4}$ & $0$ & $-\frac{1}{12}$ & $-\frac{1}{12}$ & $-\frac{1}{3}$ & $0$ & $0$ & $0$ & $\frac{1}{8}$ & $-\frac{1}{8}$ & $0$
    \\
    $\hat{Q}_{v}^{\prime}$ & $0$ & $0$ & $1$ & $-\frac{1}{4}$ & $-\frac{1}{4}$ & $-1$ & $-\frac{1}{4}$ & $-\frac{1}{4}$ & $0$ & $0$ & $0$ & $0$ & $-\frac{1}{8}$ & $\frac{1}{8}$ & $0$
    \\
    $\hat{Q}_{yz}^{\prime}$ & $0$ & $-\frac{3}{4}$ & $-\frac{1}{4}$ & $1$ & $-\frac{1}{4}$ & $-\frac{1}{4}$ & $0$ & $-\frac{1}{4}$ & $-\frac{1}{4}$ & $0$ & $0$ & $-\frac{\sqrt{3}}{8}$ & $0$ & $-\frac{1}{8}$ & $\frac{1}{8}$
    \\
    $\hat{Q}_{zx}^{\prime}$ & $0$ & $-\frac{3}{4}$ & $-\frac{1}{4}$ & $-\frac{1}{4}$ & $1$ & $-\frac{1}{4}$ & $-\frac{1}{4}$ & $0$ & $-\frac{1}{4}$ & $0$ & $0$ & $\frac{\sqrt{3}}{8}$ & $\frac{1}{8}$ & $0$ & $-\frac{1}{8}$
    \\
    $\hat{Q}_{xy}^{\prime}$ & $0$ & $0$ & $-1$ & $-\frac{1}{4}$ & $-\frac{1}{4}$ & $1$ & $-\frac{1}{4}$ & $-\frac{1}{4}$ & $0$ & $0$ & $0$ & $0$ & $-\frac{1}{8}$ & $\frac{1}{8}$ & $0$
    \\
    $\hat{G}_{x}^{\prime}$ & $-\frac{2}{3}$ & $-\frac{1}{12}$ & $-\frac{1}{4}$ & $0$ & $-\frac{1}{4}$ & $-\frac{1}{4}$ & $1$ & $-\frac{1}{4}$ & $-\frac{1}{4}$ & $-\frac{\sqrt{2}}{12}$ & $\frac{\sqrt{6}}{12}$ & $\frac{\sqrt{3}}{24}$ & $0$ & $-\frac{1}{8}$ & $\frac{1}{8}$
    \\
    $\hat{G}_{y}^{\prime}$ & $-\frac{2}{3}$ & $-\frac{1}{12}$ & $-\frac{1}{4}$ & $-\frac{1}{4}$ & $0$ & $-\frac{1}{4}$ & $-\frac{1}{4}$ & $1$ & $-\frac{1}{4}$ & $-\frac{\sqrt{2}}{12}$ & $-\frac{\sqrt{6}}{12}$ & $-\frac{\sqrt{3}}{24}$ & $\frac{1}{8}$ & $0$ & $-\frac{1}{8}$
    \\
    $\hat{G}_{z}^{\prime}$ & $-\frac{2}{3}$ & $-\frac{1}{3}$ & $0$ & $-\frac{1}{4}$ & $-\frac{1}{4}$ & $0$ & $-\frac{1}{4}$ & $-\frac{1}{4}$ & $1$ & $\frac{\sqrt{2}}{6}$ & $0$ & $0$ & $-\frac{1}{8}$ & $\frac{1}{8}$ & $0$
    \\
    \hline\hline
    \multicolumn{16}{c}{mixed Cooper pairing $\hat{\mathcal{H}}_{2}$}
    \\
    \hline\hline
    $\hat{\Lambda}$ & $\check{Q}_{0}\check{Q}_{0}^{\prime}$ & $\check{Q}_{0}\check{Q}_{u}^{\prime}$ & $\check{Q}_{0}\check{Q}_{v}^{\prime}$ & $\check{Q}_{u}\check{Q}_{0}^{\prime}$ & $\check{Q}_{u}\check{Q}_{u}^{\prime}$ & $\check{Q}_{u}\check{Q}_{v}^{\prime}$ & $\check{Q}_{v}\check{Q}_{0}^{\prime}$ & $\check{Q}_{v}\check{Q}_{u}^{\prime}$ & $\check{Q}_{v}\check{Q}_{v}^{\prime}$ & $\check{Q}_{yz}\check{Q}_{yz}^{\prime}$ & $\check{Q}_{yz}\check{G}_{x}^{\prime}$ & $\check{Q}_{zx}\check{Q}_{zx}^{\prime}$ & $\check{Q}_{zx}\check{G}_{y}^{\prime}$ & $\check{Q}_{xy}\check{Q}_{xy}^{\prime}$ & $\check{Q}_{xy}\check{G}_{z}^{\prime}$ 
    \\
    \hline
    $\hat{Q}_{0}^{\prime}$ & $-\frac{\sqrt{2}}{6}$ & $0$ & $0$ & $0$ & $-\frac{\sqrt{2}}{6}$ & $0$ & $0$ & $0$ & $-\frac{\sqrt{2}}{6}$ & $-\frac{\sqrt{2}}{6}$ & $0$ & $-\frac{\sqrt{2}}{6}$ & $0$ & $-\frac{\sqrt{2}}{6}$ & $0$ 
    \\
    $\hat{Q}_{u}^{\prime}$ & $\frac{\sqrt{2}}{12}$ & $-\frac{1}{6}$ & $0$ & $-\frac{1}{6}$ & $0$ & $0$ & $0$ & $0$ & $\frac{\sqrt{2}}{6}$ & $\frac{\sqrt{2}}{24}$ & $-\frac{\sqrt{2}}{8}$ & $\frac{\sqrt{2}}{24}$ & $\frac{\sqrt{2}}{8}$ & $\frac{\sqrt{2}}{6}$ & $0$ 
    \\
    $\hat{Q}_{v}^{\prime}$ & $\frac{\sqrt{2}}{12}$ & $\frac{1}{6}$ & $0$ & $\frac{1}{6}$ & $\frac{\sqrt{2}}{6}$ & $0$ & $0$ & $0$ & $0$ & $\frac{\sqrt{2}}{8}$ & $\frac{\sqrt{2}}{8}$ & $\frac{\sqrt{2}}{8}$ & $-\frac{\sqrt{2}}{8}$ & $0$ & $0$ 
    \\
    $\hat{Q}_{yz}^{\prime}$ & $\frac{\sqrt{2}}{12}$ & $-\frac{1}{12}$ & $\frac{\sqrt{3}}{12}$ & $-\frac{1}{12}$ & $\frac{\sqrt{2}}{24}$ & $-\frac{\sqrt{6}}{24}$ & $\frac{\sqrt{3}}{12}$ & $-\frac{\sqrt{6}}{24}$ & $\frac{\sqrt{2}}{8}$ & $0$ & $0$ & $\frac{\sqrt{2}}{8}$ & $\frac{\sqrt{2}}{8}$ & $\frac{\sqrt{2}}{8}$ & $-\frac{\sqrt{2}}{8}$ 
    \\
    $\hat{Q}_{zx}^{\prime}$ & $\frac{\sqrt{2}}{12}$ & $-\frac{1}{12}$ & $-\frac{\sqrt{3}}{12}$ & $-\frac{1}{12}$ & $\frac{\sqrt{2}}{24}$ & $\frac{\sqrt{6}}{24}$ & $-\frac{\sqrt{3}}{12}$ & $\frac{\sqrt{6}}{24}$ & $\frac{\sqrt{2}}{8}$ & $\frac{\sqrt{2}}{8}$ & $-\frac{\sqrt{2}}{8}$ & $0$ & $0$ & $\frac{\sqrt{2}}{8}$ & $\frac{\sqrt{2}}{8}$ 
    \\
    $\hat{Q}_{xy}^{\prime}$ & $\frac{\sqrt{2}}{12}$ & $\frac{1}{6}$ & $0$ & $\frac{1}{6}$ & $\frac{\sqrt{2}}{6}$ & $0$ & $0$ & $0$ & $0$ & $\frac{\sqrt{2}}{8}$ & $\frac{\sqrt{2}}{8}$ & $\frac{\sqrt{2}}{8}$ & $-\frac{\sqrt{2}}{8}$ & $0$ & $0$ 
    \\
    $\hat{G}_{x}^{\prime}$ & $-\frac{\sqrt{2}}{12}$ & $\frac{1}{12}$ & $-\frac{\sqrt{3}}{12}$ & $\frac{1}{12}$ & $-\frac{\sqrt{2}}{24}$ & $\frac{\sqrt{6}}{24}$ & $-\frac{\sqrt{3}}{12}$ & $\frac{\sqrt{6}}{24}$ & $-\frac{\sqrt{2}}{8}$ & $0$ & $0$ & $-\frac{\sqrt{2}}{8}$ & $-\frac{\sqrt{2}}{8}$ & $-\frac{\sqrt{2}}{8}$ & $\frac{\sqrt{2}}{8}$ 
    \\
    $\hat{G}_{y}^{\prime}$ & $-\frac{\sqrt{2}}{12}$ & $\frac{1}{12}$ & $\frac{\sqrt{3}}{12}$ & $\frac{1}{12}$ & $-\frac{\sqrt{2}}{24}$ & $-\frac{\sqrt{6}}{24}$ & $\frac{\sqrt{3}}{12}$ & $-\frac{\sqrt{6}}{24}$ & $-\frac{\sqrt{2}}{8}$ & $-\frac{\sqrt{2}}{8}$ & $\frac{\sqrt{2}}{8}$ & $0$ & $0$ & $-\frac{\sqrt{2}}{8}$ & $-\frac{\sqrt{2}}{8}$ 
    \\
    $\hat{G}_{z}^{\prime}$ & $-\frac{\sqrt{2}}{12}$ & $-\frac{1}{6}$ & $0$ & $-\frac{1}{6}$ & $-\frac{\sqrt{2}}{6}$ & $0$ & $0$ & $0$ & $0$ & $-\frac{\sqrt{2}}{8}$ & $-\frac{\sqrt{2}}{8}$ & $-\frac{\sqrt{2}}{8}$ & $\frac{\sqrt{2}}{8}$ & $0$ & $0$ 
    \\
  \end{longtable*}
\end{center}

\begin{center}
  \begin{longtable*}{lccccccccccccccc}
    \caption{
      The Cooper pairing $\hat{\mathcal{H}}_{\mathrm{eff}}$ in Eq.~(\ref{eq:channel_decomposition}) induced  by local M/MT-type multipole fluctuations in $p$-orbital space.
    }
    \label{tab:Pairing_p_ls_M}
    \\
    
    \hline\hline
    \multicolumn{16}{c}
    {spinless Cooper pairing $\mathcal{H}^{(0)}$}
    \\
    \hline\hline
    $\hat{\Lambda}$ & $\check{Q}_{0}$ & $\check{Q}_{u}$ &  $\check{Q}_{v}$ & $\check{Q}_{yz}$ & $\check{Q}_{zx}$ & $\check{Q}_{xy}$ &  $\check{Q}_{0}\check{Q}_{u}$ & $\check{Q}_{0}\check{Q}_{v}$ & $\check{Q}_{u}\check{Q}_{v}$ 
    \\
    \endfirsthead
    
    \multicolumn{16}{c}
    {{\tablename\ \thetable{} -- continued from previous page}}
    \\
    \endhead
    
    \multicolumn{16}{c}{{Continued on next page}}
    \\
    \endfoot
    
    \hline\hline
    \endlastfoot

    \hline 
    $\hat{\bm{M}}^{(\mathrm{s})}$ & $-\frac{2}{3}$ & $-\frac{2}{3}$ & $-\frac{2}{3}$ & $-\frac{2}{3}$ & $-\frac{2}{3}$ & $-\frac{2}{3}$ & $0$ & $0$ & $0$ 
    \\
    $\hat{M}_{x}$ & $-\frac{2}{3}$ & $\frac{2}{3}$ & $0$ & $1$ & $0$ & $0$ & $-\frac{\sqrt{2}}{12}$ & $\frac{\sqrt{6}}{12}$ & $\frac{\sqrt{3}}{6}$ 
    \\
    $\hat{M}_{y}$ & $-\frac{2}{3}$ & $\frac{2}{3}$ & $0$ & $0$ & $1$ & $0$ & $-\frac{\sqrt{2}}{12}$ & $-\frac{\sqrt{6}}{12}$ & $-\frac{\sqrt{3}}{6}$ 
    \\
    $\hat{M}_{z}$ & $-\frac{2}{3}$ & $-\frac{1}{3}$ & $1$ & $0$ & $0$ & $1$ & $\frac{\sqrt{2}}{6}$ & $0$ & $0$ 
    \\
    $\hat{M}_{x}^{\prime}$ & $-\frac{2}{3}$ & $-\frac{1}{10}$ & $-\frac{1}{30}$ & $-\frac{2}{15}$ & $-\frac{1}{30}$ & $-\frac{1}{30}$ & $\frac{7\sqrt{2}}{120}$ & $-\frac{7\sqrt{6}}{120}$ & $-\frac{\sqrt{3}}{60}$ 
    \\
    $\hat{M}_{y}^{\prime}$ & $-\frac{2}{3}$ & $-\frac{1}{10}$ & $-\frac{1}{30}$ & $-\frac{1}{30}$ & $-\frac{2}{15}$ & $-\frac{1}{30}$ & $\frac{7\sqrt{2}}{120}$ & $\frac{7\sqrt{6}}{120}$ & $\frac{\sqrt{3}}{60}$ 
    \\
    $\hat{M}_{z}^{\prime}$ & $-\frac{2}{3}$ & $0$ & $-\frac{2}{15}$ & $-\frac{1}{30}$ & $-\frac{1}{30}$ & $-\frac{2}{15}$ & $-\frac{7\sqrt{2}}{60}$ & $0$ & $0$ 
    \\
    $\hat{T}_{u}^{\prime}$ & $-\frac{2}{3}$ & $\frac{2}{3}$ & $0$ & $-\frac{1}{2}$ & $-\frac{1}{2}$ & $0$ & $-\frac{\sqrt{2}}{12}$ & $0$ & $0$ 
    \\
    $\hat{T}_{v}^{\prime}$ & $-\frac{2}{3}$ & $0$ & $\frac{2}{3}$ & $-\frac{1}{6}$ & $-\frac{1}{6}$ & $-\frac{2}{3}$ & $\frac{\sqrt{2}}{12}$ & $0$ & $0$ 
    \\
    $\hat{T}_{yz}^{\prime}$ & $-\frac{2}{3}$ & $-\frac{1}{2}$ & $-\frac{1}{6}$ & $\frac{2}{3}$ & $-\frac{1}{6}$ & $-\frac{1}{6}$ & $-\frac{\sqrt{2}}{24}$ & $\frac{\sqrt{6}}{24}$ & $-\frac{\sqrt{3}}{12}$ 
    \\
    $\hat{T}_{zx}^{\prime}$ & $-\frac{2}{3}$ & $-\frac{1}{2}$ & $-\frac{1}{6}$ & $-\frac{1}{6}$ & $\frac{2}{3}$ & $-\frac{1}{6}$ & $-\frac{\sqrt{2}}{24}$ & $-\frac{\sqrt{6}}{24}$ & $\frac{\sqrt{3}}{12}$ 
    \\
    $\hat{T}_{xy}^{\prime}$ & $-\frac{2}{3}$ & $0$ & $-\frac{2}{3}$ & $-\frac{1}{6}$ & $-\frac{1}{6}$ & $\frac{2}{3}$ & $\frac{\sqrt{2}}{12}$ & $0$ & $0$ 
    \\
    $\hat{M}_{xyz}^{\prime}$ & $-\frac{2}{3}$ & $\frac{1}{3}$ & $\frac{1}{3}$ & $-\frac{1}{3}$ & $-\frac{1}{3}$ & $-\frac{1}{3}$ & $0$ & $0$ & $0$ 
    \\
    $\hat{M}_{x}^{\alpha\prime}$ & $-\frac{2}{3}$ & $-\frac{7}{30}$ & $-\frac{3}{10}$ & $-\frac{1}{5}$ & $\frac{1}{5}$ & $\frac{1}{5}$ & $\frac{\sqrt{2}}{15}$ & $-\frac{\sqrt{6}}{15}$ & $\frac{\sqrt{3}}{60}$ 
    \\
    $\hat{M}_{y}^{\alpha\prime}$ & $-\frac{2}{3}$ & $-\frac{7}{30}$ & $-\frac{3}{10}$ & $\frac{1}{5}$ & $-\frac{1}{5}$ & $\frac{1}{5}$ & $\frac{\sqrt{2}}{15}$ & $\frac{\sqrt{6}}{15}$ & $-\frac{\sqrt{3}}{60}$ 
    \\
    $\hat{M}_{z}^{\alpha\prime}$ & $-\frac{2}{3}$ & $-\frac{1}{3}$ & $-\frac{1}{5}$ & $\frac{1}{5}$ & $\frac{1}{5}$ & $-\frac{1}{5}$ & $-\frac{2\sqrt{2}}{15}$ & $0$ & $0$ 
    \\
    $\hat{M}_{x}^{\beta\prime}$ & $-\frac{2}{3}$ & $-\frac{1}{6}$ & $\frac{1}{6}$ & $\frac{1}{3}$ & $-\frac{1}{3}$ & $-\frac{1}{3}$ & $0$ & $0$ & $-\frac{\sqrt{3}}{12}$ 
    \\
    $\hat{M}_{y}^{\beta\prime}$ & $-\frac{2}{3}$ & $-\frac{1}{6}$ & $\frac{1}{6}$ & $-\frac{1}{3}$ & $\frac{1}{3}$ & $-\frac{1}{3}$ & $0$ & $0$ & $\frac{\sqrt{3}}{12}$ 
    \\
    $\hat{M}_{z}^{\beta\prime}$ & $-\frac{2}{3}$ & $\frac{1}{3}$ & $-\frac{1}{3}$ & $-\frac{1}{3}$ & $-\frac{1}{3}$ & $\frac{1}{3}$ & $0$ & $0$ & $0$ 
    \\ 
    \hline\hline
    \multicolumn{16}{c}
    {spinful Cooper pairing $\mathcal{H}^{(1)}$}
    \\
    \hline\hline
    $\hat{\Lambda}$ & $\check{Q}_{0}^{\prime}$ & $\check{Q}_{u}^{\prime}$ & $\check{Q}_{v}^{\prime}$ & $\check{Q}_{yz}^{\prime}$ & $\check{Q}_{zx}^{\prime}$ & $\check{Q}_{xy}^{\prime}$ & $\check{G}_{x}^{\prime}$ & $\check{G}_{y}^{\prime}$ & $\check{G}_{z}^{\prime}$ & $\check{Q}_{0}^{\prime}\check{Q}_{u}^{\prime}$ & $\check{Q}_{0}^{\prime}\check{Q}_{v}^{\prime}$ & $\check{Q}_{u}^{\prime}\check{Q}_{v}^{\prime}$ & $\check{Q}_{yz}^{\prime}\check{G}_{x}^{\prime}$ & $\check{Q}_{zx}^{\prime}\check{G}_{y}^{\prime}$ & $\check{Q}_{xy}^{\prime}\check{G}_{z}^{\prime}$ 
    \\
    \hline
    $\hat{M}_{x}^{(\mathrm{s})}$ & $\frac{2}{9}$ & $\frac{4}{9}$ & $0$ & $\frac{2}{3}$ & $0$ & $0$ & $\frac{2}{3}$ & $0$ & $0$ & $\frac{\sqrt{2}}{9}$ & $-\frac{\sqrt{6}}{9}$ & $\frac{\sqrt{3}}{9}$ & $0$ & $-\frac{1}{3}$ & $\frac{1}{3}$ 
    \\
    $\hat{M}_{y}^{(\mathrm{s})}$ & $\frac{2}{9}$ & $\frac{4}{9}$ & $0$ & $0$ & $\frac{2}{3}$ & $0$ & $0$ & $\frac{2}{3}$ & $0$ & $\frac{\sqrt{2}}{9}$ & $\frac{\sqrt{6}}{9}$ & $-\frac{\sqrt{3}}{9}$ & $\frac{1}{3}$ & $0$ & $-\frac{1}{3}$ 
    \\
    $\hat{M}_{z}^{(\mathrm{s})}$ & $\frac{2}{9}$ & $-\frac{2}{9}$ & $\frac{2}{3}$ & $0$ & $0$ & $\frac{2}{3}$ & $0$ & $0$ & $\frac{2}{3}$ & $-\frac{2\sqrt{2}}{9}$ & $0$ & $0$ & $-\frac{1}{3}$ & $\frac{1}{3}$ & $0$ 
    \\
    $\hat{M}_{x}$ & $-\frac{1}{3}$ & $-\frac{1}{6}$ & $-\frac{1}{2}$ & $0$ & $-\frac{1}{2}$ & $-\frac{1}{2}$ & $0$ & $-\frac{1}{2}$ & $-\frac{1}{2}$ & $\frac{\sqrt{2}}{12}$ & $-\frac{\sqrt{6}}{12}$ & $\frac{\sqrt{3}}{12}$ & $0$ & $\frac{1}{4}$ & $-\frac{1}{4}$ 
    \\
    $\hat{M}_{y}$ & $-\frac{1}{3}$ & $-\frac{1}{6}$ & $-\frac{1}{2}$ & $-\frac{1}{2}$ & $0$ & $-\frac{1}{2}$ & $-\frac{1}{2}$ & $0$ & $-\frac{1}{2}$ & $\frac{\sqrt{2}}{12}$ & $\frac{\sqrt{6}}{12}$ & $-\frac{\sqrt{3}}{12}$ & $-\frac{1}{4}$ & $0$ & $\frac{1}{4}$ 
    \\
    $\hat{M}_{z}$ & $-\frac{1}{3}$ & $-\frac{2}{3}$ & $0$ & $-\frac{1}{2}$ & $-\frac{1}{2}$ & $0$ & $-\frac{1}{2}$ & $-\frac{1}{2}$ & $0$ & $-\frac{\sqrt{2}}{6}$ & $0$ & $0$ & $\frac{1}{4}$ & $-\frac{1}{4}$ & $0$ 
    \\
    $\hat{M}_{x}^{\prime}$ & $-\frac{8}{9}$ & $-\frac{41}{180}$ & $-\frac{3}{20}$ & $-\frac{4}{15}$ & $-\frac{3}{20}$ & $-\frac{3}{20}$ & $\frac{1}{3}$ & $\frac{1}{4}$ & $\frac{1}{4}$ & $-\frac{2\sqrt{2}}{45}$ & $\frac{2\sqrt{6}}{45}$ & $-\frac{7\sqrt{3}}{360}$ & $0$ & $-\frac{1}{24}$ & $\frac{1}{24}$ 
    \\
    $\hat{M}_{y}^{\prime}$ & $-\frac{8}{9}$ & $-\frac{41}{180}$ & $-\frac{3}{20}$ & $-\frac{3}{20}$ & $-\frac{4}{15}$ & $-\frac{3}{20}$ & $\frac{1}{4}$ & $\frac{1}{3}$ & $\frac{1}{4}$ & $-\frac{2\sqrt{2}}{45}$ & $-\frac{2\sqrt{6}}{45}$ & $\frac{7\sqrt{3}}{360}$ & $\frac{1}{24}$ & $0$ & $-\frac{1}{24}$ 
    \\
    $\hat{M}_{z}^{\prime}$ & $-\frac{8}{9}$ & $-\frac{1}{9}$ & $-\frac{4}{15}$ & $-\frac{3}{20}$ & $-\frac{3}{20}$ & $-\frac{4}{15}$ & $\frac{1}{4}$ & $\frac{1}{4}$ & $\frac{1}{3}$ & $\frac{4\sqrt{2}}{45}$ & $0$ & $0$ & $-\frac{1}{24}$ & $\frac{1}{24}$ & $0$ 
    \\
    $\hat{T}_{u}^{\prime}$ & $\frac{2}{3}$ & $\frac{1}{3}$ & $0$ & $-\frac{1}{4}$ & $-\frac{1}{4}$ & $0$ & $-\frac{1}{4}$ & $-\frac{1}{4}$ & $-1$ & $-\frac{\sqrt{2}}{6}$ & $0$ & $0$ & $-\frac{1}{8}$ & $\frac{1}{8}$ & $0$ 
    \\
    $\hat{T}_{v}^{\prime}$ & $\frac{2}{3}$ & $0$ & $\frac{1}{3}$ & $-\frac{1}{12}$ & $-\frac{1}{12}$ & $-\frac{1}{3}$ & $-\frac{3}{4}$ & $-\frac{3}{4}$ & $0$ & $\frac{\sqrt{2}}{6}$ & $0$ & $0$ & $\frac{1}{8}$ & $-\frac{1}{8}$ & $0$ 
    \\
    $\hat{T}_{yz}^{\prime}$ & $\frac{2}{3}$ & $-\frac{1}{4}$ & $-\frac{1}{12}$ & $\frac{1}{3}$ & $-\frac{1}{12}$ & $-\frac{1}{12}$ & $0$ & $-\frac{3}{4}$ & $-\frac{3}{4}$ & $-\frac{\sqrt{2}}{12}$ & $\frac{\sqrt{6}}{12}$ & $-\frac{\sqrt{3}}{24}$ & $0$ & $\frac{1}{8}$ & $-\frac{1}{8}$ 
    \\
    $\hat{T}_{zx}^{\prime}$ & $\frac{2}{3}$ & $-\frac{1}{4}$ & $-\frac{1}{12}$ & $-\frac{1}{12}$ & $\frac{1}{3}$ & $-\frac{1}{12}$ & $-\frac{3}{4}$ & $0$ & $-\frac{3}{4}$ & $-\frac{\sqrt{2}}{12}$ & $-\frac{\sqrt{6}}{12}$ & $\frac{\sqrt{3}}{24}$ & $-\frac{1}{8}$ & $0$ & $\frac{1}{8}$ 
    \\
    $\hat{T}_{xy}^{\prime}$ & $\frac{2}{3}$ & $0$ & $-\frac{1}{3}$ & $-\frac{1}{12}$ & $-\frac{1}{12}$ & $\frac{1}{3}$ & $-\frac{3}{4}$ & $-\frac{3}{4}$ & $0$ & $\frac{\sqrt{2}}{6}$ & $0$ & $0$ & $\frac{1}{8}$ & $-\frac{1}{8}$ & $0$ 
    \\
    $\hat{M}_{xyz}^{\prime}$ & $-\frac{1}{3}$ & $\frac{2}{3}$ & $\frac{2}{3}$ & $-\frac{2}{3}$ & $-\frac{2}{3}$ & $-\frac{2}{3}$ & $0$ & $0$ & $0$ & $0$ & $0$ & $0$ & $0$ & $0$ & $0$ 
    \\
    $\hat{M}_{x}^{\alpha\prime}$ & $-\frac{1}{3}$ & $-\frac{7}{15}$ & $-\frac{3}{5}$ & $-\frac{2}{5}$ & $\frac{2}{5}$ & $\frac{2}{5}$ & $0$ & $0$ & $0$ & $-\frac{\sqrt{2}}{15}$ & $\frac{\sqrt{6}}{15}$ & $\frac{\sqrt{3}}{30}$ & $0$ & $0$ & $0$ 
    \\
    $\hat{M}_{y}^{\alpha\prime}$ & $-\frac{1}{3}$ & $-\frac{7}{15}$ & $-\frac{3}{5}$ & $\frac{2}{5}$ & $-\frac{2}{5}$ & $\frac{2}{5}$ & $0$ & $0$ & $0$ & $-\frac{\sqrt{2}}{15}$ & $-\frac{\sqrt{6}}{15}$ & $-\frac{\sqrt{3}}{30}$ & $0$ & $0$ & $0$ 
    \\
    $\hat{M}_{z}^{\alpha\prime}$ & $-\frac{1}{3}$ & $-\frac{2}{3}$ & $-\frac{2}{5}$ & $\frac{2}{5}$ & $\frac{2}{5}$ & $-\frac{2}{5}$ & $0$ & $0$ & $0$ & $\frac{2\sqrt{2}}{15}$ & $0$ & $0$ & $0$ & $0$ & $0$ 
    \\
    $\hat{M}_{x}^{\beta\prime}$ & $-\frac{1}{3}$ & $-\frac{1}{3}$ & $\frac{1}{3}$ & $\frac{2}{3}$ & $-\frac{2}{3}$ & $-\frac{2}{3}$ & $0$ & $0$ & $0$ & $0$ & $0$ & $-\frac{\sqrt{3}}{6}$ & $0$ & $0$ & $0$ 
    \\
    $\hat{M}_{y}^{\beta\prime}$ & $-\frac{1}{3}$ & $-\frac{1}{3}$ & $\frac{1}{3}$ & $-\frac{2}{3}$ & $\frac{2}{3}$ & $-\frac{2}{3}$ & $0$ & $0$ & $0$ & $0$ & $0$ & $\frac{\sqrt{3}}{6}$ & $0$ & $0$ & $0$ 
    \\
    $\hat{M}_{z}^{\beta\prime}$ & $-\frac{1}{3}$ & $\frac{2}{3}$ & $-\frac{2}{3}$ & $-\frac{2}{3}$ & $-\frac{2}{3}$ & $\frac{2}{3}$ & $0$ & $0$ & $0$ & $0$ & $0$ & $0$ & $0$ & $0$ & $0$ 
    \\
    \hline\hline
    \multicolumn{16}{c}
    {mixed Cooper pairing $\mathcal{H}^{(2)}$}
    \\
    \hline\hline
    $\hat{\Lambda}$ & $\check{Q}_{0}\check{Q}_{0}^{\prime}$ & $\check{Q}_{0}\check{Q}_{u}^{\prime}$ & $\check{Q}_{0}\check{Q}_{v}^{\prime}$ & $\check{Q}_{u}\check{Q}_{0}^{\prime}$ & $\check{Q}_{u}\check{Q}_{u}^{\prime}$ & $\check{Q}_{u}\check{Q}_{v}^{\prime}$ & $\check{Q}_{v}\check{Q}_{0}^{\prime}$ & $\check{Q}_{v}\check{Q}_{u}^{\prime}$ & $\check{Q}_{v}\check{Q}_{v}^{\prime}$ & $\check{Q}_{yz}\check{Q}_{yz}^{\prime}$ & $\check{Q}_{yz}\check{G}_{x}^{\prime}$ & $\check{Q}_{zx}\check{Q}_{zx}^{\prime}$ & $\check{Q}_{zx}\check{G}_{y}^{\prime}$ & $\check{Q}_{xy}\check{Q}_{xy}^{\prime}$ & $\check{Q}_{xy}\check{G}_{z}^{\prime}$ 
    \\
    \hline
    $\hat{M}_{x}^{\prime}$ & $\frac{\sqrt{2}}{4}$ & $\frac{1}{20}$ & $-\frac{\sqrt{3}}{20}$ & $-\frac{3}{20}$ & $-\frac{3\sqrt{2}}{40}$ & $-\frac{\sqrt{6}}{40}$ & $\frac{3\sqrt{3}}{20}$ & $-\frac{\sqrt{6}}{40}$ & $-\frac{\sqrt{2}}{40}$ & $-\frac{\sqrt{2}}{10}$ & $0$ & $-\frac{\sqrt{2}}{40}$ & $-\frac{\sqrt{2}}{8}$ & $-\frac{\sqrt{2}}{40}$ & $\frac{\sqrt{2}}{8}$ 
    \\
    $\hat{M}_{y}^{\prime}$ & $\frac{\sqrt{2}}{4}$ & $\frac{1}{20}$ & $\frac{\sqrt{3}}{20}$ & $-\frac{3}{20}$ & $-\frac{3\sqrt{2}}{40}$ & $\frac{\sqrt{6}}{40}$ & $-\frac{3\sqrt{3}}{20}$ & $\frac{\sqrt{6}}{40}$ & $-\frac{\sqrt{2}}{40}$ & $-\frac{\sqrt{2}}{40}$ & $\frac{\sqrt{2}}{8}$ & $-\frac{\sqrt{2}}{10}$ & $0$ & $-\frac{\sqrt{2}}{40}$ & $-\frac{\sqrt{2}}{8}$ 
    \\
    $\hat{M}_{z}^{\prime}$ & $\frac{\sqrt{2}}{4}$ & $-\frac{1}{10}$ & $0$ & $\frac{3}{10}$ & $0$ & $0$ & $0$ & $0$ & $-\frac{\sqrt{2}}{10}$ & $-\frac{\sqrt{2}}{40}$ & $-\frac{\sqrt{2}}{8}$ & $-\frac{\sqrt{2}}{40}$ & $\frac{\sqrt{2}}{8}$ & $-\frac{\sqrt{2}}{10}$ & $0$ 
    \\
    $\hat{T}_{u}^{\prime}$ & $\frac{\sqrt{2}}{12}$ & $\frac{1}{6}$ & $0$ & $\frac{1}{6}$ & $\frac{\sqrt{2}}{6}$ & $0$ & $0$ & $0$ & $0$ & $-\frac{\sqrt{2}}{8}$ & $-\frac{\sqrt{2}}{8}$ & $-\frac{\sqrt{2}}{8}$ & $\frac{\sqrt{2}}{8}$ & $0$ & $0$ 
    \\
    $\hat{T}_{v}^{\prime}$ & $\frac{\sqrt{2}}{12}$ & $-\frac{1}{6}$ & $0$ & $-\frac{1}{6}$ & $0$ & $0$ & $0$ & $0$ & $\frac{\sqrt{2}}{6}$ & $-\frac{\sqrt{2}}{24}$ & $\frac{\sqrt{2}}{8}$ & $-\frac{\sqrt{2}}{24}$ & $-\frac{\sqrt{2}}{8}$ & $-\frac{\sqrt{2}}{6}$ & $0$ 
    \\
    $\hat{T}_{yz}^{\prime}$ & $\frac{\sqrt{2}}{12}$ & $\frac{1}{12}$ & $-\frac{\sqrt{3}}{12}$ & $\frac{1}{12}$ & $-\frac{\sqrt{2}}{8}$ & $-\frac{\sqrt{6}}{24}$ & $-\frac{\sqrt{3}}{12}$ & $-\frac{\sqrt{6}}{24}$ & $-\frac{\sqrt{2}}{24}$ & $\frac{\sqrt{2}}{6}$ & $0$ & $-\frac{\sqrt{2}}{24}$ & $\frac{\sqrt{2}}{8}$ & $-\frac{\sqrt{2}}{24}$ & $-\frac{\sqrt{2}}{8}$ 
    \\
    $\hat{T}_{zx}^{\prime}$ & $\frac{\sqrt{2}}{12}$ & $\frac{1}{12}$ & $\frac{\sqrt{3}}{12}$ & $\frac{1}{12}$ & $-\frac{\sqrt{2}}{8}$ & $\frac{\sqrt{6}}{24}$ & $\frac{\sqrt{3}}{12}$ & $\frac{\sqrt{6}}{24}$ & $-\frac{\sqrt{2}}{24}$ & $-\frac{\sqrt{2}}{24}$ & $-\frac{\sqrt{2}}{8}$ & $\frac{\sqrt{2}}{6}$ & $0$ & $-\frac{\sqrt{2}}{24}$ & $\frac{\sqrt{2}}{8}$ 
    \\
    $\hat{T}_{xy}^{\prime}$ & $\frac{\sqrt{2}}{12}$ & $-\frac{1}{6}$ & $0$ & $-\frac{1}{6}$ & $0$ & $0$ & $0$ & $0$ & $-\frac{\sqrt{2}}{6}$ & $-\frac{\sqrt{2}}{24}$ & $\frac{\sqrt{2}}{8}$ & $-\frac{\sqrt{2}}{24}$ & $-\frac{\sqrt{2}}{8}$ & $\frac{\sqrt{2}}{6}$ & $0$ 
    \\
    $\hat{M}_{xyz}^{\prime}$ & $-\frac{\sqrt{2}}{6}$ & $0$ & $0$ & $0$ & $-\frac{\sqrt{2}}{6}$ & $0$ & $0$ & $0$ & $-\frac{\sqrt{2}}{6}$ & $\frac{\sqrt{2}}{6}$ & $0$ & $\frac{\sqrt{2}}{6}$ & $0$ & $\frac{\sqrt{2}}{6}$ & $0$ 
    \\
    $\hat{M}_{x}^{\alpha\prime}$ & $-\frac{\sqrt{2}}{6}$ & $-\frac{2}{15}$ & $\frac{2\sqrt{3}}{15}$ & $\frac{1}{15}$ & $\frac{7\sqrt{2}}{60}$ & $-\frac{\sqrt{6}}{60}$ & $-\frac{\sqrt{3}}{15}$ & $-\frac{\sqrt{6}}{60}$ & $\frac{3\sqrt{2}}{20}$ & $\frac{\sqrt{2}}{10}$ & $0$ & $-\frac{\sqrt{2}}{10}$ & $0$ & $-\frac{\sqrt{2}}{10}$ & $0$ 
    \\
    $\hat{M}_{y}^{\alpha\prime}$ & $-\frac{\sqrt{2}}{6}$ & $-\frac{2}{15}$ & $-\frac{2\sqrt{3}}{15}$ & $\frac{1}{15}$ & $\frac{7\sqrt{2}}{60}$ & $\frac{\sqrt{6}}{60}$ & $\frac{\sqrt{3}}{15}$ & $\frac{\sqrt{6}}{60}$ & $\frac{3\sqrt{2}}{20}$ & $-\frac{\sqrt{2}}{10}$ & $0$ & $\frac{\sqrt{2}}{10}$ & $0$ & $-\frac{\sqrt{2}}{10}$ & $0$ 
    \\
    $\hat{M}_{z}^{\alpha\prime}$ & $-\frac{\sqrt{2}}{6}$ & $\frac{4}{15}$ & $0$ & $-\frac{2}{15}$ & $\frac{\sqrt{2}}{6}$ & $0$ & $0$ & $0$ & $\frac{\sqrt{2}}{10}$ & $-\frac{\sqrt{2}}{10}$ & $0$ & $-\frac{\sqrt{2}}{10}$ & $0$ & $\frac{\sqrt{2}}{10}$ & $0$ 
    \\
    $\hat{M}_{x}^{\beta\prime}$ & $-\frac{\sqrt{2}}{6}$ & $0$ & $0$ & $0$ & $\frac{\sqrt{2}}{12}$ & $\frac{\sqrt{6}}{12}$ & $0$ & $\frac{\sqrt{6}}{12}$ & $-\frac{\sqrt{2}}{12}$ & $-\frac{\sqrt{2}}{6}$ & $0$ & $\frac{\sqrt{2}}{6}$ & $0$ & $\frac{\sqrt{2}}{6}$ & $0$ 
    \\
    $\hat{M}_{y}^{\beta\prime}$ & $-\frac{\sqrt{2}}{6}$ & $0$ & $0$ & $0$ & $\frac{\sqrt{2}}{12}$ & $-\frac{\sqrt{6}}{12}$ & $0$ & $-\frac{\sqrt{6}}{12}$ & $-\frac{\sqrt{2}}{12}$ & $\frac{\sqrt{2}}{6}$ & $0$ & $-\frac{\sqrt{2}}{6}$ & $0$ & $\frac{\sqrt{2}}{6}$ & $0$ 
    \\
    $\hat{M}_{z}^{\beta\prime}$ & $-\frac{\sqrt{2}}{6}$ & $0$ & $0$ & $0$ & $-\frac{\sqrt{2}}{6}$ & $0$ & $0$ & $0$ & $\frac{\sqrt{2}}{6}$ & $\frac{\sqrt{2}}{6}$ & $0$ & $\frac{\sqrt{2}}{6}$ & $0$ & $-\frac{\sqrt{2}}{6}$ & $0$ 
    \\
  \end{longtable*}
\end{center}

\begin{center}
  \begin{longtable*}{lccccccccccccccc}
    \caption{
      The Cooper pairing $\hat{\mathcal{H}}_{\mathrm{eff}}$ in Eq.~(\ref{eq:channel_decomposition}) induced by local E/ET-type multipole fluctuations in $sp$-orbital space.
    }
    \label{tab:Pairing_sp_ls_E}
    \\
    
    \hline\hline
    \multicolumn{16}{c}
    {spinless Cooper pairing $\mathcal{H}^{(0)}$}
    \\
    \hline\hline
    $\hat{\Lambda}$ & $\check{Q}_{x}$ & $\check{Q}_{y}$ & $\check{Q}_{z}$ & & $\check{Q}_{0}^{s}\check{Q}_{0}$ & & $\check{Q}_{0}^{s}\check{Q}_{u}$ & & $\check{Q}_{0}^{s}\check{Q}_{v}$ 
    \\
    \endfirsthead
    
    \multicolumn{16}{c}
    {{\tablename\ \thetable{} -- continued from previous page}}
    \\
    \endhead
    
    \multicolumn{16}{c}{{Continued on next page}}
    \\
    \endfoot
    
    \hline\hline
    \endlastfoot

    \hline
    $\hat{Q}_{x}$ & $1$ & $0$ & $0$ & & $\frac{\sqrt{3}}{6}$ & & $-\frac{\sqrt{6}}{12}$ & & $\frac{\sqrt{2}}{4}$ 
    \\
    $\hat{Q}_{y}$ & $0$ & $1$ & $0$ & & $\frac{\sqrt{3}}{6}$ & & $-\frac{\sqrt{6}}{12}$ & & $-\frac{\sqrt{2}}{4}$ 
    \\
    $\hat{Q}_{z}$ & $0$ & $0$ & $1$ & & $\frac{\sqrt{3}}{6}$ & & $\frac{\sqrt{6}}{6}$ & & $0$ 
    \\
    $\hat{Q}_{x}^{\prime}$ & $0$ & $-\frac{1}{2}$ & $-\frac{1}{2}$ & & $\frac{\sqrt{3}}{6}$ & & $\frac{\sqrt{6}}{24}$ & & $-\frac{\sqrt{2}}{8}$ 
    \\
    $\hat{Q}_{y}^{\prime}$ & $-\frac{1}{2}$ & $0$ & $-\frac{1}{2}$ & & $\frac{\sqrt{3}}{6}$ & & $\frac{\sqrt{6}}{24}$ & & $\frac{\sqrt{2}}{8}$ 
    \\
    $\hat{Q}_{z}^{\prime}$ & $-\frac{1}{2}$ & $-\frac{1}{2}$ & $0$ & & $\frac{\sqrt{3}}{6}$ & & $-\frac{\sqrt{6}}{12}$ & & $0$ 
    \\
    $\hat{G}_{0}^{\prime}$ & $-\frac{1}{3}$ & $-\frac{1}{3}$ & $-\frac{1}{3}$ & & $\frac{\sqrt{3}}{6}$ & & $0$ & & $0$ 
    \\
    $\hat{G}_{u}^{\prime}$ & $-\frac{1}{6}$ & $-\frac{1}{6}$ & $-\frac{2}{3}$ & & $\frac{\sqrt{3}}{6}$ & & $\frac{\sqrt{6}}{12}$ & & $0$ 
    \\
    $\hat{G}_{v}^{\prime}$ & $-\frac{1}{2}$ & $-\frac{1}{2}$ & $0$ & & $\frac{\sqrt{3}}{6}$ & & $-\frac{\sqrt{6}}{12}$ & & $0$ 
    \\
    $\hat{G}_{yz}^{\prime}$ & $0$ & $-\frac{1}{2}$ & $-\frac{1}{2}$ & & $\frac{\sqrt{3}}{6}$ & & $\frac{\sqrt{6}}{24}$ & & $-\frac{\sqrt{2}}{8}$ 
    \\
    $\hat{G}_{zx}^{\prime}$ & $-\frac{1}{2}$ & $0$ & $-\frac{1}{2}$ & & $\frac{\sqrt{3}}{6}$ & & $\frac{\sqrt{6}}{24}$ & & $\frac{\sqrt{2}}{8}$ 
    \\
    $\hat{G}_{xy}^{\prime}$ & $-\frac{1}{2}$ & $-\frac{1}{2}$ & $0$ & & $\frac{\sqrt{3}}{6}$ & & $-\frac{\sqrt{6}}{12}$ & & $0$ 
    \\
    \hline\hline
    \multicolumn{16}{c}{spinful Cooper pairing $\hat{\mathcal{H}}_{1}$}
    \\
    \hline\hline
    $\hat{\Lambda}$ & $\check{Q}_{x}^{\prime}$ & $\check{Q}_{y}^{\prime}$ & $\check{Q}_{z}^{\prime}$ & $\check{G}_{0}^{\prime}$ & $\check{G}_{u}^{\prime}$ & $\check{G}_{v}^{\prime}$ & $\check{G}_{yz}^{\prime}$ & $\check{G}_{zx}^{\prime}$ & $\check{G}_{xy}^{\prime}$ & $\check{G}_{0}^{\prime}\check{G}_{u}^{\prime}$ & $\check{G}_{0}^{\prime}\check{G}_{v}^{\prime}$ & $\check{G}_{u}^{\prime}\check{G}_{v}^{\prime}$ & $\check{Q}_{x}^{\prime}\check{G}_{yz}^{\prime}$ & $\check{Q}_{y}^{\prime}\check{G}_{zx}^{\prime}$ & $\check{Q}_{z}^{\prime}\check{G}_{xy}^{\prime}$ 
    \\
    \hline
    $\hat{Q}_{x}$ & $0$ & $-\frac{1}{2}$ & $-\frac{1}{2}$ & $-\frac{1}{3}$ & $-\frac{1}{6}$ & $-\frac{1}{2}$ & $0$ & $-\frac{1}{2}$ & $-\frac{1}{2}$ & $\frac{\sqrt{2}}{12}$ & $-\frac{\sqrt{6}}{12}$ & $\frac{\sqrt{3}}{12}$ & $0$ & $\frac{1}{4}$ & $-\frac{1}{4}$ 
    \\
    $\hat{Q}_{y}$ & $-\frac{1}{2}$ & $0$ & $-\frac{1}{2}$ & $-\frac{1}{3}$ & $-\frac{1}{6}$ & $-\frac{1}{2}$ & $-\frac{1}{2}$ & $0$ & $-\frac{1}{2}$ & $\frac{\sqrt{2}}{12}$ & $\frac{\sqrt{6}}{12}$ & $-\frac{\sqrt{3}}{12}$ & $-\frac{1}{4}$ & $0$ & $\frac{1}{4}$ 
    \\
    $\hat{Q}_{z}$ & $-\frac{1}{2}$ & $-\frac{1}{2}$ & $0$ & $-\frac{1}{3}$ & $-\frac{2}{3}$ & $0$ & $-\frac{1}{2}$ & $-\frac{1}{2}$ & $0$ & $-\frac{\sqrt{2}}{6}$ & $0$ & $0$ & $\frac{1}{4}$ & $-\frac{1}{4}$ & $0$ 
    \\
    $\hat{Q}_{x}^{\prime}$ & $1$ & $-\frac{1}{4}$ & $-\frac{1}{4}$ & $-\frac{2}{3}$ & $-\frac{1}{12}$ & $-\frac{1}{4}$ & $0$ & $-\frac{1}{4}$ & $-\frac{1}{4}$ & $-\frac{\sqrt{2}}{12}$ & $\frac{\sqrt{6}}{12}$ & $\frac{\sqrt{3}}{24}$ & $0$ & $-\frac{1}{8}$ & $\frac{1}{8}$ 
    \\
    $\hat{Q}_{y}^{\prime}$ & $-\frac{1}{4}$ & $1$ & $-\frac{1}{4}$ & $-\frac{2}{3}$ & $-\frac{1}{12}$ & $-\frac{1}{4}$ & $-\frac{1}{4}$ & $0$ & $-\frac{1}{4}$ & $-\frac{\sqrt{2}}{12}$ & $-\frac{\sqrt{6}}{12}$ & $-\frac{\sqrt{3}}{24}$ & $\frac{1}{8}$ & $0$ & $-\frac{1}{8}$ 
    \\
    $\hat{Q}_{z}^{\prime}$ & $-\frac{1}{4}$ & $-\frac{1}{4}$ & $1$ & $-\frac{2}{3}$ & $-\frac{1}{3}$ & $0$ & $-\frac{1}{4}$ & $-\frac{1}{4}$ & $0$ & $\frac{\sqrt{2}}{6}$ & $0$ & $0$ & $-\frac{1}{8}$ & $\frac{1}{8}$ & $0$ 
    \\
    $\hat{G}_{0}^{\prime}$ & $-\frac{2}{3}$ & $-\frac{2}{3}$ & $-\frac{2}{3}$ & $1$ & $0$ & $0$ & $0$ & $0$ & $0$ & $0$ & $0$ & $0$ & $0$ & $0$ & $0$ 
    \\
    $\hat{G}_{u}^{\prime}$ & $-\frac{1}{12}$ & $-\frac{1}{12}$ & $-\frac{1}{3}$ & $0$ & $1$ & $0$ & $-\frac{3}{4}$ & $-\frac{3}{4}$ & $0$ & $0$ & $0$ & $0$ & $\frac{1}{8}$ & $-\frac{1}{8}$ & $0$ 
    \\
    $\hat{G}_{v}^{\prime}$ & $-\frac{1}{4}$ & $-\frac{1}{4}$ & $0$ & $0$ & $0$ & $1$ & $-\frac{1}{4}$ & $-\frac{1}{4}$ & $-1$ & $0$ & $0$ & $0$ & $-\frac{1}{8}$ & $\frac{1}{8}$ & $0$ 
    \\
    $\hat{G}_{yz}^{\prime}$ & $0$ & $-\frac{1}{4}$ & $-\frac{1}{4}$ & $0$ & $-\frac{3}{4}$ & $-\frac{1}{4}$ & $1$ & $-\frac{1}{4}$ & $-\frac{1}{4}$ & $0$ & $0$ & $-\frac{\sqrt{3}}{8}$ & $0$ & $-\frac{1}{8}$ & $\frac{1}{8}$ 
    \\
    $\hat{G}_{zx}^{\prime}$ & $-\frac{1}{4}$ & $0$ & $-\frac{1}{4}$ & $0$ & $-\frac{3}{4}$ & $-\frac{1}{4}$ & $-\frac{1}{4}$ & $1$ & $-\frac{1}{4}$ & $0$ & $0$ & $\frac{\sqrt{3}}{8}$ & $\frac{1}{8}$ & $0$ & $-\frac{1}{8}$ 
    \\
    $\hat{G}_{xy}^{\prime}$ & $-\frac{1}{4}$ & $-\frac{1}{4}$ & $0$ & $0$ & $0$ & $-1$ & $-\frac{1}{4}$ & $-\frac{1}{4}$ & $1$ & $0$ & $0$ & $0$ & $-\frac{1}{8}$ & $\frac{1}{8}$ & $0$ 
    \\
    \hline\hline
    \multicolumn{16}{c}{mixed Cooper pairing $\hat{\mathcal{H}}_{2}$}
    \\
    \hline\hline
    $\hat{\Lambda}$ & $\check{Q}_{0}^{s}\check{Q}_{0}^{\prime}$ & $\check{Q}_{0}^{s}\check{Q}_{u}^{\prime}$ & $\check{Q}_{0}^{s}\check{Q}_{v}^{\prime}$ & & $\check{Q}_{x}\check{Q}_{x}^{\prime}$ & $\check{Q}_{x}\check{G}_{yz}^{\prime}$ & & $\check{Q}_{y}\check{Q}_{y}^{\prime}$ & $\check{Q}_{y}\check{G}_{zx}^{\prime}$ & & $\check{Q}_{z}\check{Q}_{z}^{\prime}$ & $\check{Q}_{z}\check{G}_{xy}^{\prime}$ 
    \\
    \hline
    $\hat{Q}_{x}^{\prime}$ & $-\frac{\sqrt{6}}{12}$ & $\frac{\sqrt{3}}{12}$ & $-\frac{1}{4}$ & & $0$ & $0$ & & $-\frac{\sqrt{2}}{8}$ & $-\frac{\sqrt{2}}{8}$ & & $-\frac{\sqrt{2}}{8}$ & $\frac{\sqrt{2}}{8}$ 
    \\
    $\hat{Q}_{y}^{\prime}$ & $-\frac{\sqrt{6}}{12}$ & $\frac{\sqrt{3}}{12}$ & $\frac{1}{4}$ & & $-\frac{\sqrt{2}}{8}$ & $\frac{\sqrt{2}}{8}$ & & $0$ & $0$ & & $-\frac{\sqrt{2}}{8}$ & $-\frac{\sqrt{2}}{8}$ 
    \\
    $\hat{Q}_{z}^{\prime}$ & $-\frac{\sqrt{6}}{12}$ & $-\frac{\sqrt{3}}{6}$ & $0$ & & $-\frac{\sqrt{2}}{8}$ & $-\frac{\sqrt{2}}{8}$ & & $-\frac{\sqrt{2}}{8}$ & $\frac{\sqrt{2}}{8}$ & & $0$ & $0$ 
    \\
    $\hat{G}_{0}^{\prime}$ & $-\frac{\sqrt{6}}{6}$ & $0$ & $0$ & & $-\frac{\sqrt{2}}{6}$ & $0$ & & $-\frac{\sqrt{2}}{6}$ & $0$ & & $-\frac{\sqrt{2}}{6}$ & $0$ 
    \\
    $\hat{G}_{u}^{\prime}$ & $\frac{\sqrt{6}}{12}$ & $-\frac{\sqrt{3}}{6}$ & $0$ & & $\frac{\sqrt{2}}{24}$ & $-\frac{\sqrt{2}}{8}$ & & $\frac{\sqrt{2}}{24}$ & $\frac{\sqrt{2}}{8}$ & & $\frac{\sqrt{2}}{6}$ & $0$ 
    \\
    $\hat{G}_{v}^{\prime}$ & $\frac{\sqrt{6}}{12}$ & $\frac{\sqrt{3}}{6}$ & $0$ & & $\frac{\sqrt{2}}{8}$ & $\frac{\sqrt{2}}{8}$ & & $\frac{\sqrt{2}}{8}$ & $-\frac{\sqrt{2}}{8}$ & & $0$ & $0$ 
    \\
    $\hat{G}_{yz}^{\prime}$ & $\frac{\sqrt{6}}{12}$ & $-\frac{\sqrt{3}}{12}$ & $\frac{1}{4}$ & & $0$ & $0$ & & $\frac{\sqrt{2}}{8}$ & $\frac{\sqrt{2}}{8}$ & & $\frac{\sqrt{2}}{8}$ & $-\frac{\sqrt{2}}{8}$ 
    \\
    $\hat{G}_{zx}^{\prime}$ & $\frac{\sqrt{6}}{12}$ & $-\frac{\sqrt{3}}{12}$ & $-\frac{1}{4}$ & & $\frac{\sqrt{2}}{8}$ & $-\frac{\sqrt{2}}{8}$ & & $0$ & $0$ & & $\frac{\sqrt{2}}{8}$ & $\frac{\sqrt{2}}{8}$ 
    \\
    $\hat{G}_{xy}^{\prime}$ & $\frac{\sqrt{6}}{12}$ & $\frac{\sqrt{3}}{6}$ & $0$ & & $\frac{\sqrt{2}}{8}$ & $\frac{\sqrt{2}}{8}$ & & $\frac{\sqrt{2}}{8}$ & $-\frac{\sqrt{2}}{8}$ & & $0$ & $0$ 
    \\
  \end{longtable*}
\end{center}

\begin{center}
  \begin{longtable*}{lccccccccccccccc}
    \caption{
      The Cooper pairing $\hat{\mathcal{H}}_{\mathrm{eff}}$ in Eq.~(\ref{eq:channel_decomposition}) induced by local M/MT-type multipole fluctuations in $sp$-orbital space.
    }
    \label{tab:Pairing_sp_ls_M}
    \\
    
    \hline\hline
    \multicolumn{16}{c}
    {spinless Cooper pairing $\mathcal{H}^{(0)}$}
    \\
    \hline\hline
    $\hat{\Lambda}$ & $\check{Q}_{x}$ & $\check{Q}_{y}$ & $\check{Q}_{z}$ & & $\check{Q}_{0}^{s}\check{Q}_{0}$ & & $\check{Q}_{0}^{s}\check{Q}_{u}$ & & $\check{Q}_{0}^{s}\check{Q}_{v}$ 
    \\
    \endfirsthead
    
    \multicolumn{16}{c}
    {{\tablename\ \thetable{} -- continued from previous page}}
    \\
    \endhead
    
    \multicolumn{16}{c}{{Continued on next page}}
    \\
    \endfoot
    
    \hline\hline
    \endlastfoot

    \hline
    $\hat{T}_{x}$ & $1$ & $0$ & $0$ & & $-\frac{\sqrt{3}}{6}$ & & $\frac{\sqrt{6}}{12}$ & & $-\frac{\sqrt{2}}{4}$ 
    \\
    $\hat{T}_{y}$ & $0$ & $1$ & $0$ & & $-\frac{\sqrt{3}}{6}$ & & $\frac{\sqrt{6}}{12}$ & & $\frac{\sqrt{2}}{4}$ 
    \\
    $\hat{T}_{z}$ & $0$ & $0$ & $1$ & & $-\frac{\sqrt{3}}{6}$ & & $-\frac{\sqrt{6}}{6}$ & & $0$ 
    \\
    $\hat{T}_{x}^{\prime}$ & $0$ & $-\frac{1}{2}$ & $-\frac{1}{2}$ & & $-\frac{\sqrt{3}}{6}$ & & $-\frac{\sqrt{6}}{24}$ & & $\frac{\sqrt{2}}{8}$ 
    \\
    $\hat{T}_{y}^{\prime}$ & $-\frac{1}{2}$ & $0$ & $-\frac{1}{2}$ & & $-\frac{\sqrt{3}}{6}$ & & $-\frac{\sqrt{6}}{24}$ & & $-\frac{\sqrt{2}}{8}$ 
    \\
    $\hat{T}_{z}^{\prime}$ & $-\frac{1}{2}$ & $-\frac{1}{2}$ & $0$ & & $-\frac{\sqrt{3}}{6}$ & & $\frac{\sqrt{6}}{12}$ & & $0$ 
    \\
    $\hat{M}_{0}^{\prime}$ & $-\frac{1}{3}$ & $-\frac{1}{3}$ & $-\frac{1}{3}$ & & $-\frac{\sqrt{3}}{6}$ & & $0$ & & $0$ 
    \\
    $\hat{M}_{u}^{\prime}$ & $-\frac{1}{6}$ & $-\frac{1}{6}$ & $-\frac{2}{3}$ & & $-\frac{\sqrt{3}}{6}$ & & $-\frac{\sqrt{6}}{12}$ & & $0$ 
    \\
    $\hat{M}_{v}^{\prime}$ & $-\frac{1}{2}$ & $-\frac{1}{2}$ & $0$ & & $-\frac{\sqrt{3}}{6}$ & & $\frac{\sqrt{6}}{12}$ & & $0$ 
    \\
    $\hat{M}_{yz}^{\prime}$ & $0$ & $-\frac{1}{2}$ & $-\frac{1}{2}$ & & $-\frac{\sqrt{3}}{6}$ & & $-\frac{\sqrt{6}}{24}$ & & $\frac{\sqrt{2}}{8}$ 
    \\
    $\hat{M}_{zx}^{\prime}$ & $-\frac{1}{2}$ & $0$ & $-\frac{1}{2}$ & & $-\frac{\sqrt{3}}{6}$ & & $-\frac{\sqrt{6}}{24}$ & & $-\frac{\sqrt{2}}{8}$ 
    \\
    $\hat{M}_{xy}^{\prime}$ & $-\frac{1}{2}$ & $-\frac{1}{2}$ & $0$ & & $-\frac{\sqrt{3}}{6}$ & & $\frac{\sqrt{6}}{12}$ & & $0$ 
    \\
    \hline\hline
    \multicolumn{16}{c}{spinful Cooper pairing $\hat{\mathcal{H}}_{1}$}
    \\
    \hline\hline
    $\hat{\Lambda}$ & $\check{Q}_{x}^{\prime}$ & $\check{Q}_{y}^{\prime}$ & $\check{Q}_{z}^{\prime}$ & $\check{G}_{0}^{\prime}$ & $\check{G}_{u}^{\prime}$ & $\check{G}_{v}^{\prime}$ & $\check{G}_{yz}^{\prime}$ & $\check{G}_{zx}^{\prime}$ & $\check{G}_{xy}^{\prime}$ & $\check{G}_{0}^{\prime}\check{G}_{u}^{\prime}$ & $\check{G}_{0}^{\prime}\check{G}_{v}^{\prime}$ & $\check{G}_{u}^{\prime}\check{G}_{v}^{\prime}$ & $\check{Q}_{x}^{\prime}\check{G}_{yz}^{\prime}$ & $\check{Q}_{y}^{\prime}\check{G}_{zx}^{\prime}$ & $\check{Q}_{z}^{\prime}\check{G}_{xy}^{\prime}$ 
    \\
    \hline
    $\hat{T}_{x}$ & $0$ & $-\frac{1}{2}$ & $-\frac{1}{2}$ & $-\frac{1}{3}$ & $-\frac{1}{6}$ & $-\frac{1}{2}$ & $0$ & $-\frac{1}{2}$ & $-\frac{1}{2}$ & $\frac{\sqrt{2}}{12}$ & $-\frac{\sqrt{6}}{12}$ & $\frac{\sqrt{3}}{12}$ & $0$ & $\frac{1}{4}$ & $-\frac{1}{4}$ 
    \\
    $\hat{T}_{y}$ & $-\frac{1}{2}$ & $0$ & $-\frac{1}{2}$ & $-\frac{1}{3}$ & $-\frac{1}{6}$ & $-\frac{1}{2}$ & $-\frac{1}{2}$ & $0$ & $-\frac{1}{2}$ & $\frac{\sqrt{2}}{12}$ & $\frac{\sqrt{6}}{12}$ & $-\frac{\sqrt{3}}{12}$ & $-\frac{1}{4}$ & $0$ & $\frac{1}{4}$ 
    \\
    $\hat{T}_{z}$ & $-\frac{1}{2}$ & $-\frac{1}{2}$ & $0$ & $-\frac{1}{3}$ & $-\frac{2}{3}$ & $0$ & $-\frac{1}{2}$ & $-\frac{1}{2}$ & $0$ & $-\frac{\sqrt{2}}{6}$ & $0$ & $0$ & $\frac{1}{4}$ & $-\frac{1}{4}$ & $0$ 
    \\
    $\hat{T}_{x}^{\prime}$ & $1$ & $-\frac{1}{4}$ & $-\frac{1}{4}$ & $-\frac{2}{3}$ & $-\frac{1}{12}$ & $-\frac{1}{4}$ & $0$ & $-\frac{1}{4}$ & $-\frac{1}{4}$ & $-\frac{\sqrt{2}}{12}$ & $\frac{\sqrt{6}}{12}$ & $\frac{\sqrt{3}}{24}$ & $0$ & $-\frac{1}{8}$ & $\frac{1}{8}$ 
    \\
    $\hat{T}_{y}^{\prime}$ & $-\frac{1}{4}$ & $1$ & $-\frac{1}{4}$ & $-\frac{2}{3}$ & $-\frac{1}{12}$ & $-\frac{1}{4}$ & $-\frac{1}{4}$ & $0$ & $-\frac{1}{4}$ & $-\frac{\sqrt{2}}{12}$ & $-\frac{\sqrt{6}}{12}$ & $-\frac{\sqrt{3}}{24}$ & $\frac{1}{8}$ & $0$ & $-\frac{1}{8}$ 
    \\
    $\hat{T}_{z}^{\prime}$ & $-\frac{1}{4}$ & $-\frac{1}{4}$ & $1$ & $-\frac{2}{3}$ & $-\frac{1}{3}$ & $0$ & $-\frac{1}{4}$ & $-\frac{1}{4}$ & $0$ & $\frac{\sqrt{2}}{6}$ & $0$ & $0$ & $-\frac{1}{8}$ & $\frac{1}{8}$ & $0$ 
    \\
    $\hat{M}_{0}^{\prime}$ & $-\frac{2}{3}$ & $-\frac{2}{3}$ & $-\frac{2}{3}$ & $1$ & $0$ & $0$ & $0$ & $0$ & $0$ & $0$ & $0$ & $0$ & $0$ & $0$ & $0$ 
    \\
    $\hat{M}_{u}^{\prime}$ & $-\frac{1}{12}$ & $-\frac{1}{12}$ & $-\frac{1}{3}$ & $0$ & $1$ & $0$ & $-\frac{3}{4}$ & $-\frac{3}{4}$ & $0$ & $0$ & $0$ & $0$ & $\frac{1}{8}$ & $-\frac{1}{8}$ & $0$ 
    \\
    $\hat{M}_{v}^{\prime}$ & $-\frac{1}{4}$ & $-\frac{1}{4}$ & $0$ & $0$ & $0$ & $1$ & $-\frac{1}{4}$ & $-\frac{1}{4}$ & $-1$ & $0$ & $0$ & $0$ & $-\frac{1}{8}$ & $\frac{1}{8}$ & $0$ 
    \\
    $\hat{M}_{yz}^{\prime}$ & $0$ & $-\frac{1}{4}$ & $-\frac{1}{4}$ & $0$ & $-\frac{3}{4}$ & $-\frac{1}{4}$ & $1$ & $-\frac{1}{4}$ & $-\frac{1}{4}$ & $0$ & $0$ & $-\frac{\sqrt{3}}{8}$ & $0$ & $-\frac{1}{8}$ & $\frac{1}{8}$ 
    \\
    $\hat{M}_{zx}^{\prime}$ & $-\frac{1}{4}$ & $0$ & $-\frac{1}{4}$ & $0$ & $-\frac{3}{4}$ & $-\frac{1}{4}$ & $-\frac{1}{4}$ & $1$ & $-\frac{1}{4}$ & $0$ & $0$ & $\frac{\sqrt{3}}{8}$ & $\frac{1}{8}$ & $0$ & $-\frac{1}{8}$ 
    \\
    $\hat{M}_{xy}^{\prime}$ & $-\frac{1}{4}$ & $-\frac{1}{4}$ & $0$ & $0$ & $0$ & $-1$ & $-\frac{1}{4}$ & $-\frac{1}{4}$ & $1$ & $0$ & $0$ & $0$ & $-\frac{1}{8}$ & $\frac{1}{8}$ & $0$ 
    \\
    \hline\hline
    \multicolumn{16}{c}{mixed Cooper pairing $\hat{\mathcal{H}}_{2}$}
    \\
    \hline\hline
    $\hat{\Lambda}$ & $\check{Q}_{0}^{s}\check{Q}_{0}^{\prime}$ & $\check{Q}_{0}^{s}\check{Q}_{u}^{\prime}$ & $\check{Q}_{0}^{s}\check{Q}_{v}^{\prime}$ & & $\check{Q}_{x}\check{Q}_{x}^{\prime}$ & $\check{Q}_{x}\check{G}_{yz}^{\prime}$ & & $\check{Q}_{y}\check{Q}_{y}^{\prime}$ & $\check{Q}_{y}\check{G}_{zx}^{\prime}$ & & $\check{Q}_{z}\check{Q}_{z}^{\prime}$ & $\check{Q}_{z}\check{G}_{xy}^{\prime}$ 
    \\
    \hline
    $\hat{T}_{x}^{\prime}$ & $\frac{\sqrt{6}}{12}$ & $-\frac{\sqrt{3}}{12}$ & $\frac{1}{4}$ & & $0$ & $0$ & & $-\frac{\sqrt{2}}{8}$ & $-\frac{\sqrt{2}}{8}$ & & $-\frac{\sqrt{2}}{8}$ & $\frac{\sqrt{2}}{8}$ 
    \\
    $\hat{T}_{y}^{\prime}$ & $\frac{\sqrt{6}}{12}$ & $-\frac{\sqrt{3}}{12}$ & $-\frac{1}{4}$ & & $-\frac{\sqrt{2}}{8}$ & $\frac{\sqrt{2}}{8}$ & & $0$ & $0$ & & $-\frac{\sqrt{2}}{8}$ & $-\frac{\sqrt{2}}{8}$ 
    \\
    $\hat{T}_{z}^{\prime}$ & $\frac{\sqrt{6}}{12}$ & $\frac{\sqrt{3}}{6}$ & $0$ & & $-\frac{\sqrt{2}}{8}$ & $-\frac{\sqrt{2}}{8}$ & & $-\frac{\sqrt{2}}{8}$ & $\frac{\sqrt{2}}{8}$ & & $0$ & $0$ 
    \\
    $\hat{M}_{0}^{\prime}$ & $\frac{\sqrt{6}}{6}$ & $0$ & $0$ & & $-\frac{\sqrt{2}}{6}$ & $0$ & & $-\frac{\sqrt{2}}{6}$ & $0$ & & $-\frac{\sqrt{2}}{6}$ & $0$ 
    \\
    $\hat{M}_{u}^{\prime}$ & $-\frac{\sqrt{6}}{12}$ & $\frac{\sqrt{3}}{6}$ & $0$ & & $\frac{\sqrt{2}}{24}$ & $-\frac{\sqrt{2}}{8}$ & & $\frac{\sqrt{2}}{24}$ & $\frac{\sqrt{2}}{8}$ & & $\frac{\sqrt{2}}{6}$ & $0$ 
    \\
    $\hat{M}_{v}^{\prime}$ & $-\frac{\sqrt{6}}{12}$ & $-\frac{\sqrt{3}}{6}$ & $0$ & & $\frac{\sqrt{2}}{8}$ & $\frac{\sqrt{2}}{8}$ & & $\frac{\sqrt{2}}{8}$ & $-\frac{\sqrt{2}}{8}$ & & $0$ & $0$ 
    \\
    $\hat{M}_{yz}^{\prime}$ & $-\frac{\sqrt{6}}{12}$ & $\frac{\sqrt{3}}{12}$ & $-\frac{1}{4}$ & & $0$ & $0$ & & $\frac{\sqrt{2}}{8}$ & $\frac{\sqrt{2}}{8}$ & & $\frac{\sqrt{2}}{8}$ & $-\frac{\sqrt{2}}{8}$ 
    \\
    $\hat{M}_{zx}^{\prime}$ & $-\frac{\sqrt{6}}{12}$ & $\frac{\sqrt{3}}{12}$ & $\frac{1}{4}$ & & $\frac{\sqrt{2}}{8}$ & $-\frac{\sqrt{2}}{8}$ & & $0$ & $0$ & & $\frac{\sqrt{2}}{8}$ & $\frac{\sqrt{2}}{8}$ 
    \\
    $\hat{M}_{xy}^{\prime}$ & $-\frac{\sqrt{6}}{12}$ & $-\frac{\sqrt{3}}{6}$ & $0$ & & $\frac{\sqrt{2}}{8}$ & $\frac{\sqrt{2}}{8}$ & & $\frac{\sqrt{2}}{8}$ & $-\frac{\sqrt{2}}{8}$ & & $0$ & $0$ 
    \\
  \end{longtable*}
\end{center}

\section{Classification of pairing under $jj_{z}$ basis}
\label{app:jjz}

In this appendix, we summarize the classification of the pairing in terms of total angular momentum basis $\ket{jj_z}$ by performing the unitary transformation discussed in the main text.
First, we give the multipoles for $p$-orbitals in $j_{1}$-$j_{2}$ space $\hat{X}^{(j_{1}+j_{2})}$ by using the multipoles in Appendix~\ref{app:MP}:
\begin{subequations}
  \label{eq:jjz_MPs}
  \begin{equation}
    \begin{aligned}
      &\, \hat{Q}_{0}^{(1)}=\frac{\hat{Q}_{0}-\sqrt{2}\hat{Q}_{0}^{\prime}}{\sqrt{3}},
      \\
      &\, \hat{M}^{(1)}_{1m}=\frac{\sqrt{6}\hat{M}_{1m}-\hat{M}^{(s)}_{1m}+2\sqrt{5}\hat{M}^{\prime}_{1m}}{3\sqrt{3}}
      \\
    \end{aligned}
  \end{equation}
  for $(j_{1},j_{2})=(1/2,1/2)$, 
  \begin{equation}
    \begin{aligned}
      &\, \hat{M}_{1m}^{(2)}=\frac{\sqrt{6}\hat{M}_{1m}-4\hat{M}^{(s)}_{1m}-\sqrt{5}\hat{M}^{\prime}_{1m}}{3\sqrt{3}},
      \\
      &\, \hat{G}_{1m}^{(2)}=\hat{G}_{1m}^{\prime},
      \\
      &\, \hat{Q}_{2m}^{(2)}=\frac{\sqrt{2}\hat{Q}_{2m}+\hat{Q}_{2m}^{\prime}}{\sqrt{3}},
      \\
      &\, \hat{T}_{2m}^{(2)}=\hat{T}_{2m}^{\prime}
      \\
    \end{aligned}
  \end{equation}
  for $(j_{1},j_{2})=(1/2,3/2)$, and 
  \begin{equation}
    \begin{aligned}
      &\, \hat{Q}_{0}^{(3)}=\frac{\sqrt{2}\hat{Q}_{0}+\hat{Q}_{0}^{\prime}}{\sqrt{3}},
      \\
      &\, \hat{M}_{1m}^{(3)}=\frac{\sqrt{15}\hat{M}_{1m}+\sqrt{10}\hat{M}^{(s)}_{1m}-\sqrt{2}\hat{M}^{\prime}_{1m}}{3\sqrt{3}},
      \\
      &\, \hat{Q}_{2m}^{(3)}=\frac{\hat{Q}_{2m}-\sqrt{2}\hat{Q}_{2m}^{\prime}}{\sqrt{3}},
      \\
      &\, \hat{M}_{3m}^{(3)}=\hat{M}_{3m}^{\prime}
      \\
    \end{aligned}
  \end{equation}
  for $(j_{1},j_{2})=(3/2,3/2)$.
  Here, we have used the abbreviations such as $\hat{X}_{1m}=(X_{x},X_{y},X_{z})$, $\hat{X}_{2m}=(X_{u},X_{v},X_{yz},X_{zx},X_{xy})$, and $\hat{X}_{3m}=(X_{xyz},X_{x}^{\alpha},X_{y}^{\alpha},X_{z}^{\alpha},X_{x}^{\beta},X_{y}^{\beta},X_{z}^{\beta})$.
\end{subequations}
The pairing interaction in $p$-orbital space is summarized in Tables~\ref{tab:Pairing_p_j_E} and \ref{tab:Pairing_p_j_M}.

Second, the multipoles for $sp$-orbitals in $j_{1}$-$j_{2}$ space $\hat{X}^{(j_{1}+j_{2})}$ are as follows:
\begin{subequations}
  \begin{equation}
    \begin{aligned}
      &\, \hat{X}_{0}^{(1)}=\hat{X}_{0}^{\prime};&\, X=M,G
      \\
      &\, \hat{X}_{1m}^{(1)}=\frac{\hat{X}_{1m}+\sqrt{2}\hat{X}_{1m}^{\prime}}{\sqrt{3}};&\, X=Q,T
    \end{aligned}
  \end{equation}
  for $(j_{1},j_{2})=(1/2,1/2)$, and 
  \begin{equation}
    \begin{aligned}
      &\, \hat{X}_{1m}^{(2)}=\frac{\sqrt{2}\hat{X}_{1m}-\hat{X}_{1m}^{\prime}}{\sqrt{3}};&\, X=Q,T
      \\
      &\, \hat{X}_{2m}^{(2)}=\hat{X}_{2m}^{\prime};&\, X=M,G
      \\
    \end{aligned}
  \end{equation}
  for $(j_{1},j_{2})=(1/2,3/2)$.
\end{subequations}
The pairing interaction in $sp$-orbital space is summarized in Tables~\ref{tab:Pairing_sp_j_E} and \ref{tab:Pairing_sp_j_M}.

\begin{center}
  \begin{longtable*}{lccccccccccccccc}
    \caption{
      The Cooper pairing $\hat{\mathcal{H}}_{\mathrm{eff}}$ in Eq.~(\ref{eq:cooper_pairing}) induced  by local E/ET-type multipole fluctuations $X^{(J)}$ in $p$-orbital space.
      The first column is the multipole fluctuation $\hat{\Lambda}$ in Eq.~(\ref{eq:local_vertex}).
      The cloumns $\check{O}_{\nu}^{(J)}$ give the coefficients $c_{\nu\nu}$ for the pairing term expressed as $(V_{0}/N)c_{\nu\nu}\check{O}_{\nu}^{(J)\dag}\check{O}_{\nu}^{(J)}$, whereas $\check{O}_{\mu}^{(J)}\check{O}_{\nu}^{(J^{\prime})}$ columns represent the coefficients $c_{\mu\nu}$ of the pairing term $(V_{0}/2N)c_{\mu\nu}(\check{O}_{\mu}^{(J)\dag}\check{O}_{\nu}^{(J^{\prime})}+\check{O}_{\nu}^{(J^{\prime})\dag}\check{O}_{\mu}^{(J)})$.
    }
    \label{tab:Pairing_p_j_E}
    \\
    
    \hline\hline
    \multicolumn{16}{l}
    {$J=1$ multipole fluctuation}
    \\
    \hline\hline
    $\hat{\Lambda}$ & $\check{Q}_{0}^{(1)}$ & 
    \\
    \endfirsthead
    
    \multicolumn{16}{c}
    {{\tablename\ \thetable{} -- continued from previous page}}
    \\
    \endhead
    
    \multicolumn{16}{c}{{Continued on next page}}
    \\ 
    \endfoot
    
    \hline\hline
    \endlastfoot

    \hline
    $\hat{Q}_{0}^{(1)}$ & $2$
    \\
    \hline\hline
    \multicolumn{16}{l}{$J=2$ multipole fluctuation}
    \\
    \hline\hline
    $\hat{\Lambda}$ & $\check{Q}_{u}^{(2)}$ & $\check{Q}_{v}^{(2)}$ & $\check{Q}_{yz}^{(2)}$ & $\check{Q}_{zx}^{(2)}$ & $\check{Q}_{xy}^{(2)}$ & $\check{G}_{x}^{(2)}$ & $\check{G}_{y}^{(2)}$ & $\check{G}_{z}^{(2)}$ & $\check{Q}_{0}^{(1)}\check{Q}_{0}^{(3)}$ & $\check{Q}_{0}^{(1)}\check{Q}_{u}^{(3)}$ & $\check{Q}_{0}^{(1)}\check{Q}_{v}^{(3)}$ & $\check{Q}_{u}^{(2)}\check{Q}_{v}^{(2)}$ & $\check{Q}_{yz}^{(2)}\check{G}_{x}^{(2)}$ & $\check{Q}_{zx}^{(2)}G_{y}^{(2)}$ & $\check{Q}_{xy}^{(2)}\check{G}_{z}^{(2)}$ 
    \\
    \hline
    $\hat{Q}_{u}^{(2)}$ & $1$ & $0$ & $-\frac{3}{4}$ & $-\frac{3}{4}$ & $0$ & $-\frac{1}{4}$ & $-\frac{1}{4}$ & $-1$ & $\frac{\sqrt{2}}{4}$ & $\frac{\sqrt{2}}{4}$ & $0$ & $0$ & $-\frac{\sqrt{3}}{8}$ & $\frac{\sqrt{3}}{8}$ & $0$
    \\
    $\hat{Q}_{v}^{(2)}$ & $0$ & $1$ & $-\frac{1}{4}$ & $-\frac{1}{4}$ & $-1$ & $-\frac{3}{4}$ & $-\frac{3}{4}$ & $0$ & $\frac{\sqrt{2}}{4}$ & $-\frac{\sqrt{2}}{4}$ & $0$ & $0$ & $\frac{\sqrt{3}}{8}$ & $-\frac{\sqrt{3}}{8}$ & $0$
    \\
    $\hat{Q}_{yz}^{(2)}$ & $-\frac{3}{4}$ & $-\frac{1}{4}$ & $1$ & $-\frac{1}{4}$ & $-\frac{1}{4}$ & $0$ & $-\frac{3}{4}$ & $-\frac{3}{4}$ & $\frac{\sqrt{2}}{4}$ & $\frac{\sqrt{2}}{8}$ & $-\frac{\sqrt{6}}{8}$ & $-\frac{\sqrt{3}}{8}$ & $0$ & $\frac{\sqrt{3}}{8}$ & $-\frac{\sqrt{3}}{8}$
    \\
    $\hat{Q}_{zx}^{(2)}$ & $-\frac{3}{4}$ & $-\frac{1}{4}$ & $-\frac{1}{4}$ & $1$ & $-\frac{1}{4}$ & $-\frac{3}{4}$ & $0$ & $-\frac{3}{4}$ & $\frac{\sqrt{2}}{4}$ & $\frac{\sqrt{2}}{8}$ & $\frac{\sqrt{6}}{8}$ & $\frac{\sqrt{3}}{8}$ & $-\frac{\sqrt{3}}{8}$ & $0$ & $\frac{\sqrt{3}}{8}$
    \\
    $\hat{Q}_{xy}^{(2)}$ & $0$ & $-1$ & $-\frac{1}{4}$ & $-\frac{1}{4}$ & $1$ & $-\frac{3}{4}$ & $-\frac{3}{4}$ & $0$ & $\frac{\sqrt{2}}{4}$ & $-\frac{\sqrt{2}}{4}$ & $0$ & $0$ & $\frac{\sqrt{3}}{8}$ & $-\frac{\sqrt{3}}{8}$ & $0$
    \\
    $\hat{G}_{x}^{(2)}$ & $-\frac{1}{4}$ & $-\frac{3}{4}$ & $0$ & $-\frac{3}{4}$ & $-\frac{3}{4}$ & $1$ & $-\frac{1}{4}$ & $-\frac{1}{4}$ & $\frac{\sqrt{2}}{4}$ & $-\frac{\sqrt{2}}{8}$ & $\frac{\sqrt{6}}{8}$ & $\frac{\sqrt{3}}{8}$ & $0$ & $-\frac{\sqrt{3}}{8}$ & $\frac{\sqrt{3}}{8}$
    \\
    $\hat{G}_{y}^{(2)}$ & $-\frac{1}{4}$ & $-\frac{3}{4}$ & $-\frac{3}{4}$ & $0$ & $-\frac{3}{4}$ & $-\frac{1}{4}$ & $1$ & $-\frac{1}{4}$ & $\frac{\sqrt{2}}{4}$ & $-\frac{\sqrt{2}}{8}$ & $-\frac{\sqrt{6}}{8}$ & $-\frac{\sqrt{3}}{8}$ & $\frac{\sqrt{3}}{8}$ & $0$ & $-\frac{\sqrt{3}}{8}$
    \\
    $\hat{G}_{z}^{(2)}$ & $-1$ & $0$ & $-\frac{3}{4}$ & $-\frac{3}{4}$ & $0$ & $-\frac{1}{4}$ & $-\frac{1}{4}$ & $1$ & $\frac{\sqrt{2}}{4}$ & $\frac{\sqrt{2}}{4}$ & $0$ & $0$ & $-\frac{\sqrt{3}}{8}$ & $\frac{\sqrt{3}}{8}$ & $0$
    \\
    \hline\hline
    \multicolumn{16}{l}{$J=3$ multipole flcutuation}
    \\
    \hline\hline
    $\hat{\Lambda}$ & $\check{Q}_{0}^{(3)}$ & $\check{Q}_{u}^{(3)}$ & $\check{Q}_{v}^{(3)}$ & $\check{Q}_{yz}^{(3)}$ & $\check{Q}_{zx}^{(3)}$ & $\check{Q}_{xy}^{(3)}$ 
    \\
    \hline
    $\hat{Q}_{0}^{(3)}$ & $1$ & $1$ & $1$ & $1$ & $1$ & $1$ 
    \\
    $\hat{Q}_{u}^{(3)}$ & $1$ & $1$ & $-1$ & $-1$ & $-1$ & $-1$ 
    \\
    $\hat{Q}_{v}^{(3)}$ & $1$ & $-1$ & $1$ & $-1$ & $-1$ & $-1$ 
    \\
    $\hat{Q}_{yz}^{(3)}$ & $1$ & $-1$ & $-1$ & $1$ & $-1$ & $-1$ 
    \\
    $\hat{Q}_{zx}^{(3)}$ & $1$ & $-1$ & $-1$ & $-1$ & $1$ & $-1$ 
    \\
    $\hat{Q}_{xy}^{(3)}$ & $1$ & $-1$ & $-1$ & $-1$ & $-1$ & $1$ 
    \\
  \end{longtable*}
\end{center}

\begin{center}
  \begin{longtable*}{lccccccccccccccc}
    \caption{
      The Cooper pairing $\hat{\mathcal{H}}_{\mathrm{eff}}$ in Eq.~(\ref{eq:cooper_pairing}) induced  by local M/MT-type multipole fluctuations $X^{(J)}$ in $p$-orbital space.
    }
    \label{tab:Pairing_p_j_M}
    \\
    
    \hline\hline
    \multicolumn{16}{l}
    {$J=1$ multipole fluctuation}
    \\
    \hline\hline
    $\hat{\Lambda}$ & $\check{Q}_{0}^{(1)}$ & 
    \\
    \endfirsthead
    
    \multicolumn{16}{c}
    {{\tablename\ \thetable{} -- continued from previous page}}
    \\
    \endhead
    
    \multicolumn{16}{c}{{Continued on next page}}
    \\ 
    \endfoot
    
    \hline\hline
    \endlastfoot

    \hline
    $\hat{M}_{1m}^{(1)}$ & $-2$
    \\
    \hline\hline
    \multicolumn{16}{l}{$J=2$ multipole fluctuation}
    \\
    \hline\hline
    $\hat{\Lambda}$ & $\check{Q}_{u}^{(2)}$ & $\check{Q}_{v}^{(2)}$ & $\check{Q}_{yz}^{(2)}$ & $\check{Q}_{zx}^{(2)}$ & $\check{Q}_{xy}^{(2)}$ & $\check{G}_{x}^{(2)}$ & $\check{G}_{y}^{(2)}$ & $\check{G}_{z}^{(2)}$ & $\check{Q}_{0}^{(1)}\check{Q}_{0}^{(3)}$ & $\check{Q}_{0}^{(1)}\check{Q}_{u}^{(3)}$ & $\check{Q}_{0}^{(1)}\check{Q}_{v}^{(3)}$ & $\check{Q}_{u}^{(2)}\check{Q}_{v}^{(2)}$ & $\check{Q}_{yz}^{(2)}\check{G}_{x}^{(2)}$ & $\check{Q}_{zx}^{(2)}G_{y}^{(2)}$ & $\check{Q}_{xy}^{(2)}\check{G}_{z}^{(2)}$ 
    \\
    \hline
    $\hat{M}_{x}^{(2)}$ & $-\frac{1}{4}$ & $-\frac{3}{4}$ & $0$ & $-\frac{3}{4}$ & $-\frac{3}{4}$ & $1$ & $-\frac{1}{4}$ & $-\frac{1}{4}$ & $-\frac{\sqrt{2}}{4}$ & $\frac{\sqrt{2}}{8}$ & $-\frac{\sqrt{6}}{8}$ & $\frac{\sqrt{3}}{8}$ & $0$ & $-\frac{\sqrt{3}}{8}$ & $\frac{\sqrt{3}}{8}$
    \\
    $\hat{M}_{y}^{(2)}$ & $-\frac{1}{4}$ & $-\frac{3}{4}$ & $-\frac{3}{4}$ & $0$ & $-\frac{3}{4}$ & $-\frac{1}{4}$ & $1$ & $-\frac{1}{4}$ & $-\frac{\sqrt{2}}{4}$ & $\frac{\sqrt{2}}{8}$ & $\frac{\sqrt{6}}{8}$ & $-\frac{\sqrt{3}}{8}$ & $\frac{\sqrt{3}}{8}$ & $0$ & $-\frac{\sqrt{3}}{8}$ 
    \\
    $\hat{M}_{z}^{(2)}$ & $-1$ & $0$ & $-\frac{3}{4}$ & $-\frac{3}{4}$ & $0$ & $-\frac{1}{4}$ & $-\frac{1}{4}$ & $1$ & $-\frac{\sqrt{2}}{4}$ & $-\frac{\sqrt{2}}{4}$ & $0$ & $0$ & $-\frac{\sqrt{3}}{8}$ & $\frac{\sqrt{3}}{8}$ & $0$
    \\
    $\hat{T}_{u}^{(2)}$ & $1$ & $0$ & $-\frac{3}{4}$ & $-\frac{3}{4}$ & $0$ & $-\frac{1}{4}$ & $-\frac{1}{4}$ & $-1$ & $-\frac{\sqrt{2}}{4}$ & $-\frac{\sqrt{2}}{4}$ & $0$ & $0$ & $-\frac{\sqrt{3}}{8}$ & $\frac{\sqrt{3}}{8}$ & $0$
    \\
    $\hat{T}_{v}^{(2)}$ & $0$ & $1$ & $-\frac{1}{4}$ & $-\frac{1}{4}$ & $-1$ & $-\frac{3}{4}$ & $-\frac{3}{4}$ & $0$ & $-\frac{\sqrt{2}}{4}$ & $\frac{\sqrt{2}}{4}$ & $0$ & $0$ & $\frac{\sqrt{3}}{8}$ & $-\frac{\sqrt{3}}{8}$ & $0$
    \\
    $\hat{T}_{yz}^{(2)}$ & $-\frac{3}{4}$ & $-\frac{1}{4}$ & $1$ & $-\frac{1}{4}$ & $-\frac{1}{4}$ & $0$ & $-\frac{3}{4}$ & $-\frac{3}{4}$ & $-\frac{\sqrt{2}}{4}$ & $-\frac{\sqrt{2}}{8}$ & $\frac{\sqrt{6}}{8}$ & $-\frac{\sqrt{3}}{8}$ & $0$ & $\frac{\sqrt{3}}{8}$ & $-\frac{\sqrt{3}}{8}$
    \\
    $\hat{T}_{zx}^{(2)}$ & $-\frac{3}{4}$ & $-\frac{1}{4}$ & $-\frac{1}{4}$ & $1$ & $-\frac{1}{4}$ & $-\frac{3}{4}$ & $0$ & $-\frac{3}{4}$ & $-\frac{\sqrt{2}}{4}$ & $-\frac{\sqrt{2}}{8}$ & $-\frac{\sqrt{6}}{8}$ & $\frac{\sqrt{3}}{8}$ & $-\frac{\sqrt{3}}{8}$ & $0$ & $\frac{\sqrt{3}}{8}$
    \\
    $\hat{T}_{xy}^{(2)}$ & $0$ & $-1$ & $-\frac{1}{4}$ & $-\frac{1}{4}$ & $1$ & $-\frac{3}{4}$ & $-\frac{3}{4}$ & $0$ & $-\frac{\sqrt{2}}{4}$ & $\frac{\sqrt{2}}{4}$ & $0$ & $0$ & $\frac{\sqrt{3}}{8}$ & $-\frac{\sqrt{3}}{8}$ & $0$
    \\
    \hline\hline
    \multicolumn{16}{l}{$J=3$ multipole flcutuation}
    \\
    \hline\hline
    $\hat{\Lambda}$ & $\check{Q}_{0}^{(3)}$ & $\check{Q}_{u}^{(3)}$ & $\check{Q}_{v}^{(3)}$ & $\check{Q}_{yz}^{(3)}$ & $\check{Q}_{zx}^{(3)}$ & $\check{Q}_{xy}^{(3)}$ & & $\check{Q}_{0}^{(3)}\check{Q}_{u}^{(3)}$ & & $\check{Q}_{0}^{(3)}\check{Q}_{v}^{(3)}$ & & $\check{Q}_{u}^{(3)}\check{Q}_{v}^{(3)}$ 
    \\
    \hline
    $\hat{M}_{x}^{(3)}$ & $-1$ & $\frac{1}{5}$ & $-\frac{3}{5}$ & $\frac{3}{5}$ & $-\frac{3}{5}$ & $-\frac{3}{5}$ & & $-\frac{1}{5}$ & & $\frac{\sqrt{3}}{5}$ & & $\frac{\sqrt{3}}{5}$ 
    \\
    $\hat{M}_{y}^{(3)}$ & $-1$ & $\frac{1}{5}$ & $-\frac{3}{5}$ & $-\frac{3}{5}$ & $\frac{3}{5}$ & $-\frac{3}{5}$ & & $-\frac{1}{5}$ & & $-\frac{\sqrt{3}}{5}$ & & $-\frac{\sqrt{3}}{5}$ 
    \\
    $\hat{M}_{z}^{(3)}$ & $-1$ & $-1$ & $\frac{3}{5}$ & $-\frac{3}{5}$ & $-\frac{3}{5}$ & $\frac{3}{5}$ & & $\frac{2}{5}$ & & $0$ & & $0$ 
    \\
    $\hat{M}_{xyz}^{(3)}$ & $-1$ & $1$ & $1$ & $-1$ & $-1$ & $-1$ & & $0$ & & $0$ & & $0$ 
    \\
    $\hat{M}_{x}^{\alpha(3)}$ & $-1$ & $-\frac{7}{10}$ & $-\frac{9}{10}$ & $-\frac{3}{5}$ & $\frac{3}{5}$ & $\frac{3}{5}$ & & $\frac{1}{5}$ & & $-\frac{\sqrt{3}}{5}$ & & $\frac{\sqrt{3}}{20}$ 
    \\
    $\hat{M}_{y}^{\alpha(3)}$ & $-1$ & $-\frac{7}{10}$ & $-\frac{9}{10}$ & $\frac{3}{5}$ & $-\frac{3}{5}$ & $\frac{3}{5}$ & & $\frac{1}{5}$ & & $\frac{\sqrt{3}}{5}$ & & $-\frac{\sqrt{3}}{20}$ 
    \\
    $\hat{M}_{z}^{\alpha(3)}$ & $-1$ & $-1$ & $-\frac{3}{5}$ & $\frac{3}{5}$ & $\frac{3}{5}$ & $-\frac{3}{5}$ & & $-\frac{2}{5}$ & & $0$ & & $0$ 
    \\
    $\hat{M}_{x}^{\beta(3)}$ & $-1$ & $-\frac{1}{2}$ & $\frac{1}{2}$ & $1$ & $-1$ & $-1$ & & $0$ & & $0$ & & $-\frac{\sqrt{3}}{4}$ 
    \\
    $\hat{M}_{y}^{\beta(3)}$ & $-1$ & $-\frac{1}{2}$ & $\frac{1}{2}$ & $-1$ & $1$ & $-1$ & & $0$ & & $0$ & & $\frac{\sqrt{3}}{4}$ 
    \\
    $\hat{M}_{z}^{\beta(3)}$ & $-1$ & $1$ & $-1$ & $-1$ & $-1$ & $1$ & & $0$ & & $0$ & & $0$ 
    \\
  \end{longtable*}
\end{center}

\begin{center}
  \begin{longtable*}{lccccccccccccccc}
    \caption{
      The Cooper pairing $\hat{\mathcal{H}}_{\mathrm{eff}}$ in Eq.~(\ref{eq:cooper_pairing}) induced  by local E/ET-type multipole fluctuations $X^{(J)}$ in $sp$-orbital space.
    }
    \label{tab:Pairing_sp_j_E}
    \\
    
    \hline\hline
    \multicolumn{16}{l}
    {$J=1$ multipole fluctuation}
    \\
    \hline\hline
    $\hat{\Lambda}$ & $\check{Q}_{x}^{(1)}$ & $\check{Q}_{y}^{(1)}$ & $\check{Q}_{z}^{(1)}$ & & $\check{G}_{0}^{(1)}$ & & $\check{Q}_{0}^{s}\check{Q}_{0}^{(1)}$ 
    \\
    \endfirsthead
    
    \multicolumn{16}{c}
    {{\tablename\ \thetable{} -- continued from previous page}}
    \\
    \endhead
    
    \multicolumn{16}{c}{{Continued on next page}}
    \\ 
    \endfoot
    
    \hline\hline
    \endlastfoot

    \hline
    $\hat{Q}_{x}^{(1)}$ & $1$ & $-1$ & $-1$ & & $-1$ & & $\frac{1}{2}$ 
    \\
    $\hat{Q}_{y}^{(1)}$ & $-1$ & $1$ & $-1$ & & $-1$ & & $\frac{1}{2}$ 
    \\
    $\hat{Q}_{z}^{(1)}$ & $-1$ & $-1$ & $1$ & & $-1$ & & $\frac{1}{2}$ 
    \\
    $\hat{G}_{0}^{(1)}$ & $-1$ & $-1$ & $-1$ & & $1$ & & $\frac{1}{2}$ 
    \\
    \hline\hline
    \multicolumn{16}{l}{$J=2$ multipole fluctuation}
    \\
    \hline\hline
    $\hat{\Lambda}$ & $\check{Q}_{x}^{(2)}$ & $\check{Q}_{y}^{(2)}$ & $\check{Q}_{z}^{(2)}$ & $\check{G}_{u}^{(2)}$ & $\check{G}_{v}^{(2)}$ & $\check{G}_{yz}^{(2)}$ & $\check{G}_{zx}^{(2)}$ & $\check{G}_{xy}^{(2)}$ & $\check{Q}_{0}^{s}\check{Q}_{0}^{(3)}$ & $\check{Q}_{0}^{s}\check{Q}_{u}^{(3)}$ & $\check{Q}_{0}^{s}\check{Q}_{v}^{(3)}$ & $\check{Q}_{x}^{(2)}\check{G}_{yz}^{(2)}$ & $\check{Q}_{y}^{(2)}\check{G}_{zx}^{(2)}$ & $\check{Q}_{z}^{(2)}\check{G}_{xy}^{(2)}$ & $\check{G}_{u}^{(2)}\check{G}_{v}^{(2)}$ 
    \\
    \hline
    $\hat{Q}_{x}^{(2)}$ & $1$ & $-\frac{1}{4}$ & $-\frac{1}{4}$ & $-\frac{1}{4}$ & $-\frac{3}{4}$ & $0$ & $-\frac{3}{4}$ & $-\frac{3}{4}$ & $\frac{\sqrt{2}}{4}$ & $-\frac{\sqrt{2}}{8}$ & $\frac{\sqrt{6}}{8}$ & $0$ & $-\frac{\sqrt{3}}{8}$ & $\frac{\sqrt{3}}{8}$ & $\frac{\sqrt{3}}{8}$ 
    \\
    $\hat{Q}_{y}^{(2)}$ & $-\frac{1}{4}$ & $1$ & $-\frac{1}{4}$ & $-\frac{1}{4}$ & $-\frac{3}{4}$ & $-\frac{3}{4}$ & $0$ & $-\frac{3}{4}$ & $\frac{\sqrt{2}}{4}$ & $-\frac{\sqrt{2}}{8}$ & $-\frac{\sqrt{6}}{8}$ & $\frac{\sqrt{3}}{8}$ & $0$ & $-\frac{\sqrt{3}}{8}$ & $-\frac{\sqrt{3}}{8}$ 
    \\
    $\hat{Q}_{z}^{(2)}$ & $-\frac{1}{4}$ & $-\frac{1}{4}$ & $1$ & $-1$ & $0$ & $-\frac{3}{4}$ & $-\frac{3}{4}$ & $0$ & $\frac{\sqrt{2}}{4}$ & $\frac{\sqrt{2}}{4}$ & $0$ & $-\frac{\sqrt{3}}{8}$ & $\frac{\sqrt{3}}{8}$ & $0$ & $0$ 
    \\
    $\hat{G}_{u}^{(2)}$ & $-\frac{1}{4}$ & $-\frac{1}{4}$ & $-1$ & $1$ & $0$ & $-\frac{3}{4}$ & $-\frac{3}{4}$ & $0$ & $\frac{\sqrt{2}}{4}$ & $\frac{\sqrt{2}}{4}$ & $0$ & $-\frac{\sqrt{3}}{8}$ & $\frac{\sqrt{3}}{8}$ & $0$ & $0$ 
    \\
    $\hat{G}_{v}^{(2)}$ & $-\frac{3}{4}$ & $-\frac{3}{4}$ & $0$ & $0$ & $1$ & $-\frac{1}{4}$ & $-\frac{1}{4}$ & $-1$ & $\frac{\sqrt{2}}{4}$ & $-\frac{\sqrt{2}}{4}$ & $0$ & $\frac{\sqrt{3}}{8}$ & $-\frac{\sqrt{3}}{8}$ & $0$ & $0$ 
    \\
    $\hat{G}_{yz}^{(2)}$ & $0$ & $-\frac{3}{4}$ & $-\frac{3}{4}$ & $-\frac{3}{4}$ & $-\frac{1}{4}$ & $1$ & $-\frac{1}{4}$ & $-\frac{1}{4}$ & $\frac{\sqrt{2}}{4}$ & $\frac{\sqrt{2}}{8}$ & $-\frac{\sqrt{6}}{8}$ & $0$ & $\frac{\sqrt{3}}{8}$ & $-\frac{\sqrt{3}}{8}$ & $-\frac{\sqrt{3}}{8}$ 
    \\
    $\hat{G}_{zx}^{(2)}$ & $-\frac{3}{4}$ & $0$ & $-\frac{3}{4}$ & $-\frac{3}{4}$ & $-\frac{1}{4}$ & $-\frac{1}{4}$ & $1$ & $-\frac{1}{4}$ & $\frac{\sqrt{2}}{4}$ & $\frac{\sqrt{2}}{8}$ & $\frac{\sqrt{6}}{8}$ & $-\frac{\sqrt{3}}{8}$ & $0$ & $\frac{\sqrt{3}}{8}$ & $\frac{\sqrt{3}}{8}$ 
    \\
    $\hat{G}_{xy}^{(2)}$ & $-\frac{3}{4}$ & $-\frac{3}{4}$ & $0$ & $0$ & $-1$ & $-\frac{1}{4}$ & $-\frac{1}{4}$ & $1$ & $\frac{\sqrt{2}}{4}$ & $-\frac{\sqrt{2}}{4}$ & $0$ & $\frac{\sqrt{3}}{8}$ & $-\frac{\sqrt{3}}{8}$ & $0$ & $0$ 
    \\
  \end{longtable*}
\end{center}

\begin{center}
  \begin{longtable*}{lccccccccccccccc}
    \caption{
      The Cooper pairing $\hat{\mathcal{H}}_{\mathrm{eff}}$ in Eq.~(\ref{eq:cooper_pairing}) induced  by local M/MT-type multipole fluctuations $X^{(J)}$ in $sp$-orbital space.
    }
    \label{tab:Pairing_sp_j_M}
    \\
    
    \hline\hline
    \multicolumn{16}{l}
    {$J=1$ multipole fluctuation}
    \\
    \hline\hline
    $\hat{\Lambda}$ & $\check{Q}_{x}^{(1)}$ & $\check{Q}_{y}^{(1)}$ & $\check{Q}_{z}^{(1)}$ & & $\check{G}_{0}^{(1)}$ & & $\check{Q}_{0}^{s}\check{Q}_{0}^{(1)}$ 
    \\
    \endfirsthead
    
    \multicolumn{16}{c}
    {{\tablename\ \thetable{} -- continued from previous page}}
    \\
    \endhead
    
    \multicolumn{16}{c}{{Continued on next page}}
    \\ 
    \endfoot
    
    \hline\hline
    \endlastfoot

    \hline
    $\hat{T}_{x}^{(1)}$ & $1$ & $-1$ & $-1$ & & $-1$ & & $-\frac{1}{2}$ 
    \\
    $\hat{T}_{y}^{(1)}$ & $-1$ & $1$ & $-1$ & & $-1$ & & $-\frac{1}{2}$ 
    \\
    $\hat{T}_{z}^{(1)}$ & $-1$ & $-1$ & $1$ & & $-1$ & & $-\frac{1}{2}$ 
    \\
    $\hat{M}_{0}^{(1)}$ & $-1$ & $-1$ & $-1$ & & $1$ & & $-\frac{1}{2}$ 
    \\
    \hline\hline
    \multicolumn{16}{l}{$J=2$ multipole fluctuation}
    \\
    \hline\hline
    $\hat{\Lambda}$ & $\check{Q}_{x}^{(2)}$ & $\check{Q}_{y}^{(2)}$ & $\check{Q}_{z}^{(2)}$ & $\check{G}_{u}^{(2)}$ & $\check{G}_{v}^{(2)}$ & $\check{G}_{yz}^{(2)}$ & $\check{G}_{zx}^{(2)}$ & $\check{G}_{xy}^{(2)}$ & $\check{Q}_{0}^{s}\check{Q}_{0}^{(3)}$ & $\check{Q}_{0}^{s}\check{Q}_{u}^{(3)}$ & $\check{Q}_{0}^{s}\check{Q}_{v}^{(3)}$ & $\check{Q}_{x}^{(2)}\check{G}_{yz}^{(2)}$ & $\check{Q}_{y}^{(2)}\check{G}_{zx}^{(2)}$ & $\check{Q}_{z}^{(2)}\check{G}_{xy}^{(2)}$ & $\check{G}_{u}^{(2)}\check{G}_{v}^{(2)}$ 
    \\
    \hline
    $\hat{T}_{x}^{(2)}$ & $1$ & $-\frac{1}{4}$ & $-\frac{1}{4}$ & $-\frac{1}{4}$ & $-\frac{3}{4}$ & $0$ & $-\frac{3}{4}$ & $-\frac{3}{4}$ & $-\frac{\sqrt{2}}{4}$ & $\frac{\sqrt{2}}{8}$ & $-\frac{\sqrt{6}}{8}$ & $0$ & $-\frac{\sqrt{3}}{8}$ & $\frac{\sqrt{3}}{8}$ & $\frac{\sqrt{3}}{8}$ 
    \\
    $\hat{T}_{y}^{(2)}$ & $-\frac{1}{4}$ & $1$ & $-\frac{1}{4}$ & $-\frac{1}{4}$ & $-\frac{3}{4}$ & $-\frac{3}{4}$ & $0$ & $-\frac{3}{4}$ & $-\frac{\sqrt{2}}{4}$ & $\frac{\sqrt{2}}{8}$ & $\frac{\sqrt{6}}{8}$ & $\frac{\sqrt{3}}{8}$ & $0$ & $-\frac{\sqrt{3}}{8}$ & $-\frac{\sqrt{3}}{8}$ 
    \\
    $\hat{T}_{z}^{(2)}$ & $-\frac{1}{4}$ & $-\frac{1}{4}$ & $1$ & $-1$ & $0$ & $-\frac{3}{4}$ & $-\frac{3}{4}$ & $0$ & $-\frac{\sqrt{2}}{4}$ & $-\frac{\sqrt{2}}{4}$ & $0$ & $-\frac{\sqrt{3}}{8}$ & $\frac{\sqrt{3}}{8}$ & $0$ & $0$ 
    \\
    $\hat{M}_{u}^{(2)}$ & $-\frac{1}{4}$ & $-\frac{1}{4}$ & $-1$ & $1$ & $0$ & $-\frac{3}{4}$ & $-\frac{3}{4}$ & $0$ & $-\frac{\sqrt{2}}{4}$ & $-\frac{\sqrt{2}}{4}$ & $0$ & $-\frac{\sqrt{3}}{8}$ & $\frac{\sqrt{3}}{8}$ & $0$ & $0$ 
    \\
    $\hat{M}_{v}^{(2)}$ & $-\frac{3}{4}$ & $-\frac{3}{4}$ & $0$ & $0$ & $1$ & $-\frac{1}{4}$ & $-\frac{1}{4}$ & $-1$ & $-\frac{\sqrt{2}}{4}$ & $\frac{\sqrt{2}}{4}$ & $0$ & $\frac{\sqrt{3}}{8}$ & $-\frac{\sqrt{3}}{8}$ & $0$ & $0$ 
    \\
    $\hat{M}_{yz}^{(2)}$ & $0$ & $-\frac{3}{4}$ & $-\frac{3}{4}$ & $-\frac{3}{4}$ & $-\frac{1}{4}$ & $1$ & $-\frac{1}{4}$ & $-\frac{1}{4}$ & $-\frac{\sqrt{2}}{4}$ & $-\frac{\sqrt{2}}{8}$ & $\frac{\sqrt{6}}{8}$ & $0$ & $\frac{\sqrt{3}}{8}$ & $-\frac{\sqrt{3}}{8}$ & $-\frac{\sqrt{3}}{8}$ 
    \\
    $\hat{M}_{zx}^{(2)}$ & $-\frac{3}{4}$ & $0$ & $-\frac{3}{4}$ & $-\frac{3}{4}$ & $-\frac{1}{4}$ & $-\frac{1}{4}$ & $1$ & $-\frac{1}{4}$ & $-\frac{\sqrt{2}}{4}$ & $-\frac{\sqrt{2}}{8}$ & $-\frac{\sqrt{6}}{8}$ & $-\frac{\sqrt{3}}{8}$ & $0$ & $\frac{\sqrt{3}}{8}$ & $\frac{\sqrt{3}}{8}$ 
    \\
    $\hat{M}_{xy}^{(2)}$ & $-\frac{3}{4}$ & $-\frac{3}{4}$ & $0$ & $0$ & $-1$ & $-\frac{1}{4}$ & $-\frac{1}{4}$ & $1$ & $-\frac{\sqrt{2}}{4}$ & $\frac{\sqrt{2}}{4}$ & $0$ & $\frac{\sqrt{3}}{8}$ & $-\frac{\sqrt{3}}{8}$ & $0$ & $0$ 
    \\
  \end{longtable*}
\end{center}

\section{Molecular field dependence of pair potential}
\label{app:mf}

\begin{figure}[tbp]
  \centering
  \includegraphics[width=\linewidth]{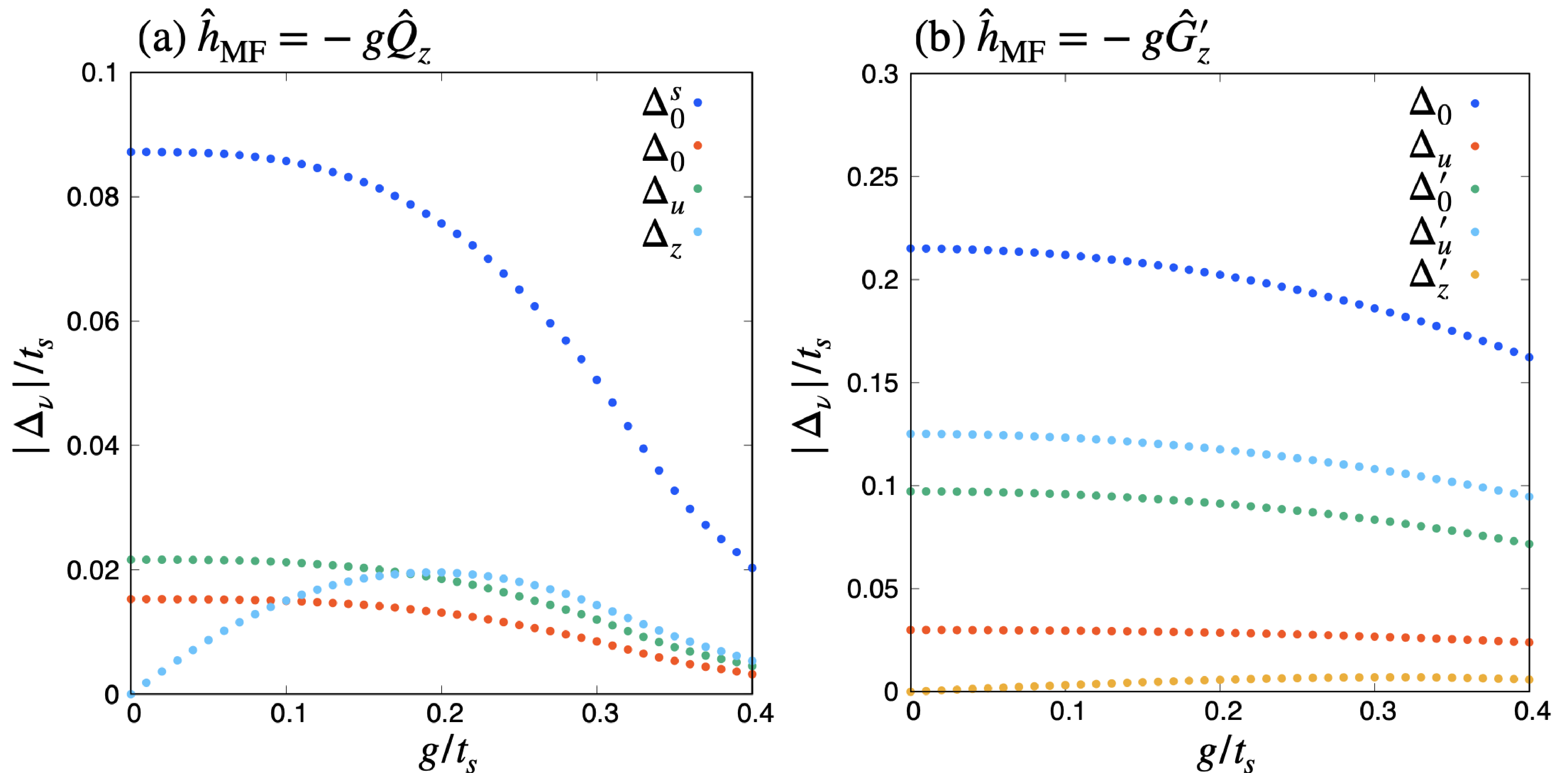}
  \caption{
    The molecular field strength dependence of the amplitude of pair potentials for (a) the polar field and (b) the axial field with density $n=0.8$ at zero temperature.
  }
  \label{fig:pair_mf}
\end{figure}

In this appendix, we show the molecular field dependence of the pair potential at zero temperature in Sec.~\ref{sec:application} in the main text.
In the weak molecular-field region, $\Delta_{z}$ grows as the molecular field increases, as shown in Fig.~\ref{fig:pair_mf}(a).
We also found that $\Delta_{z}$ has a peak at a certain value of $g$, and the SC state is no longer stable in the region for sufficiently large $g$.
A similar behavior can be seen in the case of the axial field, as shown in Fig.~\ref{fig:pair_mf}(b).

\begin{figure}[tbp]
  \centering
  \includegraphics[width=\linewidth]{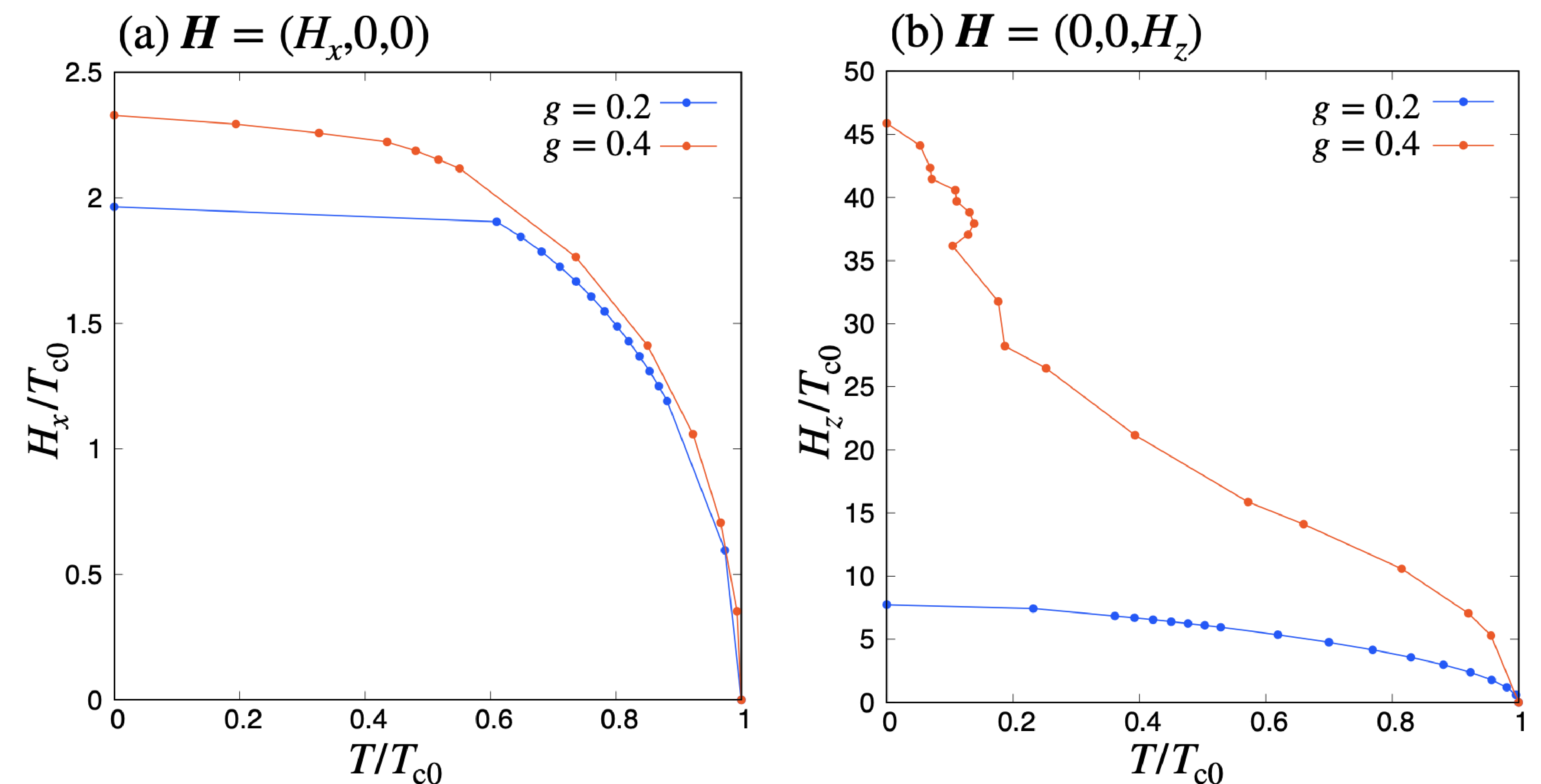}
  \caption{
    The phase boundary $(T_{\mathrm{c}},H_{\mathrm{c}})$ for (a) in-plane and (b) perpendicular magentic field under the polar molecular field at $n=0.8$.
    $T_{\mathrm{c0}}$ is the critical temperature without the magnetic field.
  }
  \label{fig:phase_mf}
\end{figure}

Figure~\ref{fig:phase_mf} shows the phase boundaries under the polar molecular field at $g=0.2$ and $0.4$.
For the in-plane magnetic field, the molecular field hardly change the phase boundary, as shown in Fig.~\ref{fig:phase_mf}(a).
On the other hand, for the perpendicular magnetic field in Fig.~\ref{fig:phase_mf}(b), the large molecular field tends to give the high critical magnetic field, which might be attributed to the larger antisymmetric spin-orbit interaction for a larger molecular field.

\bibliography{manuscript.bbl}

\end{document}